\documentclass[11pt]{article}

\usepackage{graphicx}
\usepackage{tabularx}
\usepackage{threeparttable}
\usepackage{multirow}
\usepackage{url}
\usepackage{amsmath}
\usepackage{comment}
\usepackage{nameref}

\usepackage{placeins}
\usepackage{authblk}
\usepackage{hyperref}
\usepackage{float}
\usepackage{caption}
\usepackage{subcaption}
\usepackage{amssymb}
\usepackage{amsthm}
\usepackage[english]{babel}
\usepackage{tikz}
\usepackage{lipsum,lmodern}
\usepackage[most]{tcolorbox}
\usepackage{newtxtext,newtxmath}
\usepackage{color}
\usepackage{bm}

\newcommand{\set}[1]{\left\{#1\right\}}

\title{\textbf{Modelling the spillover from online engagement to offline protest: stochastic dynamics and mean-field approximations on networks}}

\author[1]{Moyi Tian$^*$}
\author[2]{P. Jeffrey Brantingham}
\author[1]{Nancy Rodr\'iguez}

\affil[1]{Department of Applied Mathematics, University of Colorado, Boulder, CO 80309 United States}
\affil[2]{Department of Anthropology, University of California, Los Angeles, CA 90095 United States}
\affil[*]{\normalfont Corresponding author: moyi.tian@colorado.edu}

\begin{document}

\maketitle

\begin{abstract}
{Social media is transforming various aspects of offline life, from everyday decisions such as dining choices to the progression of conflicts. In this study, we propose a coupled modelling framework with an online social network layer to analyse how engagement on a specific topic spills over into offline protest activities. We develop a stochastic model and derive several mean-field models of varying complexity. These models allow us to estimate the reproductive number and anticipate when surges in activity are likely to occur. A key factor is the transmission rate between the online and offline domains; for offline outbursts to emerge, this rate must fall within a critical range, neither too low nor too high. Additionally, using synthetic networks, we examine how network structure influences the accuracy of these approximations. Our findings indicate that low-density networks need more complex approximations, whereas simpler models can effectively represent higher-density networks. When tested on two real-world networks, however, increased complexity did not enhance accuracy.

\vspace{0.5em}
\noindent\textit{Keywords:} networks, spillover effects, multilayer dynamics, protest modelling, critical transitions, mean-field approximation

\vspace{0.5em}
\noindent 2000 Math Subject Classification: 91D30, 37N25, 92D30, 34C60}
\end{abstract}

\section{Introduction}

The use of social media has expanded globally, reaching approximately two-thirds of the global population, and providing a significant segment of our population access to diverse ideas and opinions from individuals worldwide \cite{statista, DataReportal}. It is fair to say that social media has changed the world, from its effect on mental health to increased misinformation; its effects cannot be understated. Social media has also transformed how, when, and where conflicts occur \cite{proctor2021social}. A report by Mercy Corps \cite{proctor2021social} studies the effect of social media on spreading ethnic and sectarian tension. This study identifies three primary risks: the amplification of conflict drivers, the targeting of peace advocates, and the spillover of online conflict into the offline world. Of particular relevance to our work is the last, where digital tensions escalate into tangible outcomes such as protests and violence. A natural question arises, ``How is social media activity affecting uprisings?" In \cite{bastos2015tents}, the authors study the Granger causality between social media streams and onsite developments at the Indignados, Occupy, and Brazilian Vinegar protests. The authors found that contentious communication in social media predicted protest activity during the Indignados and Occupy movements. Moreover, they also found bidirectional Granger causality between online and onsite protests in the Occupy series. The role of social media has also been explored in the context of Ukraine's Euromaidan protests \cite{piechota2015role} and the 2011 Egyptian revolution \cite{kharroub2016social}.  

In recent years, several studies have developed mathematical frameworks to analyse the spatiotemporal dynamics of protest activities. In 1987, Burbeck et al. applied concepts from epidemiology to model rioting activity, though their model did not consider spatial spread \cite{burbeck1978dynamics}. This approach initially did not receive much attention, but gained traction in 2017 when Bonnasse-Gahot et al. applied it to model the spatiotemporal dynamics of the 2005 French riots, demonstrating the effectiveness of a non-local version of Burbeck's epidemiological framework in fitting empirical data \cite{bonnasse2018epidemiological}. In addition to the epidemiologically inspired models, a range of other modelling approaches have been developed to capture the dynamics of uprisings, incorporating different mechanisms and sources of influence. Berestycki and team \cite{Berestycki2010} introduced a model examining the interplay between the evolution of protest activity and social tension; this model was further analysed in \cite{berestycki2015model}. As another example, Davies et al. in \cite{davies2013mathematical} constructed a mathematical framework to analyse the London riots of 2011 and the corresponding policing strategies. This model incorporated factors such as the infectious nature of participation, distances traveled to riot locations, and the deterrent influence of law enforcement. 

While these earlier models provided important insights into the mechanisms of protest dynamics, they largely overlooked the growing influence of online social activities in shaping real-world events. Given that over 60$\%$ of the global population now uses social media \cite{statista, DataReportal}, that social media connections enhance exposure to offline events, and that the spread of information and tension through social media can also lead to activities in the physical world (offline activity), it is crucial to incorporate online dynamics into models of uprisings and vice versa. To address this, we extend the modelling framework proposed by Burbeck et al. by developing and evaluating a multilayer model that captures the interplay between the spread of tension in online social networks and the dynamics of social uprisings. In this framework, one layer represents the online network that governs information diffusion (tension), while the second layer models offline activities. Our approach builds upon recent developments in other studies, such as \cite{peng2021multilayer,eames2002modeling}. 

In particular, we construct a model that captures the coupled dynamics of online engagement with a contentious issue and corresponding offline protest using a compartmental framework \cite{Brauer2008}. Individuals transition between joint online-offline states, with transitions influenced by both social network structure and feedback between layers. The underlying online social network influences transitions from disengaged to engaged states, while only a fraction of engaged individuals take to the streets, protesting for a limited time. We also incorporate feedback: Increased offline protest activity drives further online engagement. We represent online-to-offline spillovers as a Markov process on the multilayer system and develop a range of mean-field approximations that reflect network structure to varying degrees. In simulations on both homogeneous and heterogeneous synthetic networks, we find that higher-order approximations, which capture edge-level dynamics, significantly improve performance in low-density networks but provide limited advantage in denser ones. This trend is less evident in empirical networks, suggesting that real-world structural features not included in our models may play a critical role in shaping these dynamics.

To better understand the conditions that drive offline protest outbreaks, we use the continuum models derived by these mean-field approximations and analyse the reproductive numbers, which quantify the potential for activity to spread. Our analysis highlights the importance of the transmission rate between online engagement and offline action: If this rate is too high, the reservoir of protest-susceptible individuals is quickly exhausted; if too low, online flare-ups fail to translate into offline mobilization. A critical range of this transmission rate is therefore essential for substantial offline activity.

{\it Outline}: Section \ref{sec:motivation} introduces our model and the motivation behind its development. In Section \ref{sec:fullymixed}, we explore the mean-field approximation under the fully-mixed assumption. Section \ref{sec:homo} presents two approximation methods for $k$-regular networks and compares their performance to the stochastic model. An analogous analysis for heterogeneous networks is provided in Section \ref{sec:hetero}. We conclude with a discussion in Section \ref{sec:conclusion}.

\section{Model motivation and background}\label{sec:motivation}

In this section, we introduce our two-layer model and explain its motivation. Section \ref{subsec:background} provides background context. Section \ref{subsec:Model_Sec} presents the overall structure of the model, with Section \ref{subsubsec:online} focusing on the online layer, where information and tension spread; Section \ref{subsubsec:offline} covering the offline layer, where actions such as boycotts or uprisings occur; and Section \ref{subsubsec:coupled} detailing the interactions between the two layers. Section \ref{subsec:stochastic} introduces the stochastic model that mathematically formalizes the coupled system. Our model captures the interplay between layers: Social tension spreads online and subsequently influences offline behaviour.

\subsection{Background and motivation}\label{subsec:background}

With the widespread adoption of social media worldwide, the boundaries between virtual and physical realities have become increasingly blurred. Event-based social networks have emerged, enabling individuals not only to interact online—as in traditional platforms—but also to coordinate and participate in offline activities \cite{liu2012event}. Researchers have shown that online social interactions can directly impact offline behaviour; for instance, forming new online connections has been associated with a significant $7\%$ increase in physical activity \cite{althoff2017online}.

At the same time, the role of social media in shaping mental health outcomes has drawn increasing attention. While the increasing use of social media has been linked to rising mental health issues and suicide rates \cite{twenge2018increases}, social media has also proven effective in suicide prevention, as highlighted by systematic reviews evaluating its positive interventions \cite{robinson2016social}.

Despite these findings, the connection between online activity and offline behaviour—particularly protest and collective action—remains underexplored. A notable exception is the work by Shao et al. \cite{shao2019new}, who introduced a multilayer network propagation model to characterize the coupling of online and offline communication channels. Meanwhile, a wide range of models have been developed to study information diffusion within online networks, drawing from epidemiological, threshold-based, independent cascade, and game-theoretic frameworks. Liu et al. \cite{liu2012event} provide a comprehensive survey of these and related models in the literature.

While models linking online and offline activities are beginning to emerge, a deeper understanding of the broader role of social tension in driving mass action remains essential. Social tension functions both as a signal and a catalyst for collective behaviour, influencing when and how protests arise \cite{Buzovsky2008}. Mass protest can be viewed as a form of conflict, which, as Cosner \cite{Cosner56} argues, plays a crucial role in regulating systems that are out of balance; when managed effectively, conflict can alleviate accumulated hostile emotions and thereby reduce social tension.

Although often perceived negatively, social tension can serve as an important signal of potential conflict \cite{Artemov2017}, and while effective conflict management can serve a valuable function, prevention remains preferable. It is therefore critical to monitor social tension across its developmental stages and implement systemic adjustments before it escalates into conflict. Researchers have identified distinct phases in the evolution of social tension: an early latent stage in which negative emotional states emerge unnoticed, followed by a more visible stage that can lead to conflict or protest activity \cite{Artemov2017}. 

Taken together, these points motivate our model, which captures the coupled dynamics of online and offline activity—each labelled by distinct states—and incorporates the idea that the two domains not only reinforce each other but that offline activity also serves as a means of discharging online tension. Our model provides a foundation for understanding, predicting, and ultimately managing real-world protest scenarios.

\subsection{Our model}\label{subsec:Model_Sec}

In this section, we present our model, beginning with separate discussions of the online and offline layers, followed by an explanation of how they are interconnected. We adopt a time scale corresponding to a single protest event, under which the assumption that individuals disengage from both online and offline activities afterward is reasonable.

\subsubsection{The information layer}\label{subsubsec:online}

Our opinions and judgments are shaped by the flow of information (and misinformation) and the influence of others. In the online layer (information layer), we model the contagion of engagement on a specific issue by considering online social contacts. This approach can be used to model the spread of social tension or hate speech.

We propose that engagement levels are influenced by the flow of information within a social network and awareness of offline events. To capture this dynamic, we consider a state model that operates within an online network where information (or misinformation) and opinions are exchanged. Our focus is on scenarios where, for a specific topic, individuals can be uninterested ($U$), engaged ($E$), or disengaged ($D$) after previously being engaged in the online sphere. Uninterested individuals do not post about or engage in any threads related to a given topic. An engaged individual actively participates in online conversations about the topic, while a disengaged individual has lost interest after prior engagement.

Building on this conceptual framework, our model incorporates the well-established idea that individuals’ opinions and engagement levels are shaped by their social connections \cite{das2014modeling, chen2011influence}. Specifically, we quantify how uninterested individuals become engaged through exposure to active participants within their online network. In our model, we assume that an uninterested individual becomes engaged when socially connected (online) to engaged individuals, with a transmission rate, $\tau$. Therefore, if an individual has $k$ social contacts who are engaged in a topic (actively posting on online social platforms), the rate at which this uninterested individual becomes engaged due to the influence of others is given by 
$k\tau$. This parameter varies depending on the society and the topic. For instance, societies that encourage civic engagement may have higher $\tau$ values. Conversely, in contexts where misinformation drives tension, societies that educate against misinformation may have a lower $\tau$ for that specific topic. Note that in a fully-mixed population, where every individual is assumed to be connected with everyone else, we use a population-level transmission rate, $\beta = N\tau$, with $N$ denoting the population size, to present this case. As information about an issue spreads, individuals may transition to the engaged category ($E$). After a certain period, engaged individuals may choose to disengage, moving to the disengaged category ($D$), with a disengagement rate denoted by $\gamma_i$ (where the subscript $i$ indicates a parameter specific to the information layer). The top layer of Figure \ref{fig:diag} illustrates the state model in the online layer.  

\subsubsection{The physical layer}\label{subsubsec:offline}

\begin{figure}[!htb]
    \centering
    \subcaptionbox{Diagram of the model layers\label{fig:diag}}{\includegraphics[width=.55\textwidth]{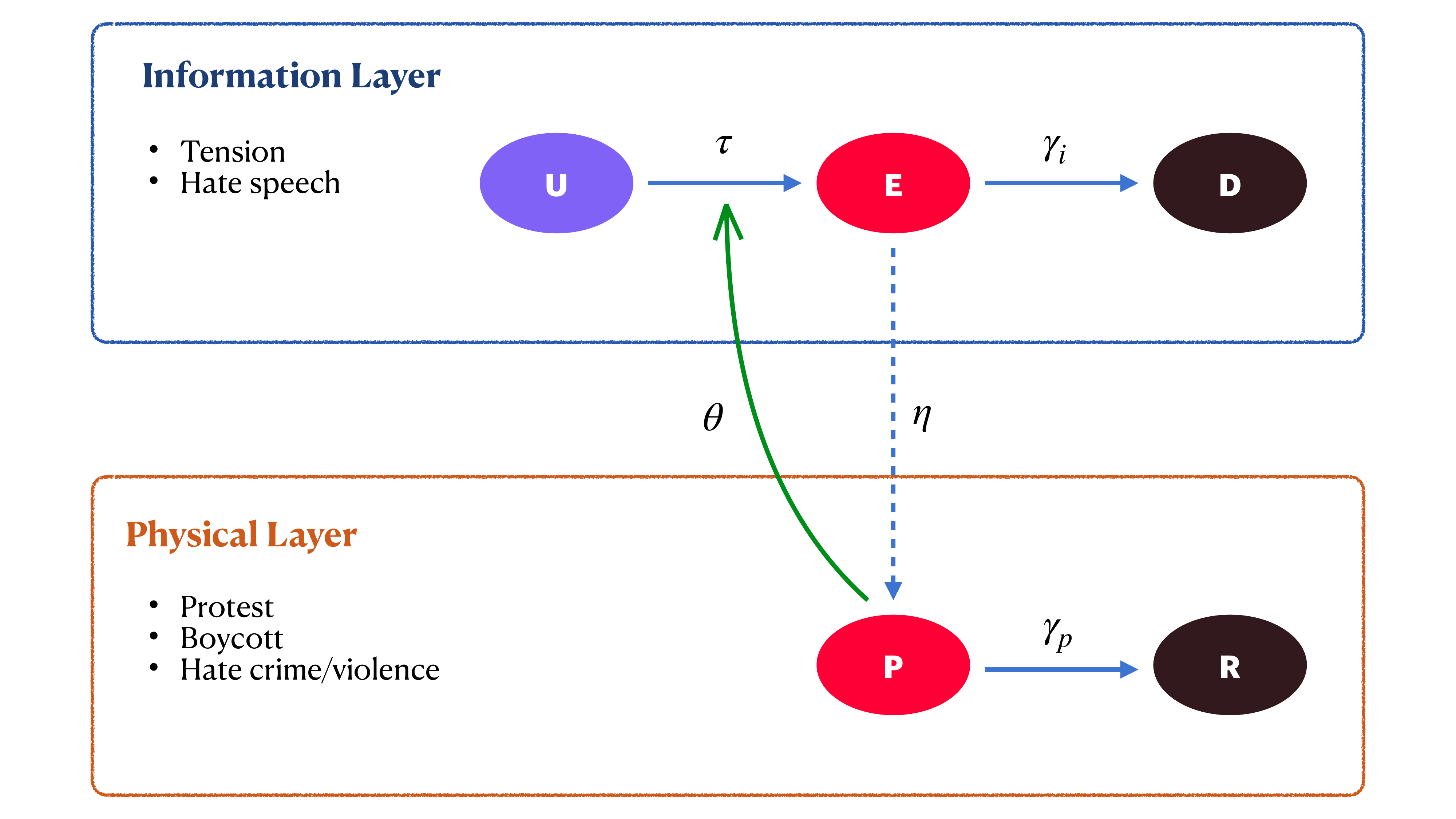}}
    \hfill
    \subcaptionbox{Markov chain states\label{fig:markov}}{\includegraphics[width=.42\textwidth]{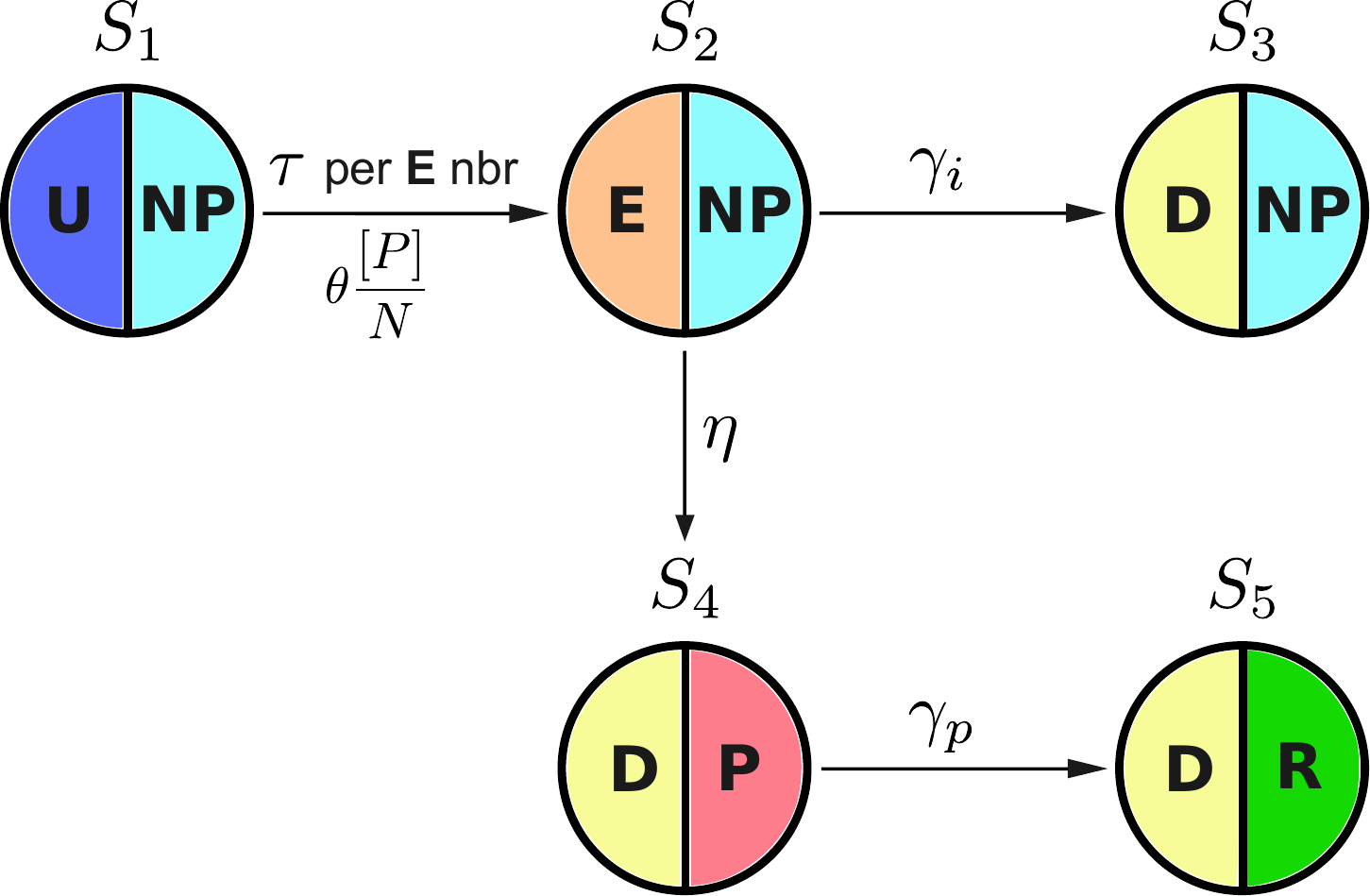}}
    \caption{Model schematics: (a) illustrates the online and offline layers, where the default offline state $NP$ is omitted, while (b) presents the corresponding mathematical model in terms of five Markovian states. Each state is composed of an online status (represented by the left half-circle) and an offline status (represented by the right half-circle).}
    \label{fig:diagrams}
\end{figure}

In the physical world, engaged individuals may choose to participate in offline activities, which can range from physically attending a protest to boycotting a product. Regardless of the form, these actions have real-world consequences. We assume that the default state prior to protest participation is a non-protesting state, denoted as $NP$. The relevant offline state in this context is that of protesters, denoted as $P$. After a certain amount of time, individuals involved in protesting may discontinue their physical engagement, transitioning to the state labelled $R$ (recovered from protesting). The rate of this transition is denoted by $\gamma_p$, where the subscript $p$ indicates that the parameter pertains to the physical layer. For simplicity, we assume that this rate is uniform across all individuals. However, we acknowledge this as a modelling simplification, as real-world offline participation can vary significantly due to constraints such as work, family responsibilities, or personal circumstances. The bottom layer of Figure \ref{fig:diag} illustrates the state model in the offline layer.

\subsubsection{The coupling of layers}\label{subsubsec:coupled}

We assume that only a fraction, $\eta$, of the individuals engaged online are willing or able to participate in offline activities. This represents the online-to-offline effect. Conversely, offline activities can also engage individuals online. There are several ways to model this coupling. One approach is to consider that the transmission rate, $\tau$, depends on the level of offline activity, represented as $\tau(P)$. Another approach is to assume that the number of individuals protesting offline can directly engage uninterested individuals online, even if they are not part of the same social network. Our framework adopts a simplified version of the latter mechanism, thereby incorporating offline-to-online feedback. Specifically, we assume that the magnitude of offline activity has a proportionally positive effect, governed by a self-excitation rate, $\theta$, which motivates an uninterested individual to transition to an engaged state online.

Additionally, we assume that we are in a \textit{tension-inhibitive regime} where individuals release tension when participating in offline activity and so are immediately moved to the disengaged state online. Moreover, we assume permanent immunity from both online and offline engagement. We leave the exploration of temporary immunity for future work.

Table \ref{table:1} summarizes the model parameters along with their physical interpretations.

\begin{table}[!htb]
\caption{Key model parameters.}
\label{table:1}
\begin{center}
\begin{tabular}{|ll|}
\hline
\multicolumn{2}{|l|}{(a) Parameters for the dyanamics on the information layer.} \\ \hline
\multicolumn{1}{|l|}{Population-level transmission rate}               & $\beta$                  \\ \hline
\multicolumn{1}{|l|}{Transmission rate per contact}   & $\tau$                  \\ \hline
\multicolumn{1}{|l|}{Recovery rate}                   & $\gamma_i$               \\ \hline
\multicolumn{2}{|l|}{(b) Parameters for the dynamics on the physical layer.}     \\ \hline
\multicolumn{1}{|l|}{Transmission rate}               & $\eta$                   \\ \hline
\multicolumn{1}{|l|}{Recovery rate}                   & $\gamma_p$               \\ \hline
\multicolumn{2}{|l|}{(c) In between layers}                                      \\ \hline
\multicolumn{1}{|l|}{Self-excitement}                 & $\theta$                 \\ \hline
\end{tabular}
\end{center}
\end{table}

\subsection{The stochastic model}\label{subsec:stochastic}

We can see this process as a Markov Chain with five possible states. Each state includes online and offline sub-states. The first state, $S_1$, includes people who are uninterested online and therefore in a non-protesting offline state. State $S_1$ can transition to $S_2$, where the individual is engaged online but remains in a non-protesting offline state. State $S_2$ can transition to one of two states: $S_3$ (disengaged online and not protesting) or $S_4$ (disengaged online but actively protesting offline). From $S_4$, individuals may transition to $S_5$ (disengaged online and done protesting). Figure \ref{fig:markov} illustrates the states and transitions between states.

We use the Gillespie algorithm \cite{Epidemics_Networks} to simulate the dynamics of the stochastic model. For each simulation run, we specify a sufficiently large maximum simulation time, $t_{\text{max}}$, such that the process terminates if the next event occurs beyond this threshold. We report the mean trajectory as well as an envelope representing the minimum and maximum values at each time point across multiple trajectories interpolated onto a uniform time grid. The number of repetitions is denoted as $\text{iter}$.

\section{Dynamics in fully-mixed population}\label{sec:fullymixed}

In this section, we examine a special case of the model assuming a fully-mixed population, which corresponds to a fully-connected network in the information layer. We also apply the ``random mixing assumption,” meaning that during each small time interval, every individual has an equal chance of interacting with any other individual, selected uniformly at random. We present the derivation of our model for this case and then discuss outburst conditions using the \textit{basic reproductive number} in Section~\ref{subsec:fullmixed_R0}. Section~\ref{subsubsec:fullmixed_ActfromE} illustrates the effects of initial online activity, while Section~\ref{subsubsec:fullmixed_R0specific} describes how to compute the reproductive number specific to the fully-mixed model, revealing how outbursts in one layer can be triggered by initial activity in the other.

We now introduce the relevant notation. Let $[X]$ denote the expected number of individuals in state $X,$ where $X\in \set{U,E,D,P,R}.$ Let $[XY]$ be the expected number of edges connecting individuals in states $X$ and $Y,$ and let $[XYZ]$ be the expected number of triplets in which a middle node in state $Y$ is connected to nodes in states $X$ and $Z$. For edges and triples, we track only the online states, so $X,Y,Z\in \set{U,E,D}.$  Table \ref{table:2} summarizes this notation.

\begin{table}[!htb]
\caption{Summary of notation}
\label{table:2}
\begin{center}
\begin{tabular}{|l|l|}
\hline
Term                          & Description                                                          \\ \hline
$[X]$                         & Expected $\#$ of individuals in state $X$                                 \\
$[XY]$                        & Expected $\#$ of edges connecting individuals in $X$ and $Y$ \\
$[XYZ]$                       & Expected $\#$ of triplets with $Y$ having both $X$ and $Z$ neighbors       \\ \hline
\end{tabular}
\end{center}
\end{table}

In the case of a fully-mixed population, if we let $\hat{[X]}=[X]/N$ and $\beta = N \tau$, we get the simplest model now in fractional form for the online and offline states in the coupled system: 
\begin{align}\label{model:fully_mixed}
\left\{\begin{array}{ll}
    \frac{d[\hat U](t)}{dt}&=-\beta[\hat U][\hat E]-\theta[\hat U][\hat P],\\[3pt]
    \frac{d[\hat E](t)}{dt}& = \beta[\hat U][\hat E]+\theta[\hat U][\hat P]-(\eta+\gamma_i)[\hat E],\\[3pt]
    \frac{d[\hat D](t)}{dt}& = (\eta+\gamma_i)[\hat E],\\[3pt]
    \frac{d[\hat P](t)}{dt}& = \eta[\hat E]-\gamma_p[\hat P],\\[3pt]
    \vspace{6pt}\frac{d[\hat R](t)}{dt}& = \gamma_p[\hat P].\end{array}\right.
\end{align}

\begin{figure}[!htb]
    \centering
    \begin{subfigure}[t]{0.49\textwidth}
        \centering
        \includegraphics[width=\linewidth]{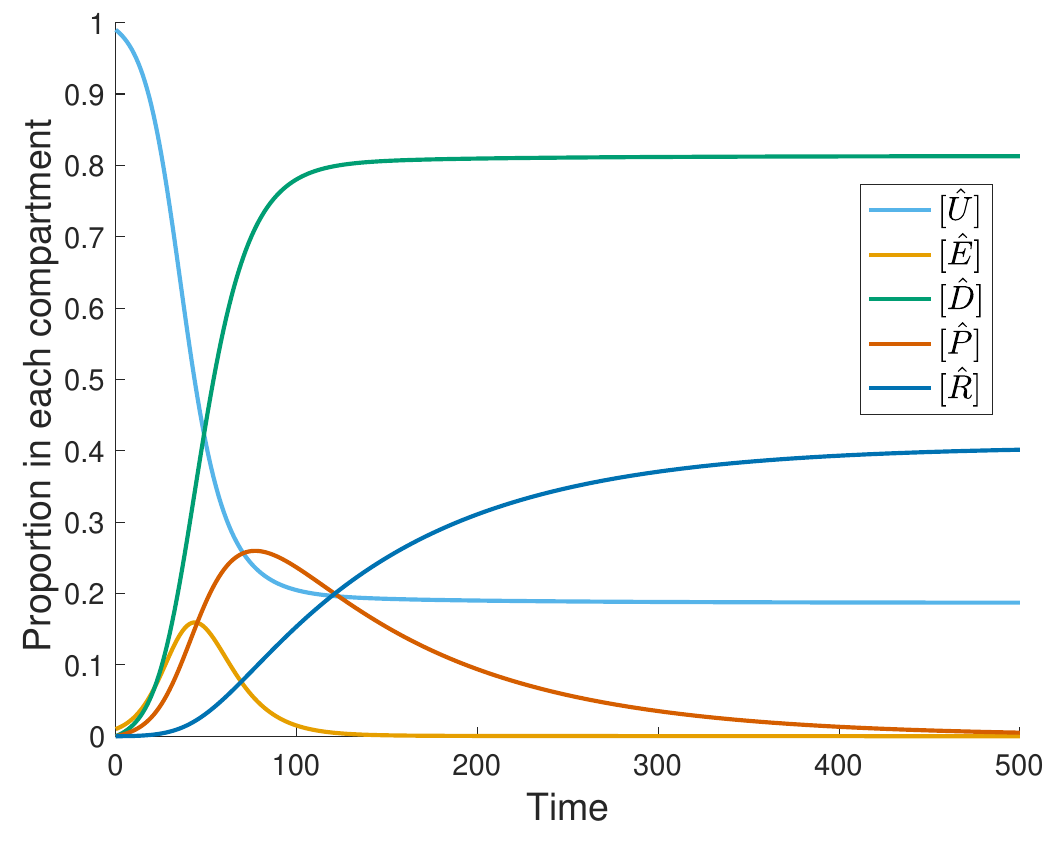} 
        \caption{With initial online $[\hat{E}](0) = 0.01$}
        \label{fig:FullyMixed_Dynamics_E0}
    \end{subfigure}
    \hfill
    \begin{subfigure}[t]{0.49\textwidth}
        \centering
        \includegraphics[width=\linewidth]{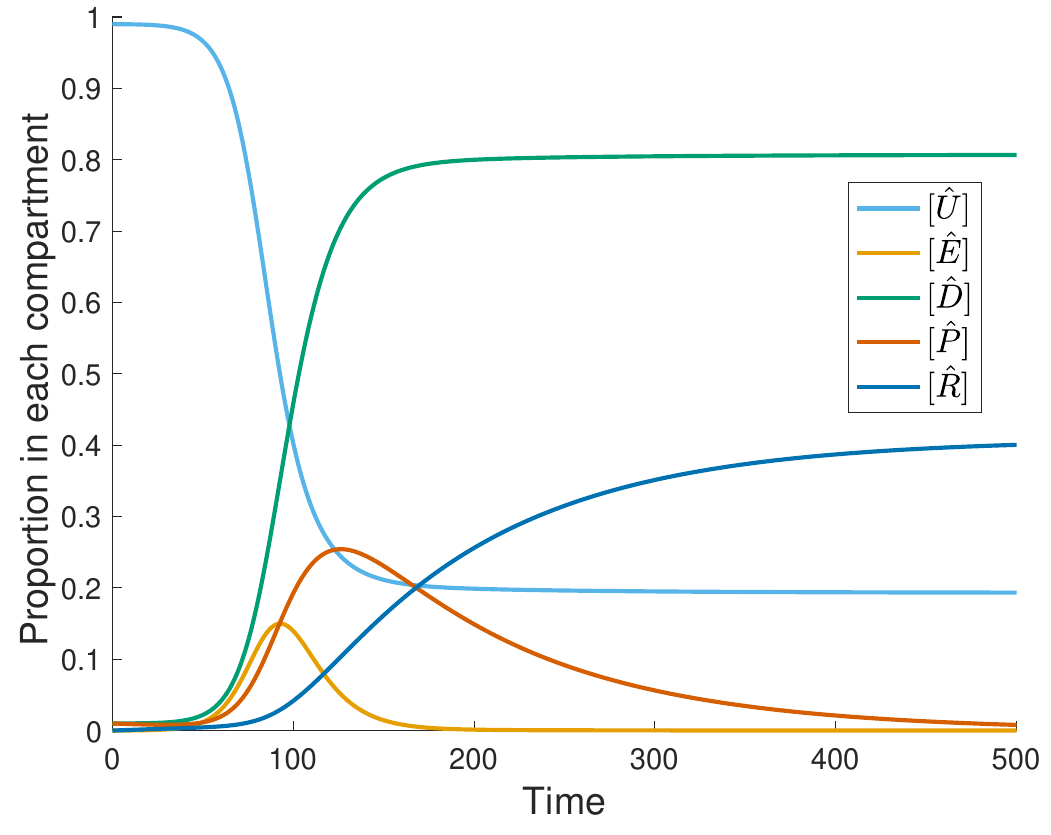} 
        \caption{With initial offline $[\hat{P}](0) = 0.01$}
        \label{fig:FullyMixed_Dynamics_P0}
    \end{subfigure}
    \caption{Dynamics of the compartment fractions in the model given by system \eqref{model:fully_mixed}. The parameters used are $\beta=0.2, \theta=0.001, \eta=0.05, \gamma_i=0.05$, and $\gamma_p=0.01$. Due to the bidirectional feedback between layers, activity can emerge in both layers over time, even if only one is initially active. (a) The initial online $[\hat{E}](0) = 0.01$, with no initial offline activity. (b) The initial offline  $[\hat{P}](0) = 0.01$, with no initial online activity.}
    \label{fig:FullyMixed_Dynamics}
\end{figure}

Figure \ref{fig:FullyMixed_Dynamics} presents example compartment dynamics when only one of the online and offline layers exhibits initial activity. Due to the feedback between the two layers, activity in one layer triggers activity in the other over time.

\subsection{Outbursts in fully-mixed model}\label{subsec:fullmixed_R0}

To understand how model parameters influence the levels of online and offline activity, we adopt the concept of the {\it basic reproductive number}, a threshold quantity widely used in epidemiology \cite{Brauer2008}. It is defined in the stochastic setting as the expected number of new cases generated by a single infectious individual in an otherwise fully susceptible population \cite{R0_ref}. Translating this idea to the context of protest dynamics driven by online-to-offline spillovers, we define the basic reproductive number, denoted by $R_0$, as the average number of individuals newly motivated by a single engaged individual in a fully uninterested (i.e. pre-engaged online) population. We will refer to this quantity as the reproductive number interchangeably throughout. As in epidemiological models, $R_0$ serves as a threshold in our setting, indicating whether small-scale unrest is likely to escalate into large-scale mobilization. This value reflects the effectiveness of both online information diffusion and the subsequent activation of offline protest.

In our model, the reservoir of individuals susceptible to protest—those currently engaged online—is shaped by parameters $\beta$ and $\gamma_i$, which govern the rates of engagement and disengagement, respectively. The parameter $\eta$ determines the rate at which engaged individuals transition to active offline protest. To adapt the epidemiological definition of $R_0$ to our setting, we consider the introduction of a single engaged individual ($E$) into an infinite population of disengaged individuals ($U$), with no initial offline activity (i.e. $[\hat{P}](0) = 0$). The reproductive number in this case is defined as the expected number of individuals that this single engaged person recruits into the $E$ state before becoming disinterested. This yields the approximation:
\begin{align} \label{eq:R0_FullyMixed}
    R_0 = \frac{\beta}{\gamma_i + \eta}. 
\end{align}
Notably, when the combined effect of $\eta$ and $\gamma_i$ is large relative to $\beta$—that is, when either parameter or both are sufficiently high—the online reservoir depletes quickly, limiting the potential for significant protest activity. Under our model assumptions, persistent offline action is necessary to sustain online engagement over time.

\subsubsection{Outbursts following initial online activity}\label{subsubsec:fullmixed_ActfromE}

An empirically important aspect of the online-to-offline dynamics is the maximum level of online and offline activity reached over time, referred to as the outburst size. We define the final size of the online outburst as $\| [\hat{E}](t) \|_{\infty}$ and the final size of the offline outburst as $\| [\hat{P}](t) \|_{\infty}$. In this study, we use the term outburst to describe a pronounced surge in activity—either online or offline—that arises over time and reflects a significant departure from the initial activity level.

\begin{figure}[!htb]
    \centering
    \begin{subfigure}[t]{0.49\textwidth}
        \centering
        \includegraphics[width=\linewidth]{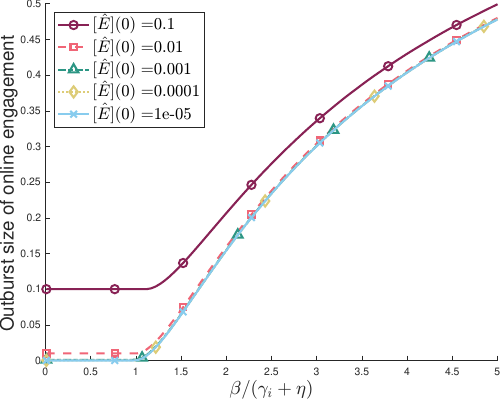} 
        \caption{Final size of the online outburst} \label{fig:FullyMixed_Outburst_vs_R0_online}
    \end{subfigure}
    \hfill
    \begin{subfigure}[t]{0.49\textwidth}
        \centering
        \includegraphics[width=\linewidth]{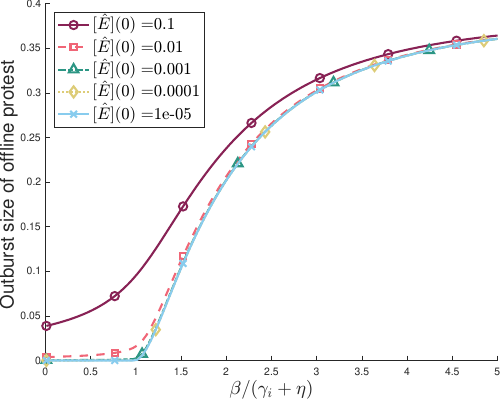} 
        \caption{Final size of the offline outburst} \label{fig:FullyMixed_Outburst_vs_R0_offline}
    \end{subfigure}
    \caption{Final sizes of online and offline outbursts as a function of the reproductive number $R_0$ in the stochastic process. Simulations are shown for varying levels of initial online $E$ and no initial offline activity. When initial online activity is below a certain threshold (e.g. $1\%$ as shown), a clear transition is observed: No outburst occurs when $R_0 < 1$, while outbursts emerge and grow as $R_0$ increases beyond 1. The parameters used are $\theta=0.001, \eta=0.1, \gamma_i=0.05, \gamma_p=0.05$ and $\beta$ is varied to obtain different $R_0$ values.}
    \label{fig:FullyMixed_Outburst_vs_R0}
\end{figure}

The reproductive number in equation~\eqref{eq:R0_FullyMixed} is useful for examining when small initial online activity can trigger online and offline outbursts. In Figure~\ref{fig:FullyMixed_Outburst_vs_R0}, we explore how the initial level of online engagement ($E$) influences the magnitude of outbursts in both layers. The parameters used are $\theta = 0.001$, $\eta = 0.1$, $\gamma_i = 0.05$, and $\gamma_p = 0.05$, while $\beta$ is varied to span regimes where $R_0 < 1$ and $R_0 > 1$. Specifically, we adjust $\beta$ so that $R_0$ ranges from 0 to 5, and we consider different initial fractions of the population engaged online.

A distinct transition is observed: When $R_0 < 1$, no outbursts occur in either layer, whereas outbursts begin to emerge and grow in size once $R_0$ exceeds 1. This threshold behaviour is visible for initial engagement levels as high as $1\%$ of the population but becomes less pronounced when the initial fraction increases to $10\%$, which is expected since the $R_0$ threshold strictly applies in the limit of small initial activity.

\subsubsection{Basic reproductive number for fully-mixed model}\label{subsubsec:fullmixed_R0specific}

The reproductive number presented in equation~\eqref{eq:R0_FullyMixed} omits certain aspects of the model, such as offline-to-online feedback and offline recovery. To more accurately characterize the conditions for outbursts in the fully-mixed population setting, we leverage the deterministic mean-field model~\eqref{model:fully_mixed} for further analysis.

Let us consider the stability of the outburst-free equilibrium solution of the system~\eqref{model:fully_mixed}, $(1,0,0,0,0)$. If we linearized the system about this equilibrium, we obtain the following Jacobian:
\begin{align*}
   J=\left[\begin{array}{ccccc}
    0&-\beta&0&-\theta&0\\
    0&\beta-(\eta+\gamma_i)& 0&\theta&0\\
    0&\eta+\gamma_i& 0&0&0\\
    0&\eta&0&-\gamma_p&0\\
    0&0&0&\gamma_p&0    
    \end{array}\right]
\end{align*}
Note that $J$ has three zero eigenvalues and two non-zero eigenvalues given by:
\begin{align}\label{eig}
 \lambda_{\pm}  = \frac{1}{2}\left[ (\beta-\eta-\gamma_i-\gamma_p)\pm \sqrt{(\beta-\eta-\gamma_i-\gamma_p)^2-4(\eta\gamma_p+\gamma_i\gamma_p-\beta\gamma_p-\eta\theta)} \right].
\end{align}

To compute the reproductive number in our deterministic model, we adopt the definition given in \cite{Epidemics_Networks}: For a deterministic model $M$ with a uniform recovery rate $\gamma$ and Malthusian parameter $\lambda$, the reproductive number is defined as $R_0^{M} = \frac{\lambda}{\gamma} + 1$. In our continuum approximation, the Malthusian parameter is $\lambda_{+}$. Therefore, the reproductive number for our model is given by
\[R_0^{F} = \frac{\lambda_{+}}{\gamma_i+\eta} + 1.\] 
Additionally, from equation \eqref{eig}, linear stability requires that
\begin{align}
&\beta<\gamma_i+\gamma_p+\eta \label{eq:stability1}\\
\text{and} \quad &\gamma_i>\frac{\beta\gamma_p+\eta\theta-\eta\gamma_p}{\gamma_p} \label{eq:stability2}. 
\end{align}
An online activity flare-up can trigger a significant offline response if either or both conditions \eqref{eq:stability1} and \eqref{eq:stability2} are violated. These conditions show that the five parameters are strongly interdependent in determining whether the inequalities hold,  indicating that the offline transmission rate, $\eta$, must lie within a critical range—a ``sweet spot"—to enable the emergence of an offline outburst.

\begin{figure}[H]
    \centering
    \begin{subfigure}[t]{0.49\textwidth}
        \centering
        \includegraphics[width=\linewidth]{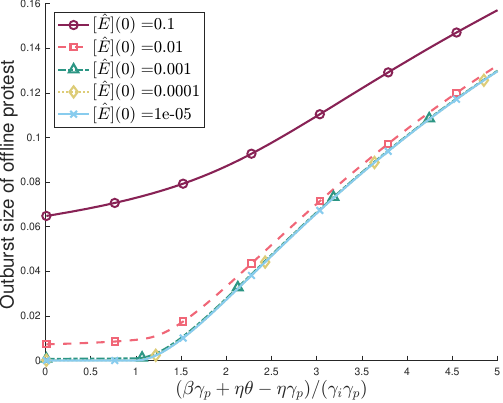} 
        \caption{Offline outburst size with varying $[\hat{E}](0)$} 
        \label{fig:FullyMixed_Outbursts_over_Stability_E0_Effect}
    \end{subfigure}
    \hfill
    \begin{subfigure}[t]{0.49\textwidth}
        \centering
        \includegraphics[width=\linewidth]{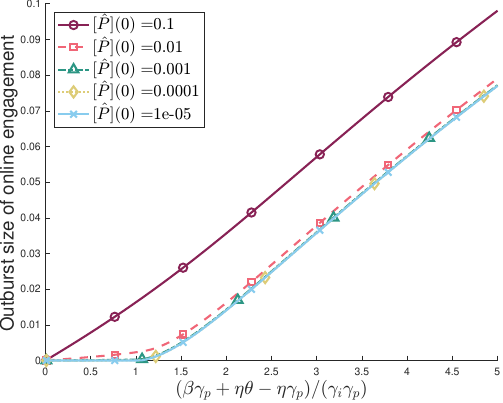} 
        \caption{Online outburst size with varying $[\hat{P}](0)$} \label{fig:FullyMixed_Outbursts_over_Stability_P0_Effect}
    \end{subfigure}
    \caption{Final sizes of offline outbursts with only initial online activity (and no initial offline activity) and the online outbursts with only initial offline activity (and no initial online activity), shown as a function of the stability condition. $1$ is the threshold value distinguishing cases where condition \eqref{eq:stability2} is satisfied (below) or violated (above). Parameters are selected such that condition \eqref{eq:stability1} is always satisfied. When only one layer is initially activated at a low level, no outburst is triggered in the other layer if both stability conditions are met. However, sufficiently high initial activation can overcome the stability condition and trigger an outburst in the other layer. Parameters are $\beta=0.1, \eta=0.1, \gamma_i=0.05, \gamma_p=0.05$ and $\theta$ is varied.}
    \label{fig:FullyMixed_Outbursts_over_Stability}
\end{figure}

In Figure \ref{fig:FullyMixed_Outbursts_over_Stability}, we illustrate the effect of initial online activity on offline outbursts, and conversely, the effect of initial offline activity on online outbursts. In each scenario, we initially activate only one of the two layers, using the set of parameters $\beta=0.1, \eta=0.1, \gamma_i=0.05$ and $\gamma_p=0.05$, so that condition \eqref{eq:stability1} is always satisfied. We then vary $\theta$ so that condition \eqref{eq:stability2} is fulfilled or violated. The results show that when the initial engagement in one layer is small and the other layer is completely inactive, no outburst occurs in the layer initialized with no activity if both stability conditions are satisfied. However, this does not hold when the initial activity level is high, indicating that large-scale initial activation in one layer can overcome the stability conditions and trigger an outburst in the other.

\section{Dynamics on homogeneous networks}\label{sec:homo}

This section explores the dynamics of the system on homogeneous networks, where all nodes have the same number of connections. In particular, we focus on $k$-regular networks, where each node has exactly degree $k$. We derive two mean-field approximations, based on the methodology in \cite{Epidemics_Networks}: a single-level (or node-level) approximation and a pairwise approximation. Section~\ref{subsec:HomogSingle} introduces the single-level approximation, while Section~\ref{subsec:homog_PW} presents the pairwise approximation. Section~\ref{subsec:homog_compare_model} compares both models to the stochastic baseline and discusses their performance. Section~\ref{subsec:homog_outbursts} analyses outbursts using reproductive numbers: Section~\ref{subsubsec:homog_stochR0} focuses on the stochastic model, Section~\ref{subsubsec:Homog_R0_SingleLevel} on the single-level approximation, and Section~\ref{subsubsec:homog_PWR0} on the pairwise approximation, including a follow-up comparison of all three and an interpretation of their empirical implications.

\subsection{Single-level approximation} \label{subsec:HomogSingle}

Let our network have $N$ nodes. The expected values of the system described in Section \ref{subsec:stochastic} satisfy the following system of equations for the online states: 
\begin{align}\label{sys:Homog_Online}
\left\{\begin{array}{ll}
    \frac{d[U](t)}{dt}&=-\tau [UE]-\frac{\theta}{N}[U][P],\\[3pt]
    \frac{d[E](t)}{dt}& = \tau [UE]+\frac{\theta}{N}[U][P]-(\eta+\gamma_i)[E],\\[3pt]
    \frac{d[D](t)}{dt}& = (\eta+\gamma_i)[E];\end{array}\right.
\end{align}
and offline states:
\begin{align}\label{sys:Homog_Offline}
\left\{\begin{array}{ll}
    \frac{d[P](t)}{dt}& = \eta[E]-\gamma_p[P],\\[3pt]
    \vspace{6pt}\frac{d[R](t)}{dt}& = \gamma_p[P].\end{array}\right.
\end{align}
Assuming that the excited nodes are distributed randomly, and the average engaged node has $k[U]/N$ uninterested neighbors. This gives a single-level approximation:
\begin{align}\label{model:Homogeneous_SingleLevel}
\left\{\begin{array}{ll}
    \frac{d[U](t)}{dt}&=-\tau \frac{k}{N}[U][E]-\frac{\theta}{N}[U][P],\\[3pt]
    \frac{d[E](t)}{dt}& = \tau \frac{k}{N}[U][E]+\frac{\theta}{N}[U][P]-(\eta+\gamma_i)[E],\\[3pt]
    \frac{d[D](t)}{dt}& = (\eta+\gamma_i)[E],\\[3pt]
    \frac{d[P](t)}{dt}& = \eta[E]-\gamma_p[P],\\[3pt]
    \frac{d[R](t)}{dt}& = \gamma_p[P].\end{array}\right.
\end{align}

\subsection{Pairwise approximation}\label{subsec:homog_PW}

Instead of approximating $[UE]$, we can derive an exact system by introducing an evolution equation for this term. The node-level dynamics remain as in system~\eqref{sys:Homog_Online}. We now also include the edge dynamics within the online layer, given by: 
\begin{align*}
\left\{\begin{array}{ll}
   \frac{d[UE](t)}{dt}&=-(\gamma_i+\eta+\tau)[UE] + \tau([UUE]-[EUE])+\frac{\theta [P]}{N}[UU]-\frac{\theta [P]}{N}[UE],\\[3pt]
   \frac{d[UU](t)}{dt}&=-\tau\left([UUE]+[EUU]\right)-\frac{2\theta [P]}{N}[UU].\end{array}\right.
\end{align*}

The dynamics of the offline states continue to be governed by the equations as in system \eqref{sys:Homog_Offline}. Note that incorporating edge dynamics improves accuracy, as we no longer assume that engaged individuals are uniformly distributed. The trade-off is that we must now close the system by approximating the triplets $[UUE]$ and $[EUE]$.
To perform this closure, observe that the total number of edges emanating from uninterested nodes is $k[U]$. Thus, the proportion of edges connecting an uninterested node to an engaged node is $[UE]/(k[U])$, and the proportion connecting to another uninterested node is $[UU]/(k[U])$. Therefore, if we choose an uninterested node $u$ and two of its neighbors, $v$ and $w$, at random, the probability that $v$ is engaged and $w$ is uninterested is given by $[UU][UE]/(k^2[U])$. Since there are $k(k-1)$ ways to choose two neighbors, we have:
\[
[UUE] \approx \frac{k-1}{k} \cdot \frac{[UU][UE]}{[U]}.
\]
Similarly, we conclude that:
\[
[EUE] \approx \frac{k-1}{k} \cdot \frac{[EU]^2}{[U]},
\]
and by symmetry, $[EUU] = [UUE]$.  Using these closures, we derive our second model, consisting of a pairwise approximation system:
\begin{align}\label{model:Homogeneous_Pairwise}
\left\{\begin{array}{ll}
    \frac{d[U](t)}{dt}&=-\tau [UE]-\frac{\theta}{N}[U][P],\\[4pt]
    \frac{d[E](t)}{dt}& = \tau [UE]+\frac{\theta}{N}[U][P]-(\eta+\gamma_i)[E],\\[4pt]
    \frac{d[D](t)}{dt}& = (\eta+\gamma_i)[E],\\[4pt]
    \frac{d[UE](t)}{dt}&=-(\gamma_i+\eta+\tau+\frac{\theta [P]}{N})[UE]+\frac{\theta [P]}{N}[UU]+\frac{\tau (k-1)}{k[U]}\left([UU][UE]-[EU]^2\right),\\[4pt]
    \frac{d[UU](t)}{dt}&=-2\tau\left(\frac{k-1}{k}\frac{[UU][UE]}{[U]}\right)-\frac{2\theta [P]}{N}[UU],\\[4pt]
    \frac{d[P](t)}{dt}& = \eta[E]-\gamma_p[P],\\[4pt]
    \frac{d[R](t)}{dt}& = \gamma_p[P]. \end{array}\right.
\end{align}

\subsection{Model comparison}\label{subsec:homog_compare_model}

We compare the dynamics of the single-level approximation \eqref{model:Homogeneous_SingleLevel} and the pairwise approximation \eqref{model:Homogeneous_Pairwise} with those of the stochastic model on $k$-regular networks. In the stochastic simulations, we find that the mean trajectories stabilize after a sufficiently large number of runs, and therefore choose $\text{iter}=100$ as a practical and sufficiently robust setting for presentation. Throughout this paper, we treat the mean trajectories from the stochastic simulations as the ground truth for evaluating the accuracy of the approximation models. In all subsequent simulations, we use initial conditions with relatively small $[E](0)$, representing low initial online engagement, and assume zero initial populations in the $D, P$, and $R$ compartments. This setup reflects the empirical motivation of examining how limited initial online engagement can lead to offline activities. 

\begin{figure}[!htbp]
    \centering
    \begin{subfigure}[b]{0.4\textwidth}  
        \centering 
        \includegraphics[width=\textwidth]{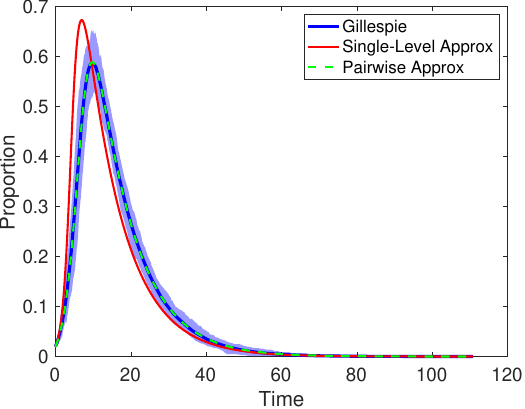}
        \caption{Online: $N=1\,000$}%
        \label{fig:e1000}
    \end{subfigure}
    \hspace{40pt}
    \begin{subfigure}[b]{0.4\textwidth}  
        \centering 
        \includegraphics[width=\textwidth]{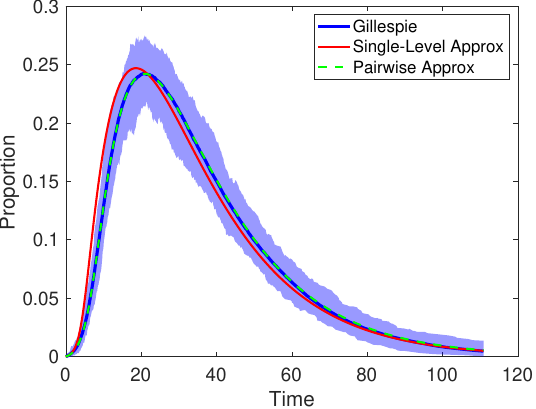}
        \caption{Offline: $N=1\,000$}%
        \label{fig:p1000}
    \end{subfigure}\\
    \begin{subfigure}[b]{0.4\textwidth}  
        \centering 
        \includegraphics[width=\textwidth]{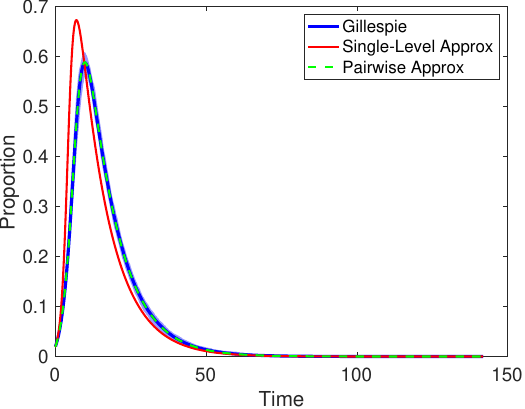}
        \caption{Online: $N=5\,000$}%
        \label{fig:e5000}
    \end{subfigure}
    \hspace{40pt}
    \begin{subfigure}[b]{0.4\textwidth}  
        \centering 
        \includegraphics[width=\textwidth]{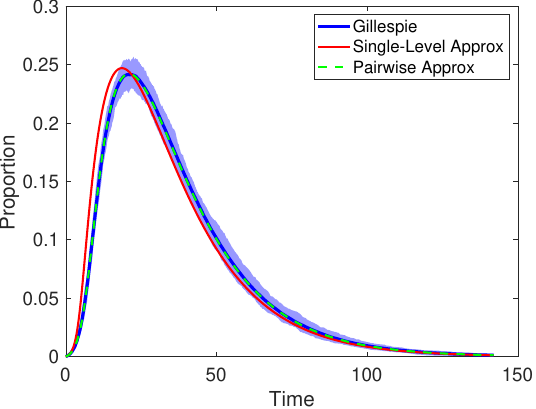}
        \caption{Offline: $N=5\,000$}%
        \label{fig:p5000}
    \end{subfigure}\\
    \begin{subfigure}[b]{0.4\textwidth}  
        \centering 
        \includegraphics[width=\textwidth]{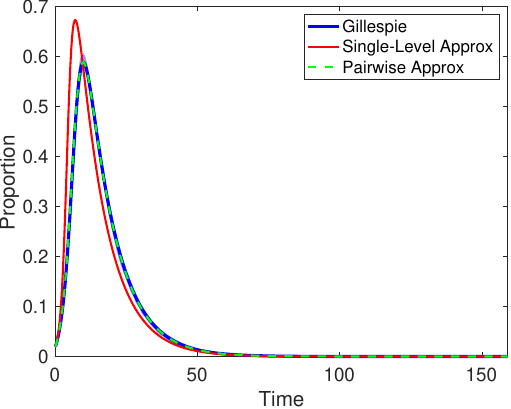}
        \caption{Online: $N=10\,000$}%
        \label{fig:e10000}
    \end{subfigure}
    \hspace{40pt}
    \begin{subfigure}[b]{0.4\textwidth}  
        \centering 
        \includegraphics[width=\textwidth]{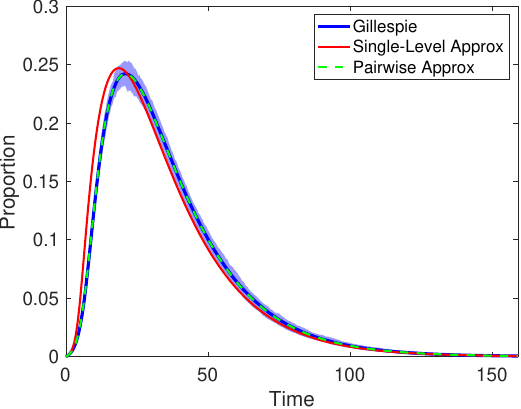}
        \caption{Offline: $N=10\,000$}%
        \label{fig:p10000}
    \end{subfigure}
    \caption{Comparison of online $E$ and offline $P$ dynamics across the stochastic model, single-level approximation, and pairwise approximation on $k$-regular networks. The left column shows online activity, while the right column depicts offline activity. The simulations use the following parameters: $k = 5$, $\tau = 0.2$, $\theta = 0.001$, $\eta = 0.05$, $\gamma_i = 0.05$, $\gamma_p = 0.05$, $t_{\text{max}} = 200$, $[E](0)/N = 0.02$, and $\text{iter} = 100$. Time is plotted up to the minimum endpoint across all Gillespie runs. Shaded area spans min–max across stochastic runs over time. With the same degree and varying network sizes, the pairwise approximation closely tracks the average behaviour of the stochastic simulations and consistently outperforms the single-level approximation.}     
    \label{fig:comp_homo}
\end{figure}

Figure \ref{fig:comp_homo} illustrates simulation results for online $E$ and offline $P$ ratios when we fix the degree to be $k = 5$ and vary the network sizes, $N$. It shows that the pairwise approximation aligns more closely with the mean trajectory of the stochastic simulations compared to the single-level approximation. Although the variance in the stochastic simulations decreases as $N$ increases, the mean trajectories and the difference between the approximation and the stochastic mean remain largely unchanged across different values of $N$.

We further examine the errors in the approximation models compared to the stochastic simulations, focusing specifically on the proportions of online $E$ and offline $P$. To compute the errors relative to degree and network size, we first identify the minimum end time across all stochastic simulation runs. We then interpolate the dynamics from both the stochastic simulations and the approximation models onto the same time grid, up to this minimum time. If we denote the mean trajectory of all interpolated stochastic simulation runs for compartment $X$ as $[X]^{\text{stoch}}(t)$, and the corresponding interpolated trajectory from an approximation model as $[X]^{\text{approx}}(t)$, then the $L^2$ error in approximating $X$ using this model is defined as $\| [X]^{\text{stoch}}(t) - [X]^{\text{approx}}(t) \|_2$. 

\begin{figure}[!htbp]
    \centering
    \begin{subfigure}[b]{0.49\textwidth}  
        \centering 
        \includegraphics[width=\textwidth]{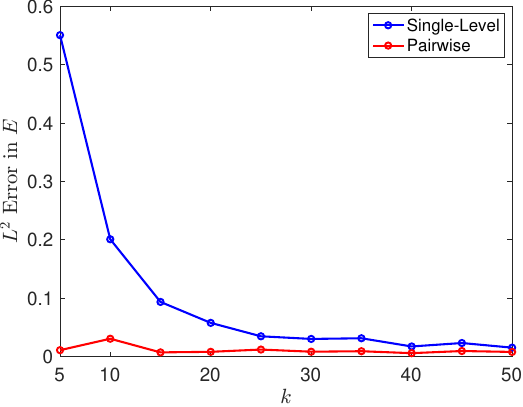}
        \caption{Online, $N=1\,000$}%
        \label{fig:onlineL2k}
    \end{subfigure}
    \hfill
    \begin{subfigure}[b]{0.49\textwidth}  
        \centering 
        \includegraphics[width=\textwidth]{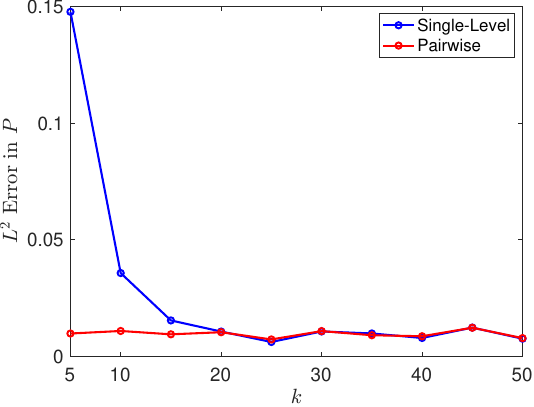}
        \caption{Offline, $N=1\,000$}%
        \label{fig:offlineL2k}
    \end{subfigure}\\
        \begin{subfigure}[b]{0.49\textwidth}  
        \centering 
        \includegraphics[width=\textwidth]{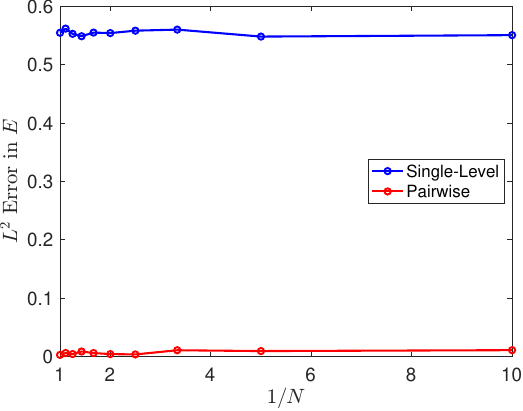}
        \caption{Online, $k=5$}%
        \label{fig:onlineL2N}
    \end{subfigure}
    \hfill
    \begin{subfigure}[b]{0.49\textwidth}  
        \centering 
        \includegraphics[width=\textwidth]{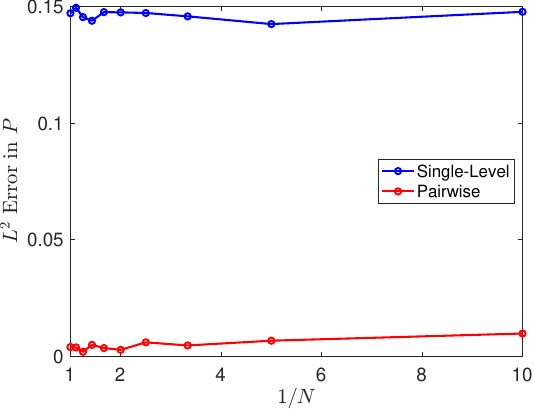}
        \caption{Offline, $k=5$}%
        \label{fig:offlineL2N}
    \end{subfigure}
    \caption{The $L^2$ errors of the single-level and pairwise approximations, relative to the stochastic model, on homogeneous (i.e. $k$-regular) networks, for $E$ and $P$ proportions. The top row shows the error as a function of the node degree $k$, and the bottom row shows the error as a function of $1/N$, the inverse of the network size. The parameters used are $N=1\,000$ (for error over $k$), $k=5$ (for error over $1/N$), $\tau=0.2, \theta=0.001, \eta=0.05, \gamma_i=0.05, \gamma_p=0.05, t_{\text{max}}=200, [E](0)/N=0.02, \text{iter}=100$. The results show that the single-level approximation improves with increasing node degree, while the accuracy of the pairwise approximation remains relatively stable. The network size does not appear to have a significant impact on the approximation error.}     
    \label{fig:L2_Error_Plots_Homogeneous}
\end{figure}

We investigate the effects of the node degree $k$ and the network size $N$. Figure \ref{fig:L2_Error_Plots_Homogeneous} shows the $L^2$ errors, indicating that increasing the node degree results in smaller errors for the single-level approximation. However, the errors for the pairwise approximation remain relatively stable across different degrees, and the network size has no significant impact.

\begin{figure}[!htbp]
    \centering
    \begin{subfigure}[b]{0.49\textwidth}  
        \centering 
        \includegraphics[width=\textwidth]{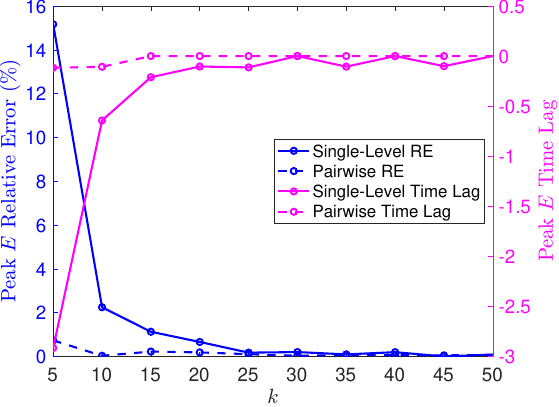}
        \caption{Online, $N=1\,000$}%
        \label{fig:onlineREk}
    \end{subfigure}
    \hfill
    \begin{subfigure}[b]{0.49\textwidth}  
        \centering 
        \includegraphics[width=\textwidth]{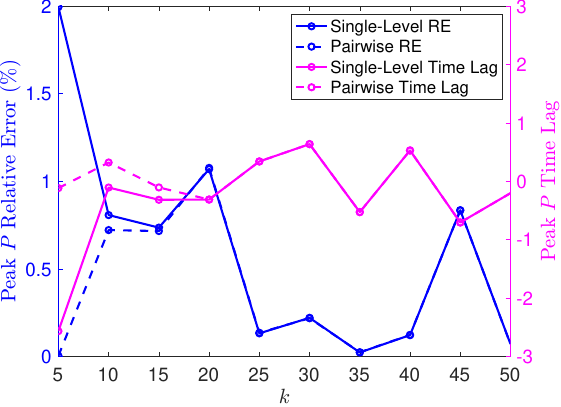}
        \caption{Offline, $N=1\,000$}%
        \label{fig:offlineREk}
    \end{subfigure}\\
        \begin{subfigure}[b]{0.49\textwidth}  
        \centering 
        \includegraphics[width=\textwidth]{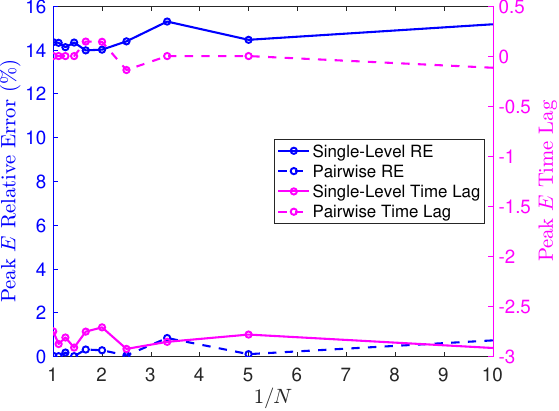}
        \caption{Online, $k=5$}%
        \label{fig:onlineREN}
    \end{subfigure}
    \hfill
    \begin{subfigure}[b]{0.49\textwidth}  
        \centering 
        \includegraphics[width=\textwidth]{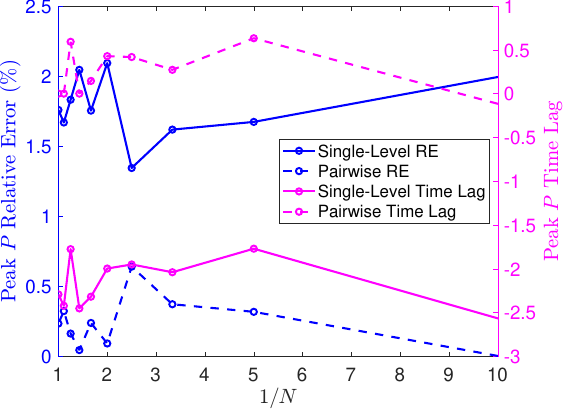}
        \caption{Offline, $k=5$}%
        \label{fig:offlineREN}
    \end{subfigure}
    \caption{Relative errors of the single-level and pairwise approximations compared to the stochastic model on homogeneous ($k$-regular) networks, for the peak values of $E$ and $P$, shown as a function of node degree $k$ (top) and inverse network size $1/N$ (bottom). The right $y$-axis shows the time lag of the predicted peaks relative to those observed in the stochastic simulations. A negative lag indicates that the peak occurs earlier than in the stochastic model; a positive lag indicates a delayed peak. The parameters used are: $N = 1\,000$ (for error over $k$), $k = 5$ (for error over $1/N$), $\tau = 0.2$, $\theta = 0.001$, $\eta = 0.05$, $\gamma_i = 0.05$, $\gamma_p = 0.05$, $t_{\text{max}} = 200$, $[E](0)/N = 0.02$, and $\text{iter} = 100$. The relative error for the single-level approximation decreases with increasing degree, while the performance of the pairwise approximation remains largely unchanged. No clear trend is observed with respect to network size. Additionally, the single-level approximation consistently predicts earlier peak times than those in the stochastic simulations, whereas the pairwise approximation shows no consistent bias and exhibits only minor deviations in peak timing.}     
    \label{fig:Relative_Error_Plots_Homogeneous}
\end{figure}

Figure \ref{fig:Relative_Error_Plots_Homogeneous} illustrates the relative errors in the peak magnitudes and their corresponding time lags. Similar to the previous observation, larger degrees yield smaller relative errors for the single-level approximation. Additionally, the single-level approximations consistently predict earlier peak times compared to the stochastic simulations.

\subsection{Outburst analysis of models on homogeneous networks}\label{subsec:homog_outbursts}

\subsubsection{Basic reproductive number: stochastic process} \label{subsubsec:homog_stochR0}

To understand how the parameters influence outburst behaviour in the stochastic process, we again turn to the reproductive number. Following the same approach used in the fully-mixed population case presented in Section~\ref{subsec:fullmixed_R0}, we approximate the reproductive number for the stochastic process on a homogeneous network by
\begin{align}\label{eq:R0_kReg}
    R_0 = \frac{k \tau}{\eta + \gamma_i}.
\end{align}

\begin{figure}[!htb]
    \centering
    \includegraphics[width=0.6\linewidth]{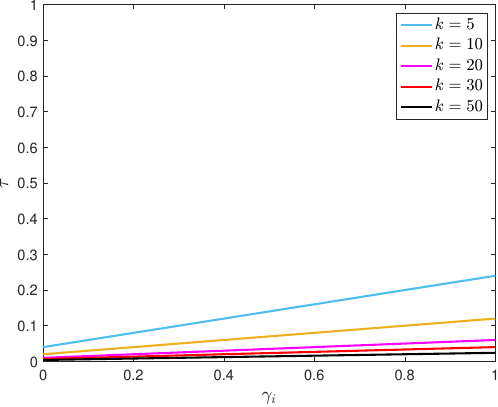}
    \caption{The analytical $R_0 = 1$ line for the stochastic model on a $k$-regular network is plotted in the $\gamma_i$-$\tau$ plane for different degree $k$. The other parameters are $\theta = 0.01$, $\eta = 0.2$, and $\gamma_p = 0.5$. As $k$ increases, the line separating the parameter space that leads to outbursts from that which does not becomes flatter. This implies that network density is a key factor, with higher density resulting in a larger region of parameter space where outbursts can occur.}
    \label{fig:StR0Lines}
\end{figure}

Equation~\eqref{eq:R0_kReg} does not depend on $\theta$ or $\gamma_p$, indicating that the online-offline feedback and the offline recovery rate do not influence this calculation. Hence, the expression serves as an approximation rather than the exact reproductive number of the process. Nevertheless, we can examine the impact of the online dynamics parameters $\tau$ and $\gamma_i$, and evaluate how the degree affects the classification threshold for an outburst using this $R_0$. Figure \ref{fig:StR0Lines} displays the $R_0 = 1$ line in the $\gamma_i$-$\tau$ plane for several representative degrees, showing that as the degree increases, the system becomes more likely to exhibit an outburst.

\begin{figure}[!htbp]
    \centering
    \includegraphics[width=0.98\linewidth]{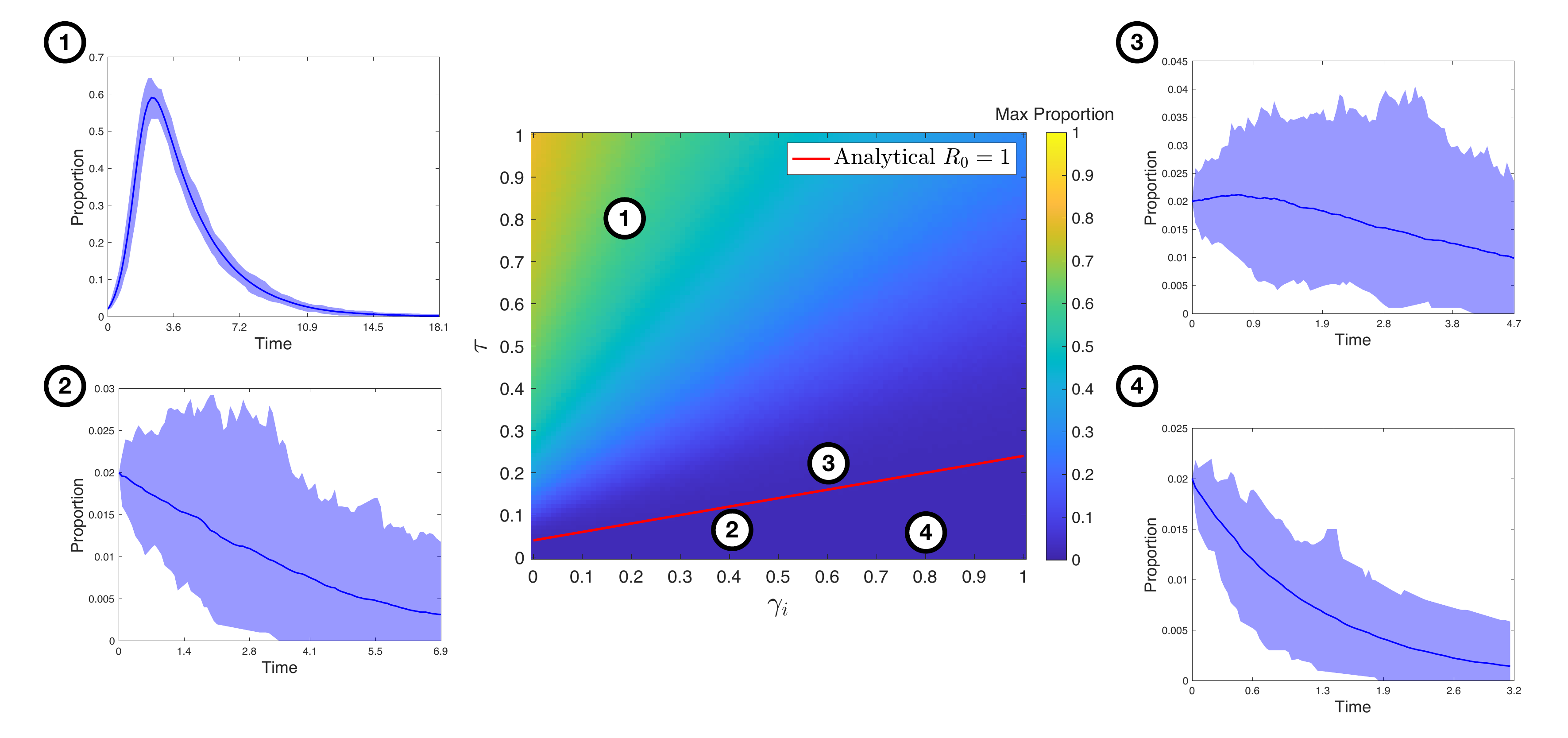}
    \caption{Heatmap from stochastic simulations on homogeneous networks for the proportions of the mean final sizes of online outbursts in $E$, averaged over $\text{iter} = 100$ runs across $(\gamma_i, \tau)$. The grid resolution is $100 \times 100$. Parameters are $N=1\,000, k=5, \theta=0.01, \eta=0.2, \gamma_p=0.5, t_{\text{max}}=500$, and $[E](0)=20$. The red line indicates the threshold $R_0 = 1$. Temporal dynamics of $E$ are illustrated for selected parameter values. Shaded area spans min–max across stochastic runs over time. In the region where $R_0 > 1$, online activity initially grows before decaying, indicating an outburst. In contrast, for $R_0 < 1$, the engaged population strictly decays over time.}
    \label{fig:stochastic_tau_gamma_i_E}
\end{figure}

\begin{figure}[!htbp]
    \centering
    \includegraphics[width=0.98\linewidth]{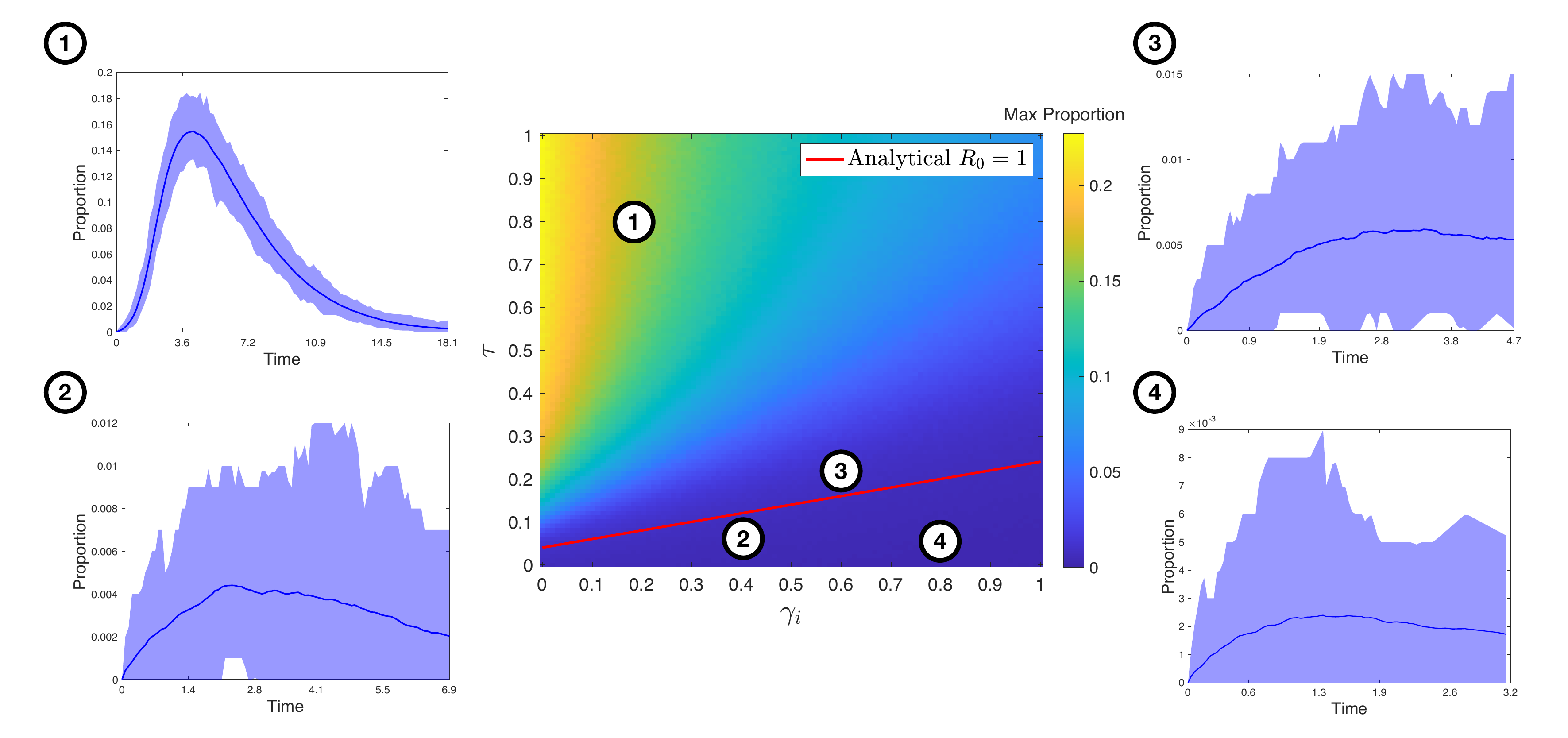}
    \caption{Heatmap from stochastic simulations on homogeneous networks showing the proportions of mean final offline outburst sizes in $P$, swept over $\tau$ and $\gamma_i$. Parameter values match those in Figure~\ref{fig:stochastic_tau_gamma_i_E}. Since only online activity is initialized, $P$ dynamics generally show an initial increase followed by decay. The threshold $R_0 = 1$ classifies offline activity levels: Values well above it yield pronounced offline outbursts, while those below lead to negligible activity.}
    \label{fig:stochastic_tau_gamma_i_P}
\end{figure}

To illustrate how the $R_0 = 1$ threshold relates to the sizes and temporal dynamics of online $E$ and offline $P$, we perform parameter sweeps across the $\gamma_i$-$\tau$ plane. We present the averaged final sizes of outbursts as heatmaps, along with representative time dynamics in different regions. Figure \ref{fig:stochastic_tau_gamma_i_E} displays the results for the $E$ compartment, while Figure \ref{fig:stochastic_tau_gamma_i_P} shows the corresponding results for the $P$ compartment.

In Appendix \ref{app:heatmap_stochastic}, we present additional heatmaps and corresponding discussions on online and offline outbursts across other parameter combinations. In particular, we examine pairs involving the offline transmission rate, specifically $(\gamma_p, \eta)$ and $(\tau, \eta)$. The results provide supporting evidence that the emergence of intense offline protest requires $\eta$ to lie within a moderate range relative to other parameters in the system.

\subsubsection{Basic reproductive number: single-level approximation}\label{subsubsec:Homog_R0_SingleLevel}

To derive the reproductive number for the single-level approximation model on homogeneous networks, we begin by considering an equilibrium solution of the system, $([U], [E], [D], [P], [R]) = ([U]^*, 0, N-[U]^*, 0, [R]^*)$ where $[U]^*, [R]^* \in [0,N]$. The Jacobian at this equilibrium is
\begin{align*}
   J^{MF}=\left[\begin{array}{ccccc}
    0&-\frac{k\tau}{N}[U]^*&0&-\frac{\theta}{N}[U]^*&0\\
    0&\frac{k\tau}{N}[U]^*-(\eta+\gamma_i)&0&\frac{\theta}{N}[U]^*&0\\
    0&\eta+\gamma_i&0&0&0\\
    0&\eta&0&-\gamma_p&0\\
    0&0&0&\gamma_p&0    
    \end{array}\right]
\end{align*}
Note that $J^{MF}$ has three zero eigenvalues and the other two eigenvalues are
\begin{align*} 
     \lambda_{\pm}^{MF}  = \frac{1}{2}\Bigg[ & \left(\frac{k\tau}{N}[U]^*-\eta-\gamma_i-\gamma_p \right)\\
     & \pm \sqrt{\left(\frac{k\tau}{N}[U]^*-\eta-\gamma_i-\gamma_p \right)^2-4 \left(\eta\gamma_p+\gamma_i\gamma_p-\frac{k\tau\gamma_p}{N}[U]^*-\frac{\theta\eta}{N}[U]^* \right)} \Bigg].
\end{align*}

To find the reproductive number for the homogeneous single-level model, $R_0^{MF}$, we consider the online/offline engagement-free steady state when $[U]^* = N$. Here, the Malthusian parameter is
\begin{align*} 
 \lambda^{MF}  = \frac{1}{2}\left[ \left(k\tau -\eta-\gamma_i-\gamma_p \right) + \sqrt{\left(k\tau-\eta-\gamma_i-\gamma_p \right)^2-4 \left(\eta\gamma_p+\gamma_i\gamma_p-k\tau\gamma_p-\theta\eta \right)} \right].
\end{align*}
Then we derive that $\displaystyle R_0^{MF} = \frac{\lambda^{MF}}{\eta + \gamma_i} + 1$. 
We also compare $R_0^{MF}$ with the reproductive number calculated from the next generation matrix \cite{Next_Gen_R0}. The results are identical with a numerically negligible difference. 

Observe that $\operatorname{Re}(R_0^{MF}) < 1$ when
\begin{equation} \label{eq:stability_MF}
    k\tau -\eta-\gamma_i-\gamma_p < 0 \quad \mbox{and} \quad \eta\gamma_p+\gamma_i\gamma_p-k\tau\gamma_p-\theta\eta > 0.
\end{equation}
Notice that this stability condition corresponds to stability conditions for the fully-connected network \eqref{eq:stability1} and \eqref{eq:stability2} with $\beta$ replaced by $k\tau$.

Similar to the analysis performed on the stochastic model, we examine the impact of the degree on the $R_0^{\text{MF}} = 1$ threshold and how parameters influence online and offline outbursts in the homogeneous single-level model. The results are consistent with those shown in Figures \ref{fig:StR0Lines}, \ref{fig:stochastic_tau_gamma_i_E}, and \ref{fig:stochastic_tau_gamma_i_P}. Further details of these results are provided in Appendix \ref{app:R0_Heatmap_Homogeneous_SingleLevel}.

Beyond the engagement- and protest-free equilibrium, the system also admits more general equilibrium solutions. For completeness, we discuss these solutions and outline a reduced-dimensional stability analysis in Appendix \ref{app:stability_homogeneous_SingleLevel}. Additionally, a useful property of the homogeneous single-level model is that it allows for explicit solutions for $[U]$ and $[E]$, treating $[D]$ and $[R]$ as parameters. This approach provides another perspective to examine the dynamics and stability of solutions in the online layer, as detailed in Appendix \ref{app:solve_homogeneous_single}.

\subsubsection{Basic reproductive number: pairwise approximation}\label{subsubsec:homog_PWR0}

To find the reproductive number for the homogeneous pairwise model, we consider the online and offline engagement-free equilibrium solution of the homogeneous pairwise system 
\[ ([U], [E], [D], [UE], [UU][P], [R]) = (N, 0, 0, 0, Nk,0,0). \]
The Jacobian at this equilibrium is
\begin{align*}
   J^{PW}=\left[\begin{array}{ccccccc}
    0&0&0&-\tau&0&-\theta&0\\
    0&-(\eta+\gamma_i)&0&\tau&0&\theta&0\\
    0&\eta+\gamma_i&0&0&0&0&0\\
    0&0&0&(k-2)\tau-\eta-\gamma_i&0&k\theta&0\\
    0&0&0&-2(k-1)\tau&0&-2k\theta&0\\
    0&\eta&0&0&0&-\gamma_p&0\\
    0&0&0&0&0&\gamma_p&0    
    \end{array}\right]
\end{align*}
There are seven eigenvalues of $J^{PW}$, where four of them are zeros and the other three correspond to the first, second, and third roots of the following polynomial, respectively.
\begin{align*}
    f(x) = & x^3 + \left( 2\tau-k\tau+2\eta+2\gamma_i+\gamma_p \right)x^2 \\
           & + \left( 2\tau\eta-k\tau\eta+2\tau\gamma_i-k\tau\gamma_i+2\tau\gamma_p-k\tau\gamma_p+\eta^2-\eta\theta+2\eta\gamma_i+2\eta\gamma_p+\gamma_i^2+2\gamma_i\gamma_p \right)x \\
           & -2\tau\eta\theta+2\tau\eta\gamma_p-k\tau\eta\gamma_p+2\tau\gamma_{i}\gamma_p-k\tau\gamma_{i}\gamma_p-\eta^2\theta-\eta\theta\gamma_i+2\eta\gamma_{i}\gamma_p+\eta^2\gamma_p+\gamma_{i}^2\gamma_p.
\end{align*}
We numerically find the largest eigenvalue, which is the Malthusian parameter $\lambda^{PW}$ to the system. Then the reproductive number of the homogeneous pairwise model is $\displaystyle R_0^{PW} = \frac{\lambda^{PW}}{\eta+\gamma_i}+1$.

In Appendix \ref{app:R0_Heatmap_Homogeneous_Pairwise}, we examine the effect of degree on the $R_0^{PW} = 1$ threshold and explore how the parameter pair $(\gamma_i, \tau)$ influences online and offline outbursts in the homogeneous pairwise model.

\begin{figure}[!htb]
    \centering
    \begin{subfigure}[t]{0.32\textwidth}
        \centering
        \includegraphics[width=\linewidth]{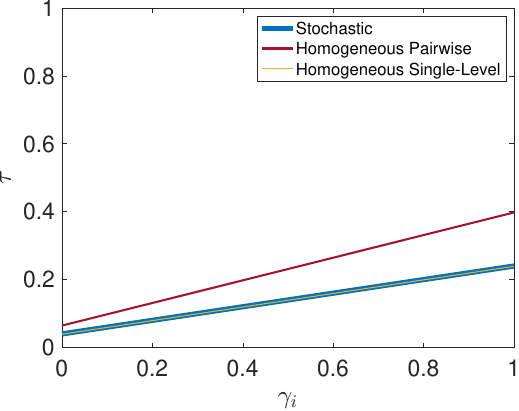} 
        \caption{$k = 5$} \label{fig:3R0-1}
    \end{subfigure}
    \hfill
    \begin{subfigure}[t]{0.32\textwidth}
        \centering
        \includegraphics[width=\linewidth]{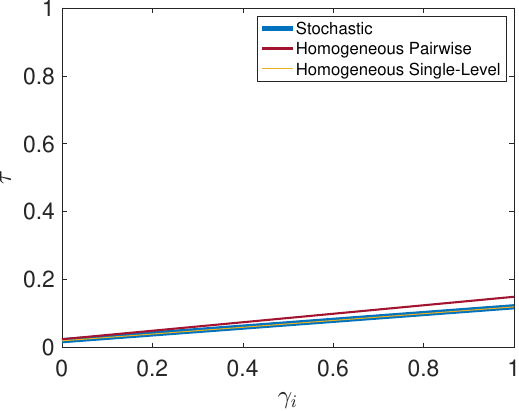} 
        \caption{$k = 10$} \label{fig:3R0-2}
    \end{subfigure}
    \hfill
    \begin{subfigure}[t]{0.32\textwidth}
        \centering
        \includegraphics[width=\linewidth]{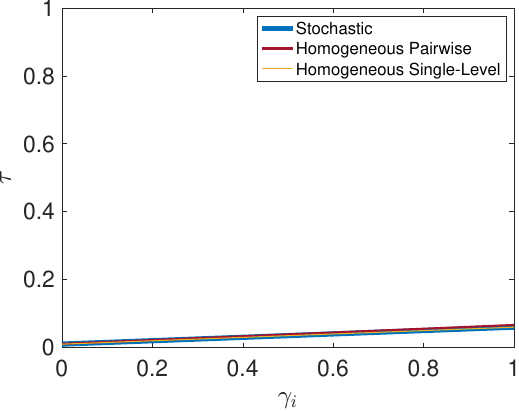} 
        \caption{$k = 20$} 
        \label{fig:3R0-3}
    \end{subfigure}
    \caption{Plot of the $R_0 = 1$, $R_0^{MF} = 1$, and $R_0^{PW} = 1$ thresholds for different degrees. The pairwise approximation predicts a smaller parameter region that will lead to outbursts. As $k$ increases, the $R_0^{PW} = 1$ line converges to those of $R_0 = 1$ and $R_0^{MF} = 1$.
}
    \label{fig:compare3R0s}
\end{figure}

We now have analytical formulas for the reproductive numbers of all three models on homogeneous networks—the stochastic model, the single-level approximation, and the pairwise approximation. We can compare how closely the lines $R_0 = 1$, $R_0^{MF} = 1$, and $R_0^{PW} = 1$ align with each other. Figure \ref{fig:compare3R0s} shows this comparison for the different degree $k$. The results indicate that the $R_0 = 1$ and $R_0^{MF} = 1$ thresholds closely overlap across different degrees, suggesting that the computation of $R_0^{MF}$ largely aligns with the approximation of $R_0$ in the stochastic model. On the other hand, the $R_0^{PW} = 1$ line lies farther from these thresholds at lower node degrees but converges toward them as the node degree increases.

\begin{figure}[!htb]
    \centering
    \begin{subfigure}[t]{0.46\textwidth}
        \centering
        \includegraphics[width=\linewidth]{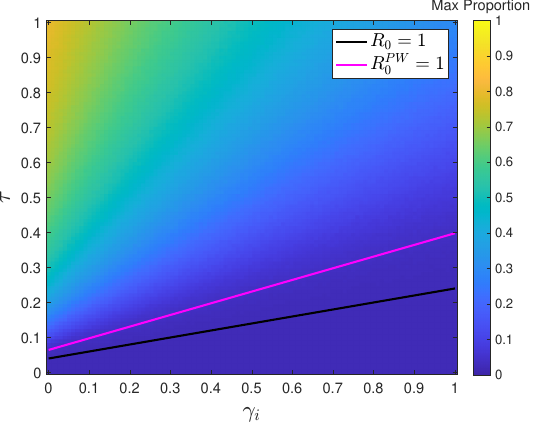} 
        \caption{Online $E$} \label{fig:stochastic_tau_gamma_i_E_with_Homogeneous_PairwiseR0}
    \end{subfigure}
    \hfill
    \begin{subfigure}[t]{0.46\textwidth}
    \centering
        \includegraphics[width=\linewidth]{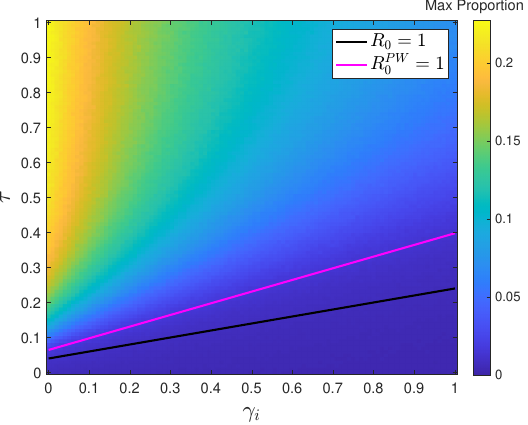} 
        \caption{Offline $P$} \label{fig:stochastic_tau_gamma_i_P_with_Homogeneous_PairwiseR0}
    \end{subfigure}
    \caption{Heatmaps from stochastic simulation on homogeneous networks for the proportions of the mean final size of outbursts in online $E$ and offline $P$ swept in $\tau$ and $\gamma_i$. Grid is $100 \times 100$. Parameters are $N=1\,000, k=5, \theta=0.01, \eta=0.2, \gamma_p=0.5, t_{\text{max}}=500, [E](0)=20$, and $\text{iter}=100$. The $R_0 = 1$ and $R_0^{PW} = 1$ lines are both plotted in black and pink, respectively. Compared to the $R_0^{PW}$ threshold, the $R_0$ threshold tends to classify a broader range of parameter values as leading to outbursts, making it a more aggressive criterion.}
    \label{fig:stochastic_tau_gamma_i_with_Homogeneous_PairwiseR0}
\end{figure}

It is important to note that $R_0^{PW}$ should not be interpreted as a poor approximation of the true reproductive number, since our $R_0$ expression for the stochastic model, equation~\eqref{eq:R0_kReg}, is itself an approximation—it does not capture the effects of the coupled feedback $\theta$ or the offline recovery rate $\gamma_p$. On the other hand, $R_0^{PW}$ may provide a more accurate approximation for classifying outbursts, particularly because it reflects additional structural dynamics absent from the simplified $R_0$ estimate. How accurate is the $R_0^{PW} = 1$ prediction empirically when compared to stochastic simulations? We can compare the $R_0^{PW} = 1$ threshold to the $R_0 = 1$ threshold within the $\gamma_i$-$\tau$ parameter sweep from the stochastic simulations, previously shown in Figures \ref{fig:stochastic_tau_gamma_i_E} and \ref{fig:stochastic_tau_gamma_i_P}. Figure \ref{fig:stochastic_tau_gamma_i_with_Homogeneous_PairwiseR0} presents this comparison, indicating that $R_0$ predicts outbursts more readily than $R_0^{PW}$.

\begin{figure}[!htb]
    \centering
    \begin{subfigure}[t]{0.425\textwidth}
        \centering
        \includegraphics[width=\linewidth]{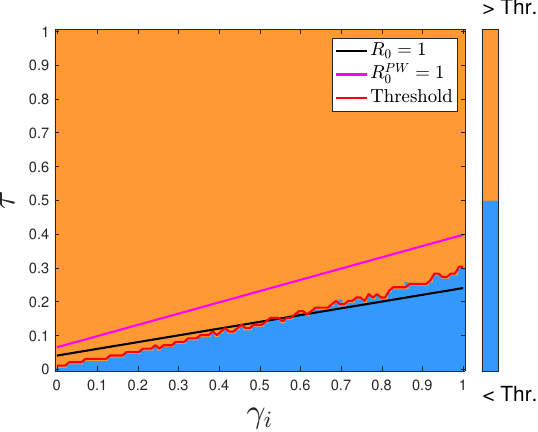} 
        \caption{Threshold $= 0.007$} \label{fig:PCategorization_stochastic_tau_gamma_i_2R-1}
    \end{subfigure}
    \hfill
    \begin{subfigure}[t]{0.425\textwidth}
        \centering
        \includegraphics[width=\linewidth]{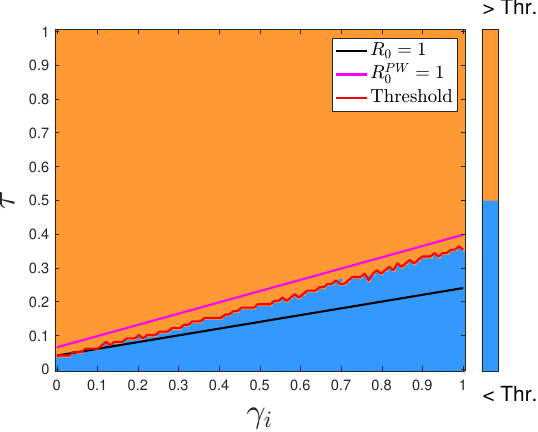} 
        \caption{Threshold $= 0.01$} \label{fig:PCategorization_stochastic_tau_gamma_i_2R-2}
    \end{subfigure}
    \begin{subfigure}[t]{0.425\textwidth}
        \centering
        \includegraphics[width=\linewidth]{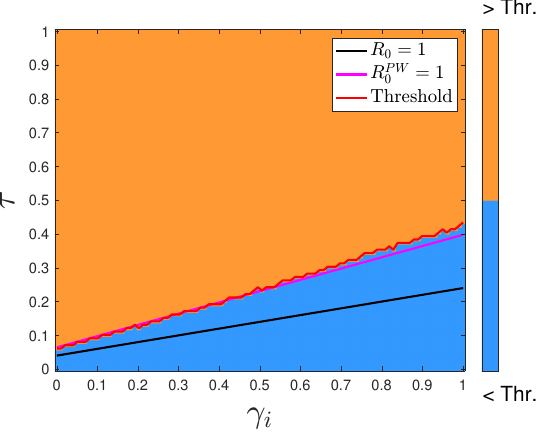} 
        \caption{Threshold $= 0.015$} 
        \label{fig:PCategorization_stochastic_tau_gamma_i_2R-3}
    \end{subfigure}
    \hfill
    \begin{subfigure}[t]{0.425\textwidth}
        \centering
        \includegraphics[width=\linewidth]{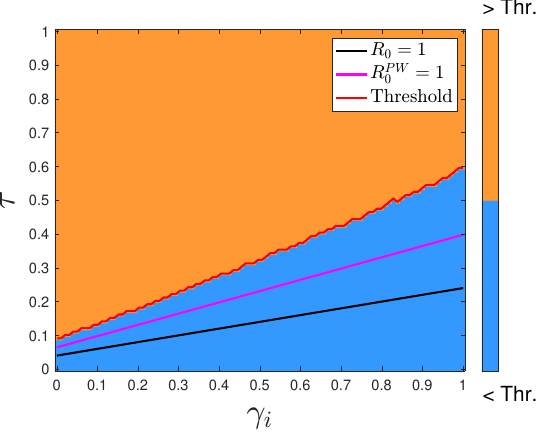} 
        \caption{Threshold $= 0.035$} \label{fig:PCategorization_stochastic_tau_gamma_i_2R-4}
    \end{subfigure}
    \caption{Categorization of protesting outbursts based on threshold for $P$ proportions swept in $\gamma_i$ and $\tau$ of stochastic model on homogeneous networks. Grid resolution is $100 \times 100$. Parameters are $N=1\,000, k=5, \theta=0.01, \eta=0.2, \gamma_p=0.5, t_{\text{max}}=500, [E](0)=20$, and $\text{iter}=100$. The $R_0 = 1$ and $R_0^{PW} = 1$ lines are plotted in black and pink, respectively. As the selected classification threshold increases, our predicted $R_0$ and $R_0^{PW}$ become more conservative.}
    \label{fig:PCategorization_stochastic_tau_gamma_i_2R0}
\end{figure}

Two questions of interest are: How can we determine whether the $R_0 = 1$ threshold is empirically meaningful, and how can we identify a suitable cutoff? Considering offline protesting outbursts, we can define a classification threshold and compare it against the $R_0 = 1$ line. Figure \ref{fig:PCategorization_stochastic_tau_gamma_i_2R0} provides illustrative examples comparing the $R_0 = 1$ threshold with various cutoff values for the proportions of offline outburst sizes obtained from stochastic simulations across the parameter sweep in $(\gamma_i, \tau)$ shown in Figure~\ref{fig:stochastic_tau_gamma_i_P_with_Homogeneous_PairwiseR0}. We also superimpose $R_0^{PW} = 1$ threshold to examine how it differs from $R_0 = 1$. The results suggest that the analytical $R_0 = 1$ threshold yields a relatively conservative prediction when categorizing social protests as outbursts, compared to $R_0^{PW} = 1$. Notably, according to \cite{chenoweth}, social protests with at least 3.5\% active participation during peak events have historically tended to succeed, with very few observed exceptions, suggesting this may represent a critical empirical threshold. Compared to this value, both $R_0 = 1$ and $R_0^{PW} = 1$ appear overly conservative in classifying outbursts.

\section{Dynamics on heterogeneous networks}\label{sec:hetero}

The mean-field approximations derived in Section~\ref{sec:homo} are not reliable when the network structure deviates significantly from homogeneity in node degrees. This limitation arises because the closure assumptions used in the homogeneous case are violated in heterogeneous networks. To address this, we extend our approach by explicitly tracking the evolution of node states conditioned on degree. In particular, we develop mean-field models tailored to heterogeneous networks and compare their performance on both synthetic and empirical networks. Section~\ref{subsec:hetero_single} introduces the single-level approximation for heterogeneous networks, while Section~\ref{subsec:hetero_PW} presents the pairwise model. Section~\ref{subsec:hetero_compare} compares both approximations against the stochastic model: Section~\ref{subsubsec:hetero_synthetic} focuses on synthetic networks, while Section~\ref{subsubsec:hetero_empirical} examines two real-world Facebook networks.

In our heterogeneous framework, we track the expected number of individuals in state $X$ with degree $n.$ We denote by $[X_nY_m]$ the expected number of edges connecting individuals in state $X$ with degree $n$ to individuals in state $Y$ with degree $m$. Similarly, $[X_nY_mZ_p]$ denotes the expected number of triples in which a degree-$m$ node in state $Y$ is connected to both a degree-$n$ node in state $X$ and a degree-$p$ node in state $Z$. These notations are summarized in Table~\ref{table:3}. 
\begin{table}[!htb]
\caption{Description of terms used in the heterogeneous model}
\label{table:3}
\begin{center}
\resizebox{\textwidth}{!}{%
\begin{tabular}{|l|l|}
\hline
Term                          & Description                                                          \\ \hline
$K$                           & Maximum degree of any node                                \\
$N$                           & Total number of nodes                                \\
$[X]$                         & Expected $\#$ of individuals in state $X$                                 \\
$[X_n]$                       & Expected $\#$ of individuals in state $X$ with degree $n$                  \\
$[n]$                         & Expected $\#$ of individuals with degree $n$                                   \\
$[X_nY_m]$                    & Expected $\#$ of edges connecting individuals in $X_n$ with individuals $Y_m$ \\
$[nm]$                        & Expected $\#$ of edges between individuals with $n$ and $m$ edges             \\
$[X_nY]=\sum_{m}X_nY_m$       & Expected total $\#$ of individuals in $Y$ connected to $X_n$                  \\
$[XY]=\sum_n\sum_{m}X_nY_m$   & Expected total $\#$ of edges between $X$ and $Y$                            \\
$[X_nY_mZ_p]$                 & Expected $\#$ of triples with $Y_m$ having both $X_n$ and $Z_p$ neighbors       \\ \hline
\end{tabular}
}
\end{center}
\end{table}

In the information layer, we have the following system of equations for the number of individuals in each online state with a given number of edges, $n$,
\begin{align*}
\left\{\begin{array}{ll}
\frac{d[U_n]}{dt}&=-\tau[U_nE] - \frac{\theta}{N}[U_n][P],\vspace{4pt}\\
\frac{d[E_n]}{dt}&=\tau[U_nE] + \frac{\theta}{N}[U_n][P]-(\eta+\gamma_i)[E_n],\vspace{4pt}\\
\frac{d[D_n]}{dt}&=(\eta+\gamma_i)[E_n], 
\end{array}\right.
\end{align*}
and in the physical layer, we obtain
\begin{align*}
\left\{\begin{array}{ll}
\frac{d[P_n]}{dt}&= \eta[E_n] -\gamma_p[P_n]\vspace{4pt}\\
\frac{d[R_n]}{dt}&= \gamma_p[P_n] 
\end{array}\right.
\end{align*}
where $n = 1,\ldots, K$.

\subsection{Heterogeneous mean-field at the single level}\label{subsec:hetero_single}

Inspired by \cite{eames2002modeling}, we use the closure 
$$
[X_nY_m]\approx [nm]\frac{X_n}{[n]}\frac{Y_m}{[m]}
$$
which forms the basis for our single-level mean-field approximation model on heterogeneous networks. Assuming, furthermore, that 
$$
[nm]=nm[n][m]/\sum_{j=1}^{K}j[j],
$$
where $K$ is the maximum degree any node has, we get a heterogeneous single-level approximation for the online dynamics and our system becomes
\begin{align}\label{model:Heterogeneous_SingleLevel}
\left\{\begin{array}{ll}
    \frac{d[U_n]}{dt}&=-\tau n[U_n]\pi_E - \frac{\theta}{N}[U_n][P],\\[4pt]
    \frac{d[E_n]}{dt}&=\tau n[U_n]\pi_E + \frac{\theta}{N}[U_n][P]-(\eta+\gamma_i)[E_n],\\[4pt]
    \frac{d[D_n]}{dt}&=(\eta+\gamma_i)[E_n],\\[4pt]
    \frac{d[P](t)}{dt}& = \eta[E]-\gamma_p[P],\\[4pt]
    \frac{d[R](t)}{dt}& = \gamma_p[P].
\end{array}\right.
\end{align}
where $\pi_E$ is the probability that a randomly chosen stub is connected to a stub of an engaged node, and is equal to
$$
\pi_E = \frac{\sum_{j=1}^Kj[E_j]}{\sum_{j=1}^Kj[j]}.
$$
Note that in the derivation of the model, we also use the approximation that the average number of edges connecting uninterested degree-$n$ nodes to engaged nodes is approximated by
$$
[U_nE]\approx n[U_n]\pi_E.
$$
See \cite{SingleLevelApprox} for a derivation of this approximation. 

\begin{figure}[!htb]
    \centering
    \begin{subfigure}[b]{0.49\textwidth}  
        \centering 
        \includegraphics[width=\textwidth]{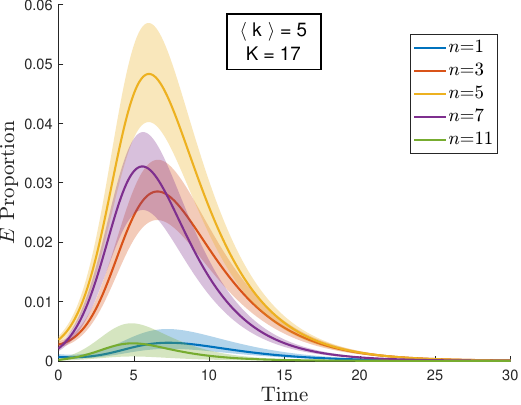}
        \caption{Dynamics with lower average degree}%
        \label{fig:HeteroSingle_k_dynamics_Online_k=5}
    \end{subfigure}
    \hfill
    \begin{subfigure}[b]{0.49\textwidth}  
        \centering 
        \includegraphics[width=\textwidth]{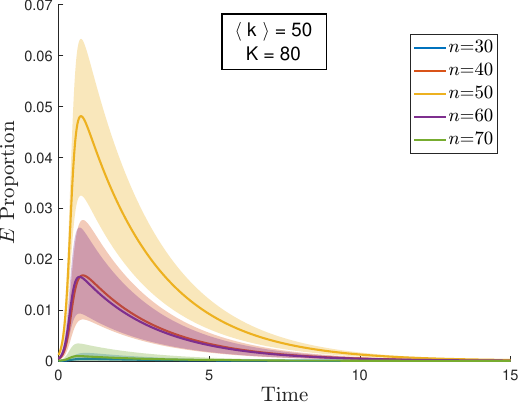}
        \caption{Dynamics with higher average degree}%
        \label{fig:HeteroSingle_k_dynamics_Online_k=50}
    \end{subfigure}
    \caption{Dynamics of the heterogeneous single-level model~\eqref{model:Heterogeneous_SingleLevel}: The proportion of number of nodes in the online state $E$, grouped by selected node degrees, is shown over time. Results are based on simulations across $100$ Erd\H{o}s-R\'{e}nyi random networks. Shaded regions represent the range between the maximum and minimum values across all simulations over time. Parameters used are $N = 1\,000$, $\tau = 0.2$, $\theta = 0.01$, $\eta = 0.2$, $\gamma_i = 0.2$, $\gamma_p = 0.5$, $t_{\text{max}} = 200$, and $ [E](0) = 20$. Based on the two examples shown, the degrees of engaged nodes are most concentrated near the average, with fewer nodes exhibiting degrees at the extremes.}      
    \label{fig:HeteroSingle_k_dynamics}
\end{figure}

The approximated model for heterogeneous networks groups the number of nodes in each online state by degree. In Figure \ref{fig:HeteroSingle_k_dynamics}, we present examples of online $E$ dynamics on Erd\H{o}s-R\'{e}nyi random networks, grouped by selected degrees, $n$, centred around the average degree, denoted as $\langle k \rangle$. At each time point, we observe that most engaged nodes have degrees near the average, while the number of engaged nodes with degrees farther from the average decreases. This pattern is consistent with the expected degree distribution of Erd\H{o}s-R\'{e}nyi random networks.

\subsection{Heterogeneous pairwise approximation}\label{subsec:hetero_PW}

We look at a pairwise model closure at the level of triplets and keep track of how the edges evolve (including the effect of $\eta$ and $P$):
\begin{align*}
\left\{\begin{array}{ll}
\frac{d[U_nE_m]}{dt}=&\tau\sum_j\left([U_nU_mE_j] - [E_jU_nE_m]\right)\vspace{4pt}\\
&-(\tau +\eta+\gamma_i)[U_nE_m]+\frac{\theta [P]}{N}[U_nU_m]-\frac{\theta [P]}{N}[U_nE_m], \vspace{4pt}\\
\frac{d[U_nU_m]}{dt}=&-\tau\sum_j\left([U_nU_mE_j] + [E_jU_nU_m]\right)-\frac{2\theta [P]}{N}[U_nU_m].\vspace{4pt}\\
\end{array}\right.
\end{align*}
Using the approximation proposed by House and Keeling \cite{House_Keeling}, we have
\begin{align}\label{approx:compact_pair2}
    [U_n E] \approx [UE] \frac{n[U_n]}{\sum_{l=1}^K l[U_l]} \mbox{ and } [U_n U] \approx [UU] \frac{n[U_n]}{\sum_{l=1}^K l[U_l]}.
\end{align}
This assumes that the neighbors of all uninterested people are interchangeable. Using this approximation, the triple closure
\[[U_n U_m E] \approx \frac{m-1}{m} \frac{[U_n U_m][U_m E]}{[U_m]}\]
takes the form
\[[U_n U_m E] \approx (m-1)[U_n U_m] \frac{[UE]}{[UX]}\]
where $[UX] = \sum_{l=1}^K l[U_l]$. Applying approximation \eqref{approx:compact_pair2} yields
\begin{align*}
    \sum_{m=1}^K \sum_{n=1}^K [U_n U_m E] &\approx \sum_{m=1}^K (m-1)[UU_m]\frac{[UE]}{[UX]}\\
    &\approx \frac{[UU][UE]}{[UX]^2} \sum_{m=1}^K (m-1)m[U_m].
\end{align*}
Similarly, we can derive approximations
\[\sum_{n=1}^K \sum_{m=1}^K [E U_n U_m] \approx \frac{[UE][UU]}{[UX]^2} \sum_{n=1}^K (n-1)n[U_n]\]
and 
\[\sum_{n=1}^K \sum_{m=1}^K [E U_n E_m] \approx \frac{[UE][UE]}{[UX]^2} \sum_{n=1}^K (n-1)n[U_n].\]
For simplicity, we define $Q_c = \frac{1}{[UX]^2} \sum_{n=1}^K (n-1)n[U_n]$.  This gives the pairwise approximation:
\begin{align}\label{model:Heterogeneous_Pairwise}
\left\{\begin{array}{ll}
    \frac{d[U_n]}{dt}&=-\tau n[U_n]\frac{[UE]}{[UX]} - \frac{\theta}{N} [U_n][P],\\[4pt]
    \frac{d[E_n]}{dt}&= \tau n[U_n]\frac{[UE]}{[UX]} + \frac{\theta}{N} [U_n][P]-(\eta+\gamma_i)[E_n],\\[4pt]
    \frac{d[D_n]}{dt}&=(\eta+\gamma_i)[E_n],\\[4pt]
    \frac{d[UE]}{dt}=&\tau Q_c[UE]([UU]-[UE]) - (\tau+\eta+\gamma_i+\frac{\theta}{N}[P])[UE]\\[4pt] 
    & +\frac{\theta}{N}[P][UU],\\[4pt]
    \frac{d[UU]}{dt}=&-2\tau Q_c[UU][UE] - \frac{2\theta}{N}[P][UU],\\[4pt]
    \frac{d[P](t)}{dt}& = \eta[E]-\gamma_p[P],\\[4pt]
    \frac{d[R](t)}{dt}& = \gamma_p[P].
\end{array}\right.
\end{align}
for $n=1,\dots, K$.

\begin{figure}[!htb]
    \centering
    \begin{subfigure}[b]{0.49\textwidth}  
        \centering 
        \includegraphics[width=\textwidth]{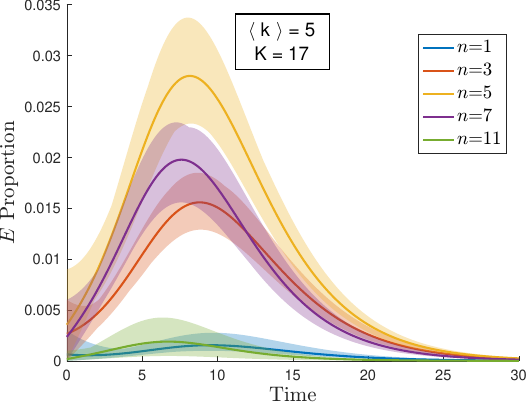}
        \caption{Dynamics with lower average degree}%
        \label{fig:HeteroPW_k_dynamics_Online_k=5}
    \end{subfigure}
    \hfill
    \begin{subfigure}[b]{0.49\textwidth}  
        \centering 
        \includegraphics[width=\textwidth]{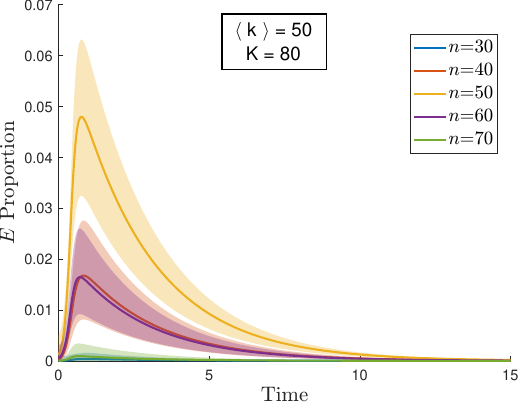}
        \caption{Dynamics with higher average degree}%
        \label{fig:HeteroPW_k_dynamics_Online_k=50}
    \end{subfigure}
    \caption{Dynamics of the heterogeneous pairwise model \eqref{model:Heterogeneous_Pairwise} in $E$. Parameters and figure attributes match those in Figure~\ref{fig:HeteroSingle_k_dynamics}. Compared to the single-level model, the pairwise model shows generally lower engagement across degree groups when $\langle k \rangle$ is small, while its degree-based group dynamics more closely resemble those of the single-level model as $\langle k \rangle$ increases.}      
    \label{fig:HeteroPW_k_dynamics}
\end{figure}

Similar to the heterogeneous single-level model, we present in Figure \ref{fig:HeteroPW_k_dynamics} examples of online $E$ dynamics on Erd\H{o}s–R\'{e}nyi random networks, grouped by selected degrees. Compared to the dynamics shown in Figure \ref{fig:HeteroSingle_k_dynamics} for the heterogeneous single-level model, we observe that when the network has a smaller average degree $\langle k \rangle$, the engaged population tends to be generally lower across degree groups in the heterogeneous pairwise model.

The dynamics of models on heterogeneous networks are more challenging to examine analytically. However, we can use numerical simulations and leverage analytical results from homogeneous-network models to interpret and understand outburst phenomena in heterogeneous cases. Further discussions related to these results are provided in Appendix \ref{app:heatmap_heterogeneous}.

\subsection{Comparison of results}\label{subsec:hetero_compare}
We present and compare simulation results between the stochastic, single-level, and pairwise models on heterogeneous networks. We test both on synthetic networks and empirical networks.

\subsubsection{Synthetic networks}\label{subsubsec:hetero_synthetic}

\begin{figure}[!htb]
    \centering
    \begin{subfigure}[b]{0.24\textwidth}  
        \centering 
        \includegraphics[width=\textwidth]{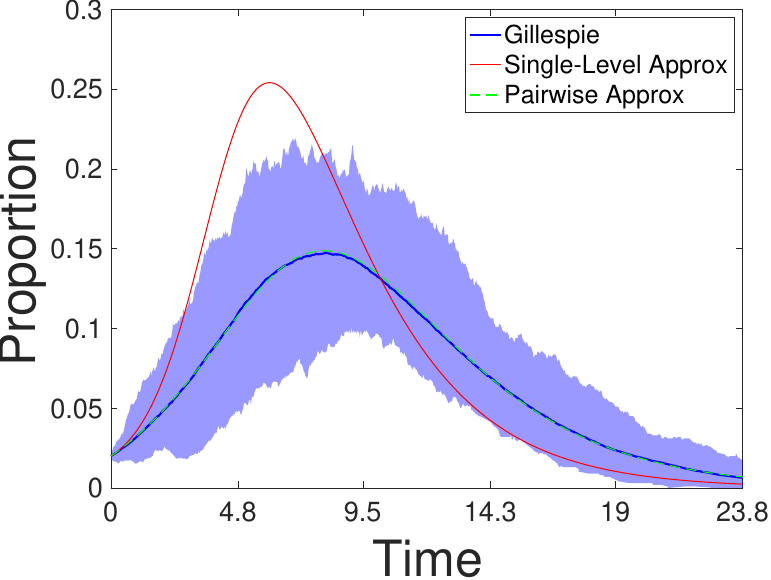}
        \caption{Online $\langle k \rangle=5$}%
        \label{fig:ERonline5}
    \end{subfigure}
    \begin{subfigure}[b]{0.24\textwidth}  
        \centering 
        \includegraphics[width=\textwidth]{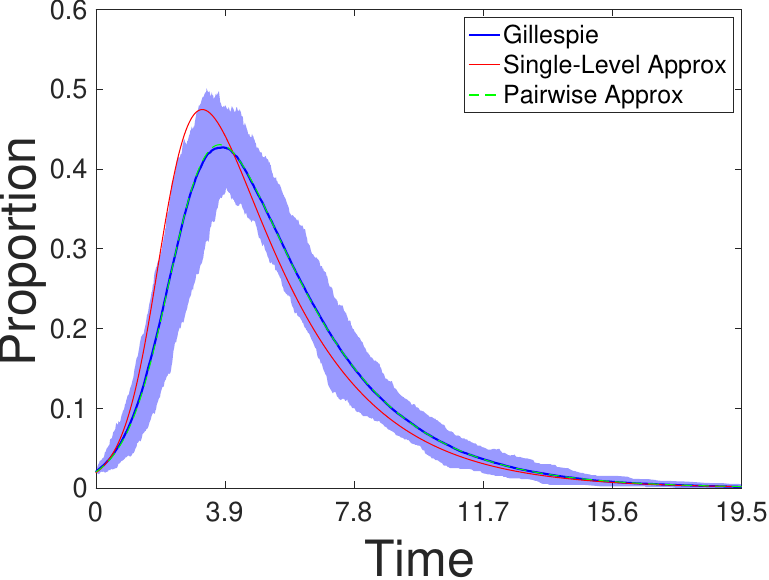}
        \caption{Online $\langle k \rangle=10$}%
        \label{fig:ERonline10}
    \end{subfigure}
    \begin{subfigure}[b]{0.24\textwidth}  
        \centering 
        \includegraphics[width=\textwidth]{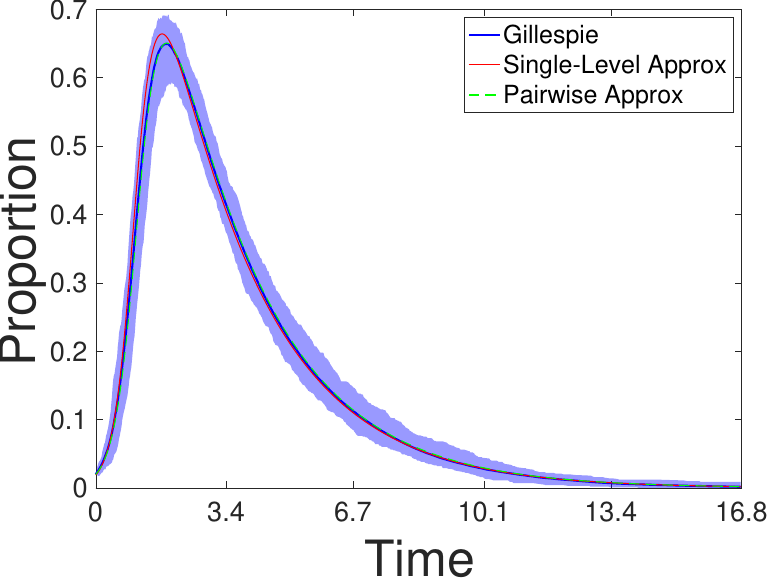}
        \caption{Online $\langle k \rangle=20$}%
        \label{fig:ERonline20}
    \end{subfigure}
    \begin{subfigure}[b]{0.24\textwidth}  
        \centering 
        \includegraphics[width=\textwidth]{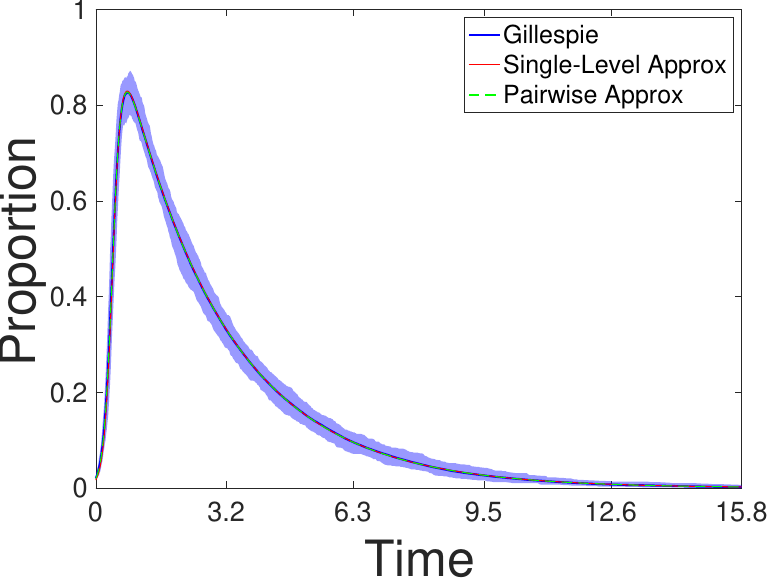}
        \caption{Online $\langle k \rangle=50$}%
        \label{fig:ERonline50}
    \end{subfigure}\\
        \begin{subfigure}[b]{0.24\textwidth}  
        \centering 
        \includegraphics[width=\textwidth]{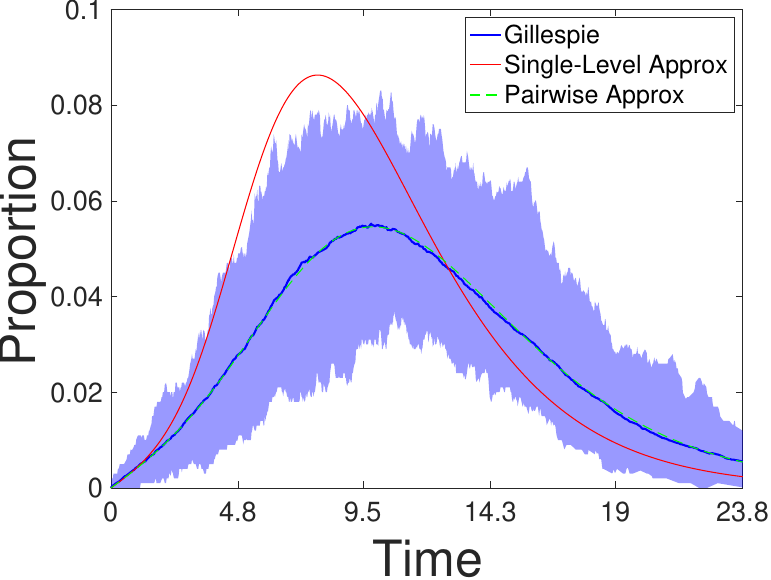}
        \caption{Offline $\langle k \rangle=5$}%
        \label{fig:ERoffline5}
    \end{subfigure}
    \begin{subfigure}[b]{0.24\textwidth}  
        \centering 
        \includegraphics[width=\textwidth]{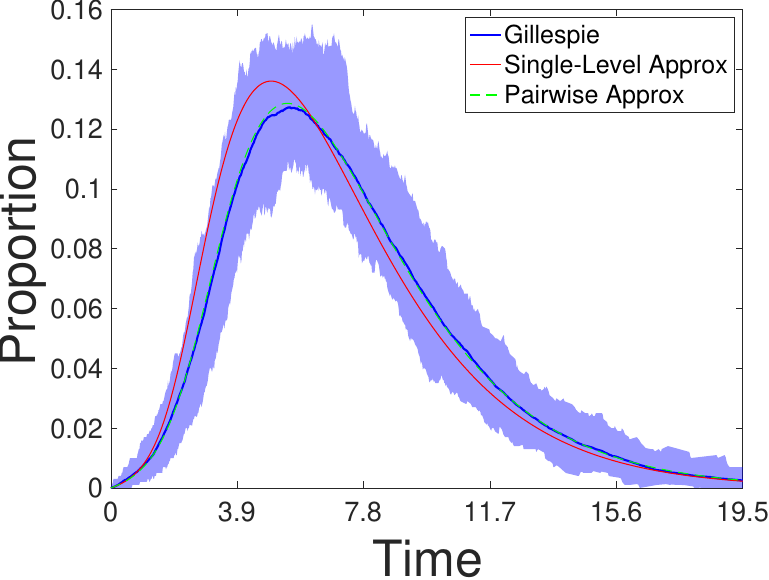}
        \caption{Offline $\langle k \rangle=10$}%
        \label{fig:ERoffline10}
    \end{subfigure}
    \begin{subfigure}[b]{0.24\textwidth}  
        \centering 
        \includegraphics[width=\textwidth]{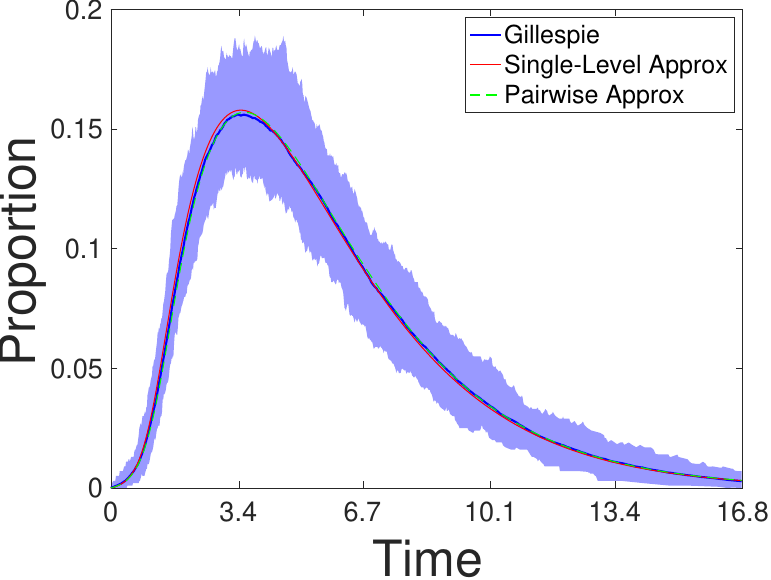}
        \caption{Offline $\langle k \rangle=20$}%
        \label{fig:ERoffline20}
    \end{subfigure}
    \begin{subfigure}[b]{0.24\textwidth}  
        \centering 
        \includegraphics[width=\textwidth]{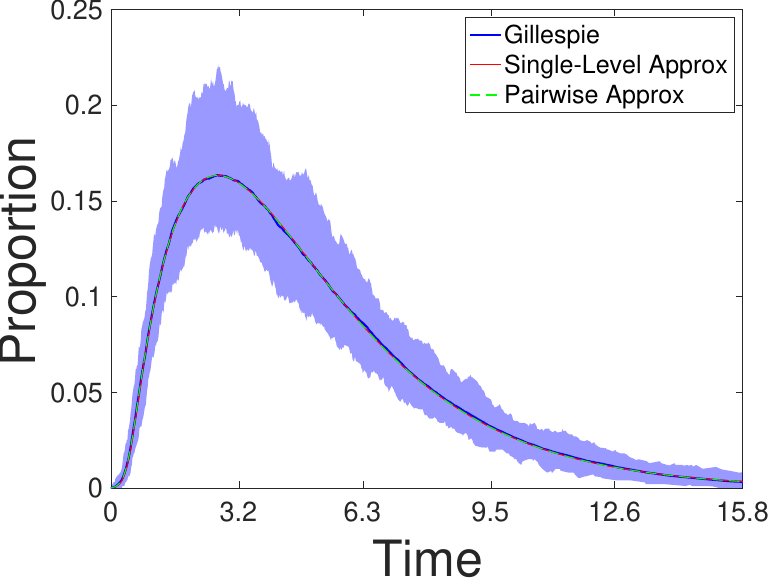}
        \caption{Offline $\langle k \rangle=50$}%
        \label{fig:ERoffline50}
    \end{subfigure}
    \caption{Dynamics of online $E$ (top row) and offline $P$ (bottom row) on Erd\H{o}s-R\'{e}nyi networks with different average degree $\langle k \rangle$. $N = 1\,000, \tau=0.2, \theta=0.01, \eta=0.2, \gamma_i=0.2, \gamma_p=0.5, t_{\text{max}}=200, [E](0)=20$, and $\text{iter}=100$. Time shown up to the shortest stochastic run. Shaded area spans min–max across runs. The pairwise approximation aligns more closely with the stochastic mean dynamics than the single-level approximation; the latter improves as $\langle k \rangle$ increases.}     
    \label{fig:EREgs}
\end{figure}

We examine the dynamics on Erd\H{o}s-R\'{e}nyi networks. Figure \ref{fig:EREgs} compares the $E$ compartment and $P$ compartment dynamics across our different models, varying average degree $\langle k \rangle$ while keeping $N = 1\,000$ fixed. The plots show that the pairwise approximation aligns more closely with the mean dynamics of the stochastic model than the single-level approximation. Moreover, as the average degree of the network increases, the performance of the single-level approximation improves. This is because the closure assumption underlying the single-level model becomes more accurate as network connectivity becomes denser.

In Appendix~\ref{app:heter_model_on_other_syn_nets}, we compare model dynamics on additional synthetic networks, including bimodal random networks, scale-free random networks, and Watts-Strogatz networks.

\subsubsection{Real-world networks}\label{subsubsec:hetero_empirical}

To explore model behaviour on real-world networks of different scales, we present simulation results on two Facebook networks. 

\begin{figure}[!htb]
    \centering
    \begin{subfigure}[b]{0.32\textwidth}  
        \centering 
        \includegraphics[width=\textwidth]{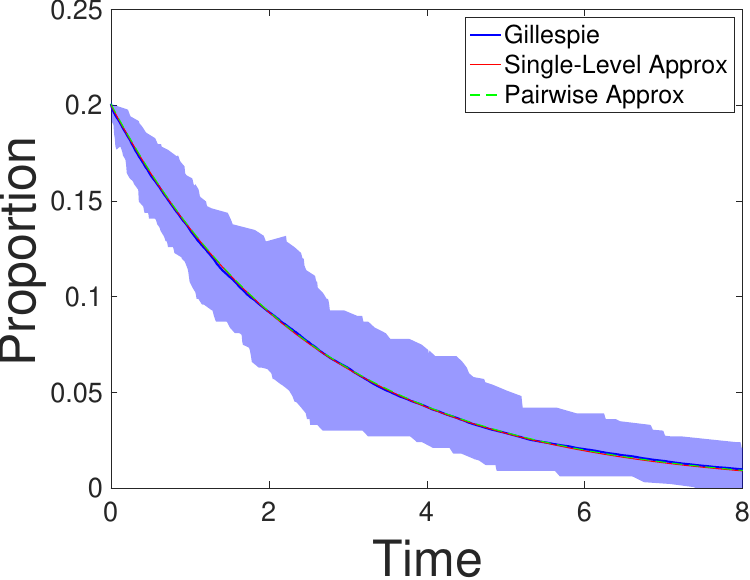}
        \caption{Online}%
    \end{subfigure}
    \hfill
    \begin{subfigure}[b]{0.32\textwidth}  
        \centering 
        \includegraphics[width=\textwidth]{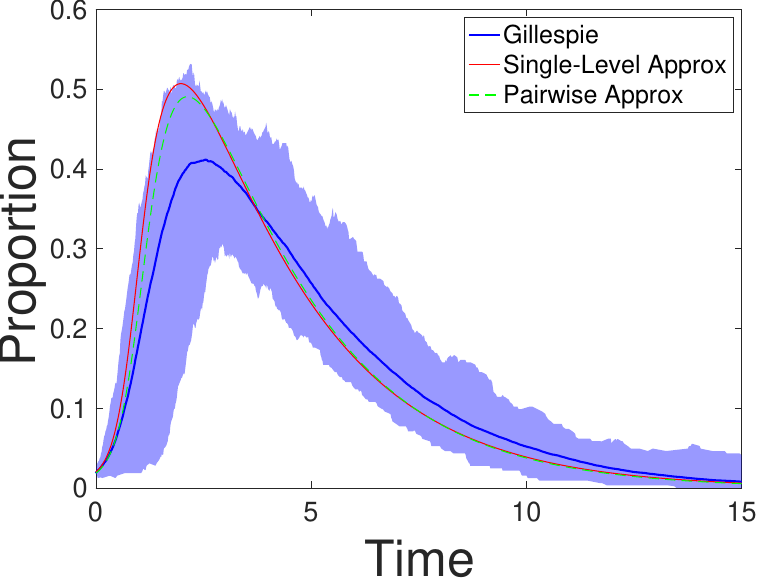}
        \caption{Online}%
    \end{subfigure}
    \hfill
    \begin{subfigure}[b]{0.32\textwidth}  
        \centering 
        \includegraphics[width=\textwidth]{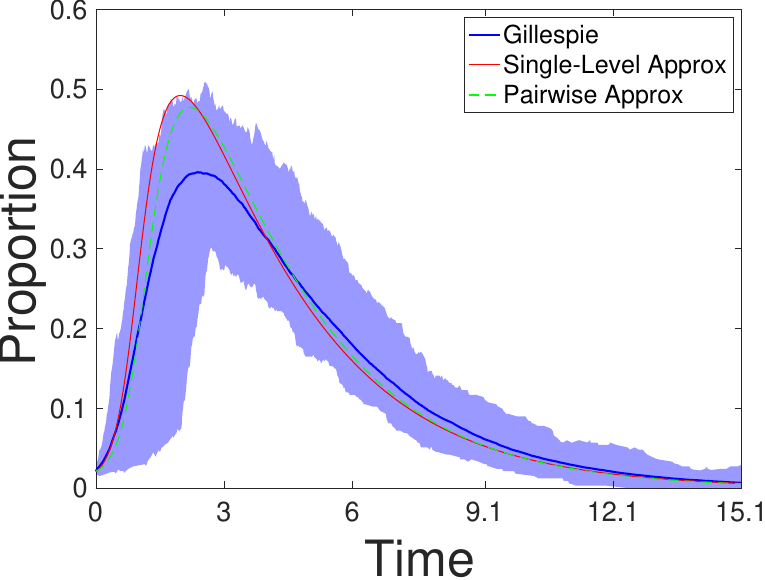}
        \caption{Online}%
    \end{subfigure}\\
        \begin{subfigure}[b]{0.32\textwidth}  
        \centering 
        \includegraphics[width=\textwidth]{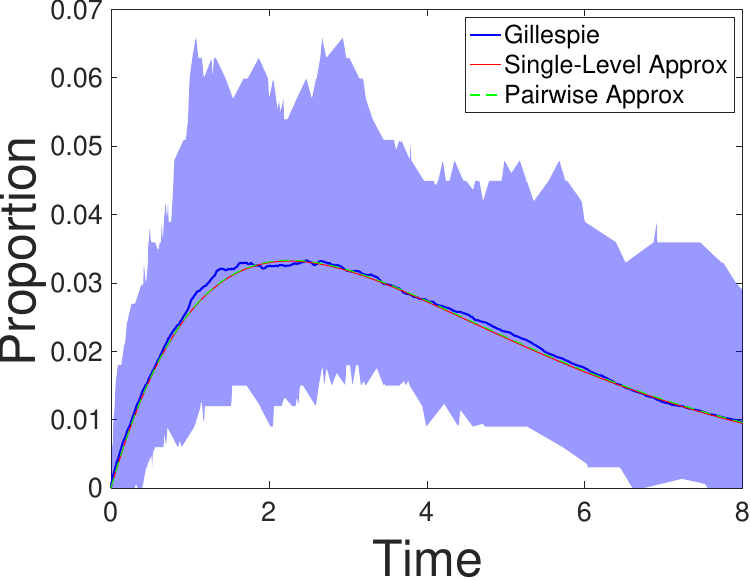}
        \caption{Offline}%
    \end{subfigure}
    \hfill
    \begin{subfigure}[b]{0.32\textwidth}  
        \centering 
        \includegraphics[width=\textwidth]{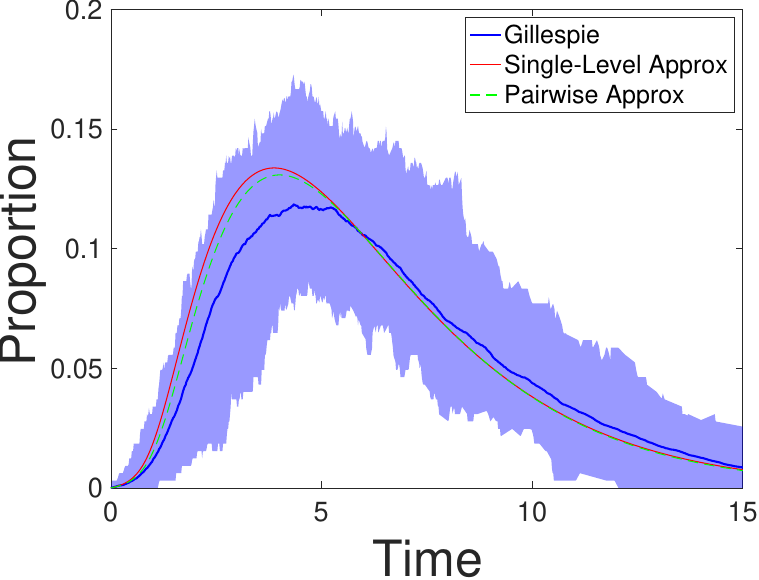}
        \caption{Offline}%
    \end{subfigure}
    \hfill
    \begin{subfigure}[b]{0.32\textwidth}  
        \centering 
        \includegraphics[width=\textwidth]{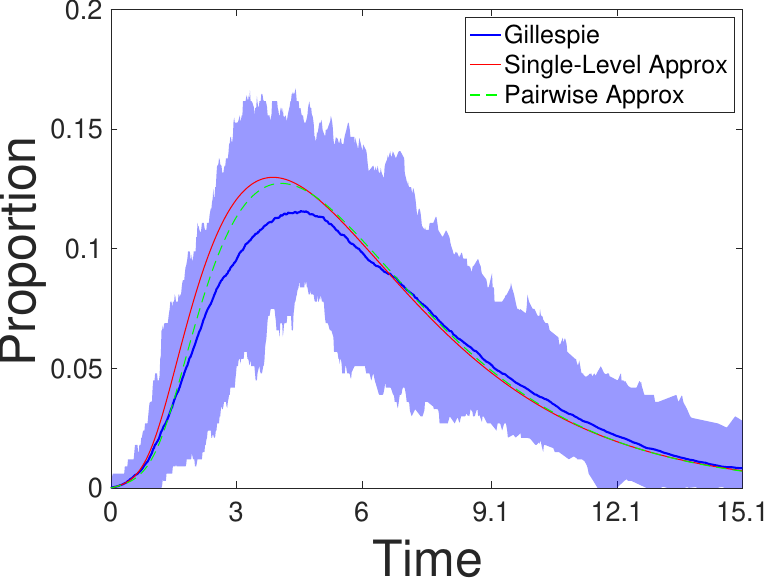}
        \caption{Offline}%
    \end{subfigure}
    \caption{Comparison of the solutions to the stochastic model, single-level and pairwise approximations. The top row illustrates online activity, and the bottom row the offline activity. (a), (c), (d), and (f) are on the original Facebook Friendship network with all $334$ nodes. (b) and (e) are on the giant cluster of the Facebook Friendship network, which contains $324$ nodes. The parameters used are (a), (d) $\tau=0.001, \theta=0.01, \eta=0.2, \gamma_i=0.2, \gamma_p=0.5, t_{\text{max}}=200, [E](0)/N=0.2$, and $\text{iter}=100$; (b), (c), (e), (f) $\tau=0.2, \theta=0.01, \eta=0.2, \gamma_i=0.2, \gamma_p=0.5, t_{\text{max}}=200, [E](0)/N=0.02$, and $\text{iter}=100$. Both approximation models closely match the stochastic mean when $E$ strictly decays. When online outbursts occur, approximation errors become more pronounced, with the pairwise model showing only marginal improvement over the single-level. The inclusion or exclusion of isolated components has negligible effect on the dynamics.}     \label{fig:real_network_example_FB}
\end{figure}

Figure \ref{fig:real_network_example_FB} compares model simulations on a Facebook Friendship network \cite{FF_network} consisting of $334$ nodes and $2\,218$ edges. The mean node degree is approximately $13.3$, with a standard deviation of about $10.4$, reflecting substantial variability in node connectivity. The average clustering coefficient is approximately $0.45$,  which is notably higher than those observed in the synthetic networks we examined, except for the Watts-Strogatz networks (see Appendix~\ref{app:heter_model_on_other_syn_nets}), which are designed to capture small-world effects. To eliminate the potential influence of isolated nodes, we also consider simulations restricted to the giant component. For this network, the giant component contains $324$ nodes. The simulations show little difference in the overall dynamics between the full network, including isolated nodes, and the giant component. Additionally, we observe that both approximation models align more closely with the stochastic mean when the online $E$ exhibits strictly decaying behaviour.

\begin{figure}[!htb]
    \centering
    \begin{subfigure}[b]{0.32\textwidth}  
        \centering 
        \includegraphics[width=\textwidth]{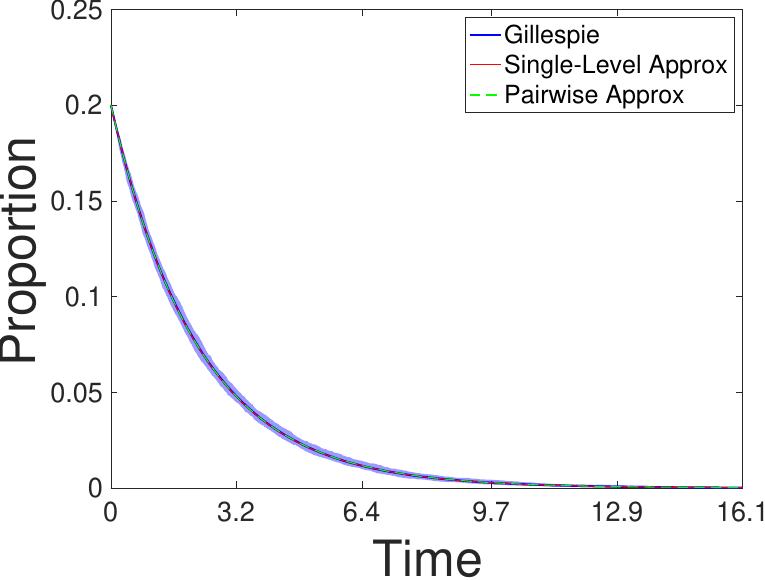}
        \caption{Online}%
    \end{subfigure}
    \hfill
    \begin{subfigure}[b]{0.32\textwidth}  
        \centering 
        \includegraphics[width=\textwidth]{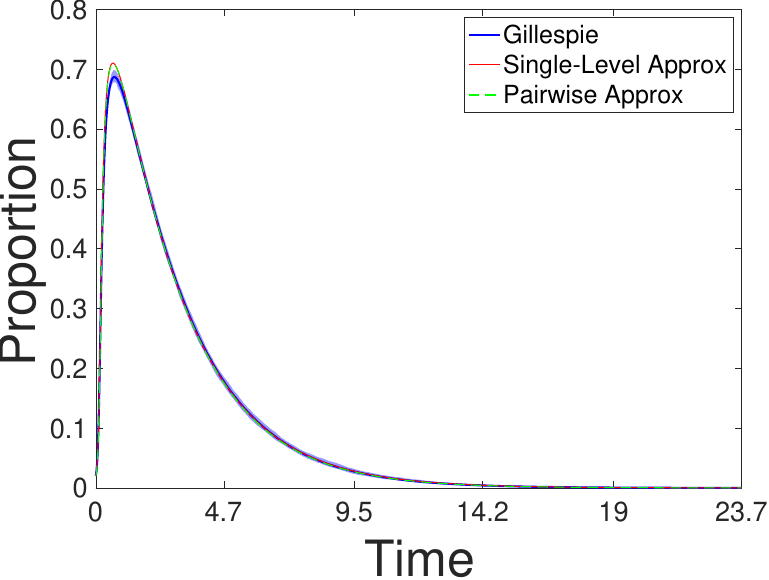}
        \caption{Online}%
    \end{subfigure}
    \hfill
    \begin{subfigure}[b]{0.32\textwidth}  
        \centering 
        \includegraphics[width=\textwidth]{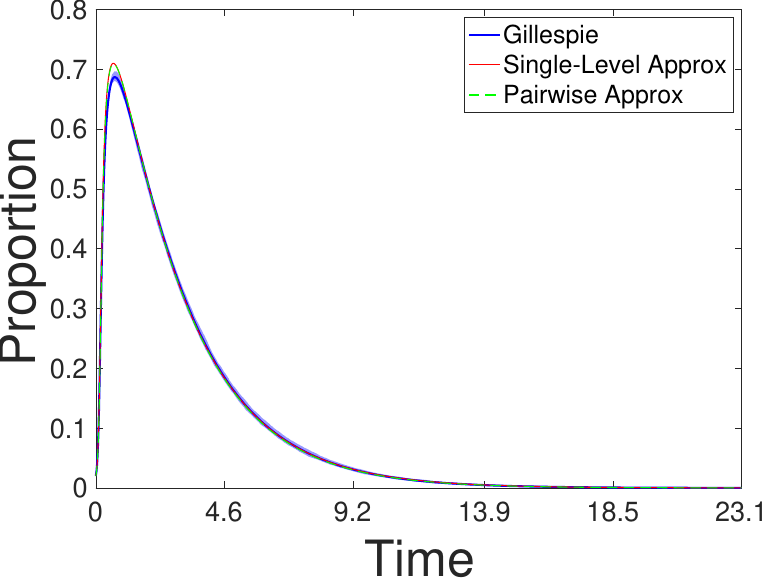}
        \caption{Online}%
    \end{subfigure}\\
        \begin{subfigure}[b]{0.32\textwidth}  
        \centering 
        \includegraphics[width=\textwidth]{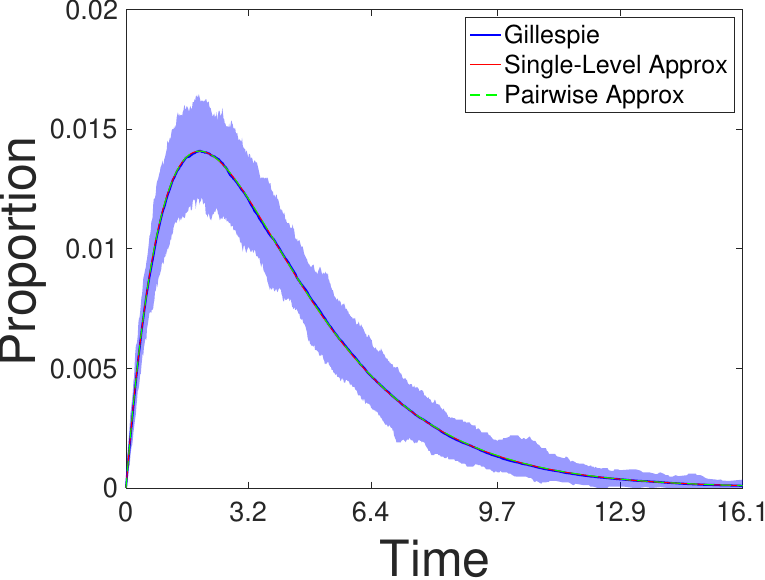}
        \caption{Offline}%
    \end{subfigure}
    \hfill
    \begin{subfigure}[b]{0.32\textwidth}  
        \centering 
        \includegraphics[width=\textwidth]{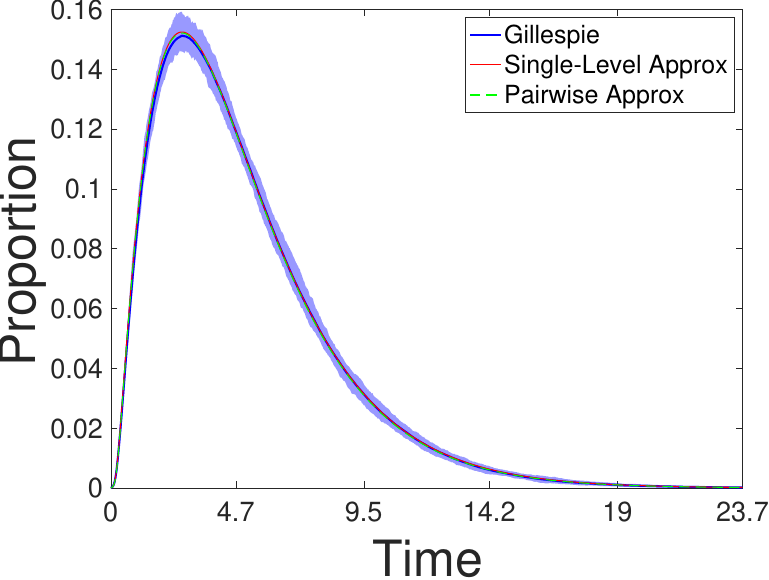}
        \caption{Offline}%
    \end{subfigure}
    \hfill
    \begin{subfigure}[b]{0.32\textwidth}  
        \centering 
        \includegraphics[width=\textwidth]{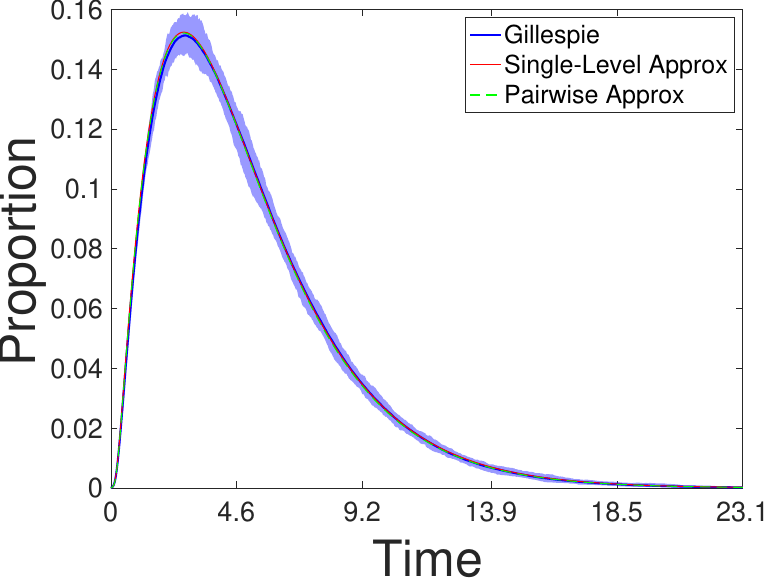}
        \caption{Offline}%
    \end{subfigure}
    \caption{Comparison of the solutions to the stochastic model, single-level and pairwise approximations. The top row illustrates online activity, and the bottom row the offline activity. (a), (c), (d), and (f) are on the UC 33 network. (b) and (e) are on the giant cluster of the UC 33 network. The parameters used are (a), (d) $\tau=0.001, \theta=0.01, \eta=0.1, \gamma_i=0.4, \gamma_p=0.6, t_{\text{max}}=200, [E](0)/N=0.2$, and $\text{iter}=100$; (b), (c), (e), (f) $\tau=0.2, \theta=0.01, \eta=0.2, \gamma_i=0.2, \gamma_p=0.5, t_{\text{max}}=200, [E](0)/N=0.02$, and $\text{iter}=100$. These results on the larger empirical network confirm the earlier findings reported in Figure~\ref{fig:real_network_example_FB}: Both approximation models closely follow the stochastic mean when $E$ decays monotonically. In scenarios with online outbursts, approximation accuracy declines, and the pairwise model offers only limited improvement over the single-level model. Restricting the simulation to the giant component versus using the full network has minimal effect on the results.}     \label{fig:real_network_example_UC_33}
\end{figure}

Figure \ref{fig:real_network_example_UC_33} presents a similar comparison using the UC 33 Facebook network, originally introduced in \cite{UC33_original}. This network includes $16\,808$ nodes and $522\,147$ edges. The mean node degree is approximately $62.1$, with a standard deviation of about $62.6$, again indicating high variability in degree distribution. The average clustering coefficient is approximately $0.23$, which, although lower than that of the previous Facebook friendship network, is still higher than those observed in the synthetic networks we tested, with the exception of the Watts–Strogatz networks. As above, we consider the giant component, which comprises $16\,800$ nodes and $522\,141$ edges. As in the previous example, the approximation models demonstrate better agreement with the stochastic mean when the online $E$ decays monotonically.

Both examples shown in Figures~\ref{fig:real_network_example_FB} and~\ref{fig:real_network_example_UC_33} demonstrate that both the single-level and pairwise approximations align more closely with the stochastic mean dynamics of online $E$ when the engagement trajectory is strictly decaying, corresponding to scenarios without outbursts. The reason is that in outburst-free cases, the recovery processes dominate the dynamics, and since recovery is independent of the network structure, the network-based approximations in our models contribute little to the overall approximation error.

Examining subfigures (b), (c), (e), and (f) in both Figures~\ref{fig:real_network_example_FB} and~\ref{fig:real_network_example_UC_33}, we observe that the pairwise approximation offers a slight advantage over the single-level approximation in matching the stochastic means. However, this advantage is less pronounced than in synthetic networks, see, for instance, Figure~\ref{fig:EREgs}. In particular, for the UC 33 Facebook network example in Figure~\ref{fig:real_network_example_UC_33}, the difference between the two approximations is nearly negligible. This may be due to real-world network characteristics, such as clustering, that are not captured by mean-field approximations, which rely solely on the degrees. As a result, increasing the order of interactions in the approximation does not necessarily lead to substantial improvements in accuracy.

These findings suggest that for real-world networks, more complex models are not always required to achieve satisfactory performance. In many cases, single-level approximations may provide a reasonable balance between computational efficiency and modelling accuracy.

\section{Conclusion and discussion} \label{sec:conclusion}

In this work, we proposed a two-layer compartmental model on networks as a mathematical framework for capturing the coupled dynamics of online engagement and offline social activity. The model highlights the crucial role of social media in facilitating spillover effects from information to the physical layers of social movements. Specifically, we assume that online interactions occur over a network structure and model a bidirectional influence between the online and offline layers: Online engagement can initiate offline protests, while offline participation feeds back to enhance online engagement.

To analyse the model, we apply mean-field approximations at varying levels of complexity to derive deterministic continuum systems from the underlying stochastic process. We evaluate the accuracy of these approximations on synthetic networks, such as $k$-regular graphs and Erd\H{o}s-R\'{e}nyi random networks, as well as on empirical networks, comparing them to the true stochastic dynamics. The resulting continuum systems enable us to apply tools from dynamical systems theory to analyse and predict protest outbursts under varying parameter regimes.

Our findings show that lower-density networks require more detailed approximation models to achieve accuracy comparable to that of simpler models applied to high-density networks. Therefore, model selection in practical applications should consider both the structural characteristics of the underlying network and the desired level of accuracy. Furthermore, we find that the parameters governing the interaction between the online and offline layers introduce nontrivial effects, with a critical range of cross-layer transmission rates significantly influencing the magnitude of offline outbursts.

Our model is based on simplified assumptions and is intended as a first step toward understanding the interplay between online and offline social dynamics. Several natural extensions include the removal of tension-inhibitive regimes and permanent immunity, incorporation of offline interactions on a network, and modelling of correlated transmission and recovery parameters. From a methodological perspective on approximations, enhancements to the mean-field approximation, such as accounting for clustering \cite{CommunityStruct,Hetero_Cluster}, or the adoption of alternative approaches such as percolation-based methods \cite{Epidemics_Networks,EBCM_SIR} present promising avenues for future work. A particularly compelling direction enabled by our approximated continuum models is the integration with empirical data to perform parameter estimation \cite{GP_ML, WSINDy} and system identification \cite{SINDy,WSINDy,IDENT,NeuralODE}. Such efforts would bring the model closer to real-world applicability and deepen our understanding of how social media influences collective offline behaviour, such as protesting.

\section*{Funding}  
This work was supported by the U.S. National Science Foundation Division of Mathematical Sciences [grant number 2042413]; and the Air Force Office of Scientific Research Multidisciplinary University Research Initiative [grant number FA9550-22-1-0380].

\section*{Acknowledgment}
We would like to thank our funders, the NSF and AFOSR. We would also like to thank Timothy Wessler for his initial contribution to this work.

\appendix
\section{\label{app:heatmap_stochastic}Effects from parameters on online and offline outbursts in stochastic model}

This appendix presents additional heatmaps exploring the effects of parameter combinations, $(\gamma_p, \eta)$ and $(\tau, \eta)$, on online and offline outbursts in the stochastic model.

\begin{figure}[!htbp]
    \centering
    \begin{subfigure}[t]{0.3\textwidth}
        \centering
        \includegraphics[width=\linewidth]{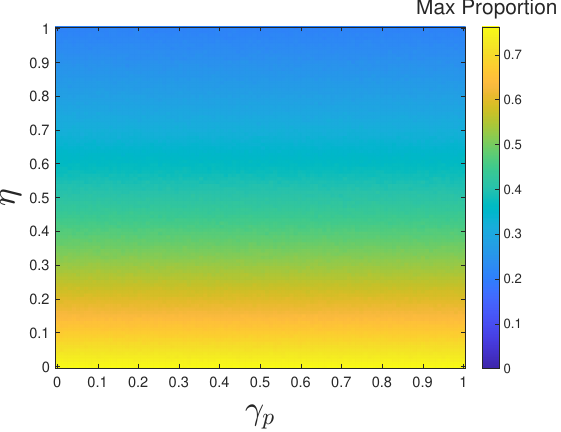}
        \vspace{-18pt}
        \caption{Online $E$} \label{fig:stochastic_eta_gamma_p_E}
    \end{subfigure}
     \hfill
    \begin{subfigure}[t]{0.3\textwidth}
    \centering
        \includegraphics[width=\linewidth]{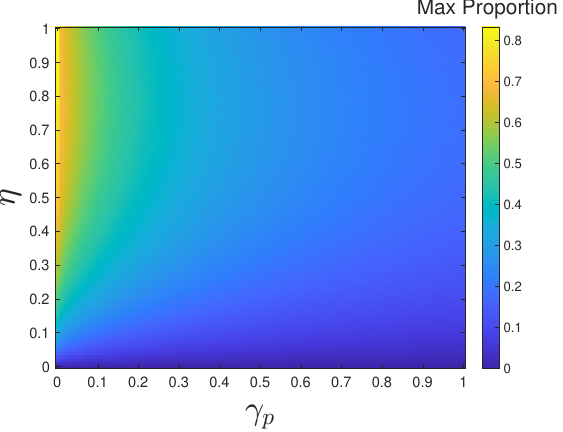}
        \vspace{-18pt}
        \caption{Offline $P$ with $\theta = 0.01$} \label{fig:stochastic_eta_gamma_p_P}
    \end{subfigure}
    \hfill
    \begin{subfigure}[t]{0.3\textwidth}
    \centering
        \includegraphics[width=\linewidth]{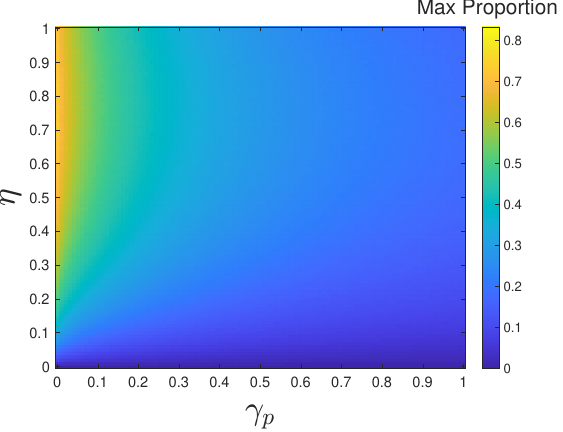}
        \vspace{-18pt}
        \caption{Offline $P$ with $\theta = 0$} \label{fig:stochastic_eta_gamma_p_theta0_P}
    \end{subfigure}
    \vspace{-6pt}
    \caption{Heatmaps from stochastic simulations on homogeneous networks showing the proportions of mean final outburst sizes in $E$ and $P$, averaged over $\text{iter} = 100$ runs per parameter combination swept over $\eta$ and $\gamma_p$. Grid resolution: $100 \times 100$. Parameters: $N=1\,000, k=5, \tau=0.8, \gamma_i=0.2, t_{\text{max}}=500$, and $ [E](0)=20$. (a) and (b) use $\theta = 0.01$, and (c) uses $\theta = 0$. The colour bar range in (c) is aligned with that in (b) for comparison. The $E$ heatmap for $\theta = 0$ closely matches (a) and is therefore omitted. Low $\eta$ leads to high online engagement, while high offline activity emerges when $\gamma_p$ is small and $\eta$ is large. With offline-to-online feedback removed ($\theta = 0$), the narrow region with very small $\gamma_p$ and large $\eta$ still produces the highest protest ratios in the $(\gamma_p, \eta)$ space, though with reduced intensity compared to the $P$ heatmap with nonzero $\theta$ in (b).}
    \label{fig:stochastic_eta_gamma_p}
\end{figure}

First, we examine the roles of the offline parameters $\eta$ and $\gamma_p$. A relatively high online transmission rate, $\tau$, is chosen to keep the $E$ reservoir sufficiently populated, enabling clearer observations of how the offline parameters influence outbursts across a broad sweep in the $(\gamma_p, \eta)$ space. To assess the impact of the offline-to-online feedback, we also perform the same sweep with $\theta = 0$, thereby removing this feedback mechanism entirely. Figure~\ref{fig:stochastic_eta_gamma_p} presents the resulting heatmaps. As the heatmap for $E$ with $\theta = 0$ shows little difference from that with $\theta = 0.01$, we omit it to avoid redundancy. One notable difference appears in the heatmaps for $P$: With $\theta = 0.01$, a narrow region with very small $\gamma_p$ and large $\eta$ yields significantly higher protest ratios, whereas this region shows lower ratios when $\theta = 0$. We conjecture that because $\theta = 0.01$ is small, the offline-to-online feedback becomes visibly influential only in this specific region, where offline recovery is minimal and cumulative feedback effects are strong.

\begin{figure}[!htbp]
    \centering
    \begin{subfigure}[t]{0.3\textwidth}
        \centering
        \includegraphics[width=\linewidth]{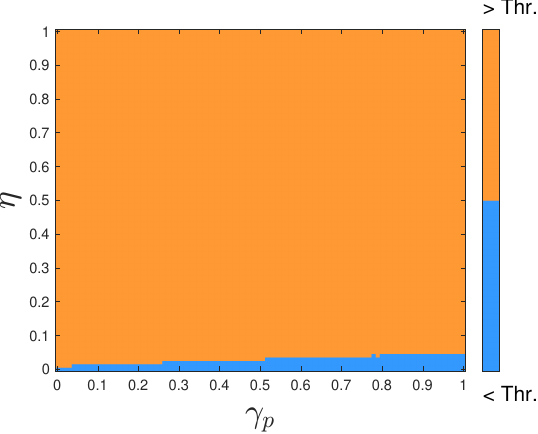} 
        \vspace{-18pt}
        \caption{Threshold $= 0.035$} \label{fig:PCategorization_stochastic_eta_gamma_p-1}
    \end{subfigure}
    \hfill
    \begin{subfigure}[t]{0.3\textwidth}
        \centering
        \includegraphics[width=\linewidth]{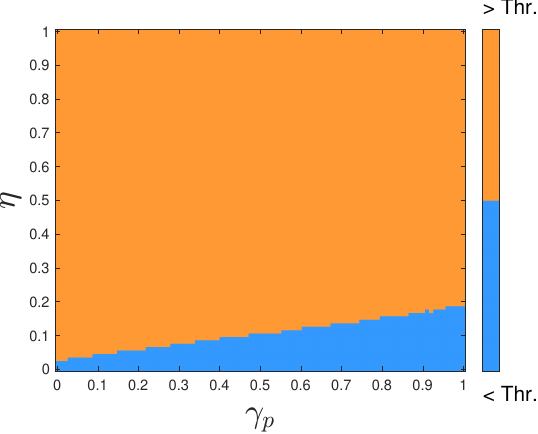} 
        \vspace{-18pt}
        \caption{Threshold $= 0.1$} \label{fig:PCategorization_stochastic_eta_gamma_p-2}
    \end{subfigure}
    \hfill
    \begin{subfigure}[t]{0.3\textwidth}
        \centering
        \includegraphics[width=\linewidth]{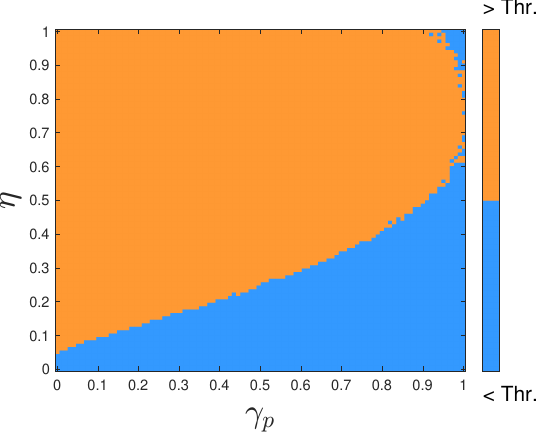} 
        \vspace{-18pt}
        \caption{Threshold $= 0.185$} 
        \label{fig:PCategorization_stochastic_eta_gamma_p-3}
    \end{subfigure}
    \hfill
    \begin{subfigure}[t]{0.3\textwidth}
        \centering
        \includegraphics[width=\linewidth]{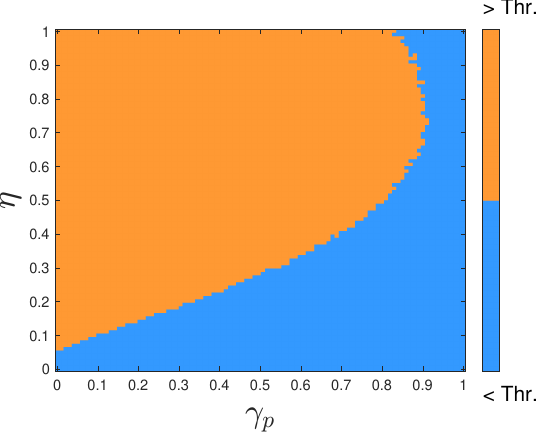} 
        \vspace{-18pt}
        \caption{Threshold $= 0.2$} \label{fig:PCategorization_stochastic_eta_gamma_p-4}
    \end{subfigure}
    \hfill
    \begin{subfigure}[t]{0.3\textwidth}
        \centering
        \includegraphics[width=\linewidth]{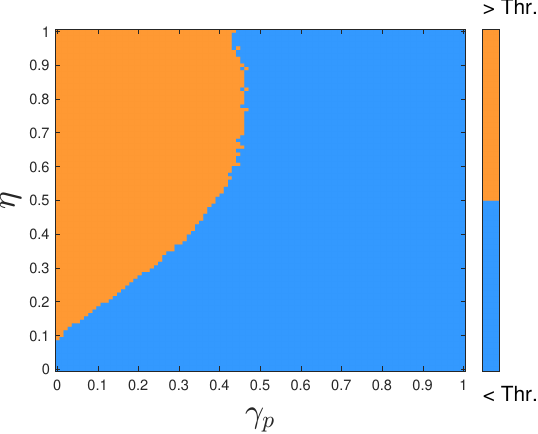}
        \vspace{-18pt}
        \caption{Threshold $= 0.3$} \label{fig:PCategorization_stochastic_eta_gamma_p-5}
    \end{subfigure}
    \hfill
    \begin{subfigure}[t]{0.3\textwidth}
        \centering
        \includegraphics[width=\linewidth]{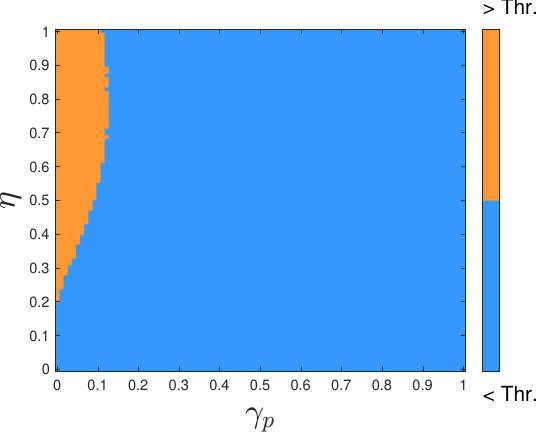}
        \vspace{-18pt}
        \caption{Threshold $= 0.5$} 
        \label{fig:PCategorization_stochastic_eta_gamma_p-6}
    \end{subfigure}
    \vspace{-6pt}
    \caption{Categorization of protest outbursts in the stochastic model on homogeneous networks, based on thresholding the proportion of the mean final outburst size of $P$ over $\text{iter}=100$ simulations at each grid point across sweeps in $\gamma_p$ and $\eta$. Grid resolution is $100 \times 100$. Parameters are $N=1\,000, k=5, \theta=0.01, \tau=0.8, \gamma_i=0.2, t_{\text{max}}=500$, and $[E](0)=20$. Varying the thresholds yields classification boundaries that are non-monotonic in the $\gamma_p$–$\eta$ space, indicating that the threshold for proportions of offline outbursts in this system reflects a nonlinear interaction between these parameters.}
    \label{fig:PCategorization_stochastic_eta_gamma_p}
\end{figure}

Figure \ref{fig:PCategorization_stochastic_eta_gamma_p} presents a categorization heatmap in the $(\gamma_p, \eta)$ plane, obtained by applying final outburst size thresholds to $P$ on the result shown in Figure \ref{fig:stochastic_eta_gamma_p_P}. The series of thresholds used in the plots is set relatively high to more clearly demonstrate the dynamics, as lower thresholds would fail to capture meaningful variation given the high $\tau$ value in this example.

\begin{figure}[!htbp]
    \centering
    \begin{subfigure}[t]{0.45\textwidth}
        \centering
        \includegraphics[width=\linewidth]{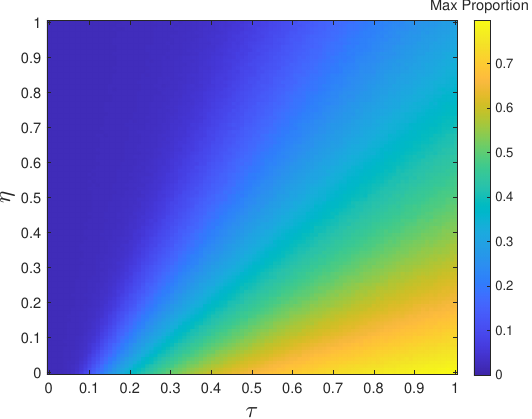}
        \caption{Online $E$} \label{fig:stochastic_tau_eta_E}
    \end{subfigure}
    \hfill
    \begin{subfigure}[t]{0.45\textwidth}
    \centering
        \includegraphics[width=\linewidth]{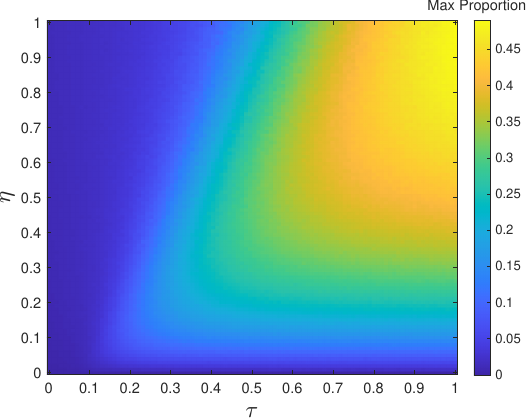}
        \caption{Offline $P$} \label{fig:stochastic_tau_eta_P}
    \end{subfigure}
    \caption{Heatmaps from stochastic simulations on homogeneous networks showing the proportions of mean final outburst sizes in $E$ and $P$, averaged over $\text{iter}=100$ runs for each parameter combination swept over $\tau$ and $\eta$. Grid resolution is $100 \times 100$. Parameters are $N=1\,000, k=5, \theta=0, \gamma_i=0.2, \gamma_p=0.2, t_{\text{max}}=500$, and $[E](0)=20$. Even in the absence of offline-to-online feedback, the results reveal a trade-off between the online and offline transmission parameters. High online activity is favored by larger $\tau$ and smaller $\eta$, while substantial offline outbursts require a large $\tau$ and an $\eta$ that falls within a moderate range relative to $\tau$.}
    \label{fig:stochastic_tau_eta}
\end{figure}

The effect of the feedback parameter $\theta$ is complex and can obscure the individual influence of other model parameters. To isolate the roles of the transmission parameters, we set $\theta=0$, removing offline-to-online feedback. We then perform a parameter sweep over $\tau$ and $\eta$ to investigate their effects as the online and offline transmission parameters on outbursts. Figure~\ref{fig:stochastic_tau_eta} presents heatmaps of the average final sizes of online $E$ and offline $P$ outbursts. These results suggest that $\tau$ must be sufficiently large to sustain the engaged population ($E$), which serves as the reservoir for potential protesters. In contrast, $\eta$ governs the rate at which this reservoir is converted into offline activity ($P$), thereby influencing whether substantial online or offline outbursts can emerge. Notably, for a wide range of $\tau$, intense protest occurs when $\eta$ lies in a ``sweet spot": If $\eta$ is too low, few individuals join offline protests; if too high, $E$ is depleted too quickly to contribute to large-scale offline activity.

\begin{figure}[!htbp]
    \centering
    \begin{subfigure}[t]{0.3\textwidth}
        \centering
        \includegraphics[width=\linewidth]{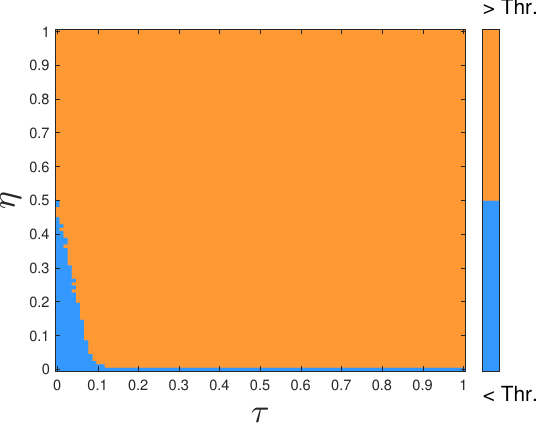}
        \caption{Threshold $= 0.01$} \label{fig:PCategorization_stochastic_tau_eta-1}
    \end{subfigure}
    \hfill
    \begin{subfigure}[t]{0.3\textwidth}
        \centering
        \includegraphics[width=\linewidth]{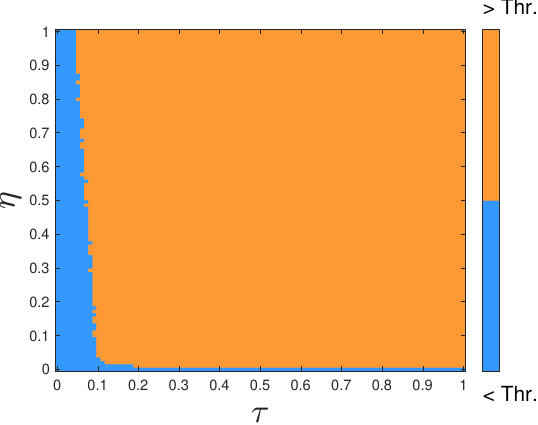}
        \caption{Threshold $= 0.015$} \label{fig:PCategorization_stochastic_tau_eta-2}
    \end{subfigure}
    \hfill
    \begin{subfigure}[t]{0.3\textwidth}
        \centering
        \includegraphics[width=\linewidth]{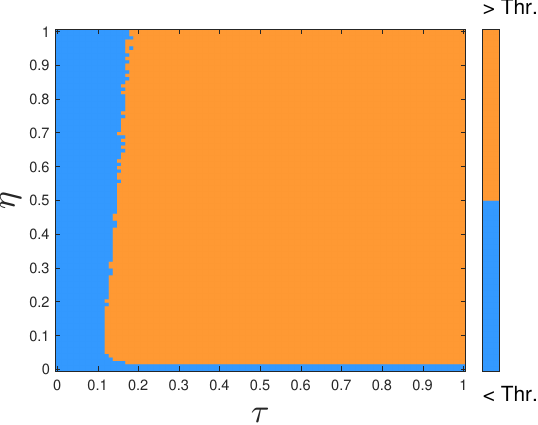}
        \caption{Threshold $= 0.025$}\label{fig:PCategorization_stochastic_tau_eta-3}
    \end{subfigure}
    \hfill
    \begin{subfigure}[t]{0.3\textwidth}
        \centering
        \includegraphics[width=\linewidth]{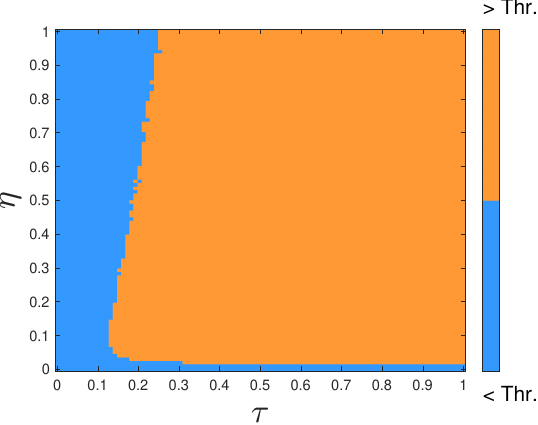}
        \caption{Threshold $= 0.035$} \label{fig:PCategorization_stochastic_tau_eta-4}
    \end{subfigure}
    \hfill
    \begin{subfigure}[t]{0.3\textwidth}
        \centering
        \includegraphics[width=\linewidth]{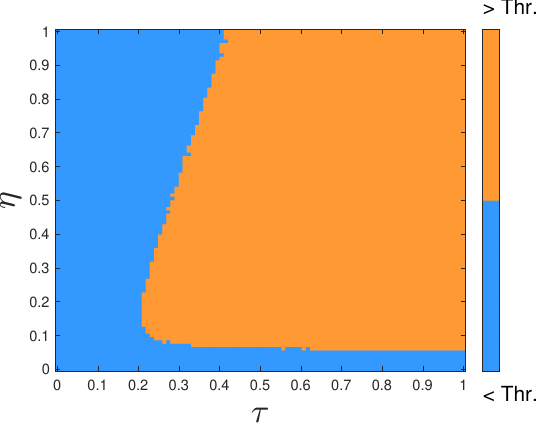} 
        \caption{Threshold $= 0.1$} \label{fig:PCategorization_stochastic_tau_eta-5}
    \end{subfigure}
    \hfill
    \begin{subfigure}[t]{0.3\textwidth}
        \centering
        \includegraphics[width=\linewidth]{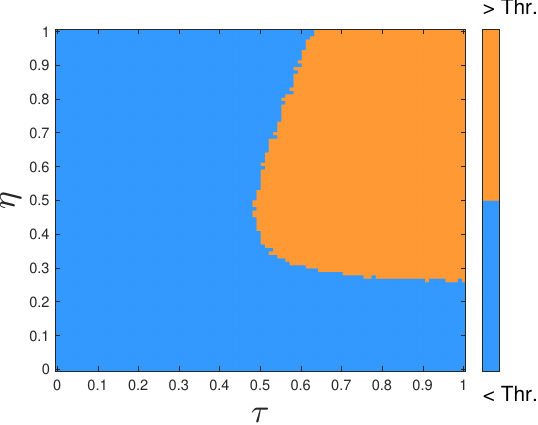}
        \caption{Threshold $= 0.3$}\label{fig:PCategorization_stochastic_tau_eta-6}
    \end{subfigure}
    \caption{Categorization of protest outbursts in the stochastic model on homogeneous networks, based on thresholding the mean final outburst size of $P$ across $\text{iter}=100$ simulations for each $(\tau, \eta)$ pair. Grid resolution is $100 \times 100$. Parameters are $N=1\,000, k=5, \theta=0, \gamma_i=0.2, \gamma_p=0.2, t_{\text{max}}=500$, and $[E](0)=20$. The resulting classification boundaries highlight a trade-off between $\tau$ and $\eta$ in enabling offline outbursts and suggest a nonlinear interplay in determining the effective threshold for protest emergence.}
    \label{fig:PCategorization_stochastic_tau_eta}
\end{figure}

Figure \ref{fig:PCategorization_stochastic_tau_eta} shows the categorization of protest outbursts in the $\tau$–$\eta$ parameter space by applying thresholds to the proportions of the final offline outburst sizes shown in Figure \ref{fig:stochastic_tau_eta_P}.

\section{\label{app:R0_Heatmap_Homogeneous_SingleLevel} Outbursts in single-level model on homogeneous networks}

This appendix presents additional results and discussion regarding the $R_0^{MF} = 1$ threshold and the online and offline outbursts in the single-level approximation model on homogeneous networks.

\begin{figure}[!htbp]
    \centering
    \includegraphics[width=0.5\linewidth]{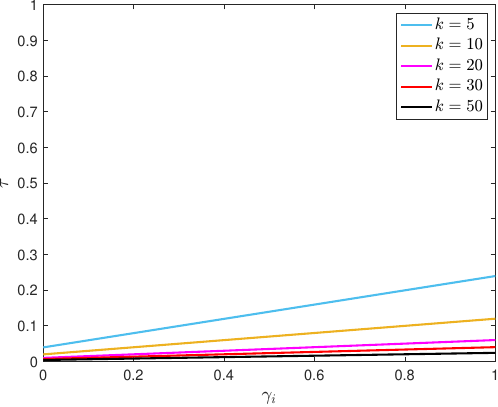}
    \caption{The analytical $R_0^{MF} = 1$ lines in the $\gamma_i$-$\tau$ plane for different degree $k$. Parameters are $\theta=0.01, \eta=0.2$, and $\gamma_p=0.5$. As the degree increases, the single-level approximation model becomes more prone to outbursts.}
    \label{fig:HomogSingleR0Lines}
\end{figure}

Figure \ref{fig:HomogSingleR0Lines} displays the $R_0^{MF} = 1$ lines in the $\gamma_i$-$\tau$ plane for several example degrees. Consistent with the insights from Figure~\ref{fig:StR0Lines} for the stochastic model, a higher degree is associated with a greater tendency for outbursts in the single-level model when using $R_0^{MF} = 1$ as a threshold.

\begin{figure}[!htbp]
    \centering
    \begin{subfigure}[t]{0.49\textwidth}
        \centering
        \includegraphics[width=\linewidth]{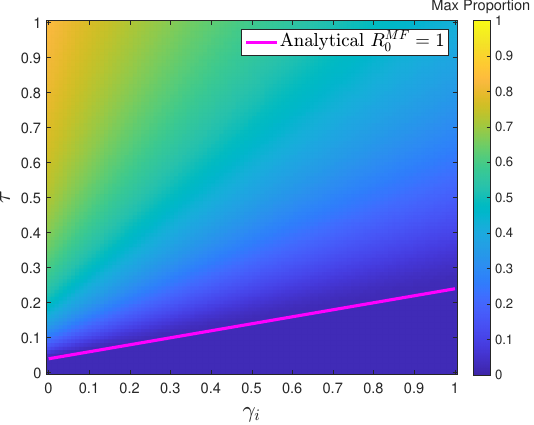} 
        \caption{Online $E$} \label{fig:HomogSingle_tau_gamma_i_E}
    \end{subfigure}
    \hfill
    \begin{subfigure}[t]{0.49\textwidth}
    \centering
        \includegraphics[width=\linewidth]{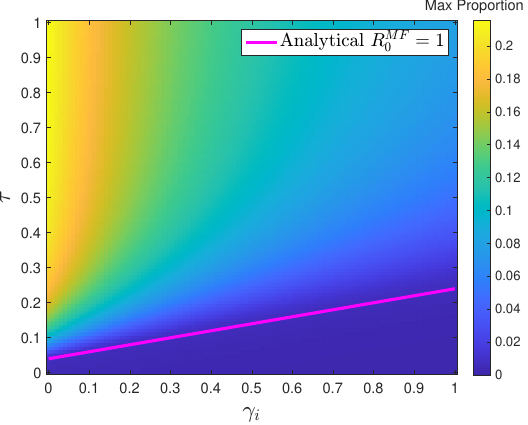} 
        \caption{Offline $P$} \label{fig:HomogSingle_tau_gamma_i_P}
    \end{subfigure}
    \caption{Heatmaps from simulations of the single-level approximation model on homogeneous networks showing the proportions of mean final outburst sizes in $E$ and $P$ across a sweep of $\tau$ and $\gamma_i$. Grid resolution is $100 \times 100$. Parameters are $N=1\,000, k=5, \theta=0.01, \eta=0.2, \gamma_p=0.5, t_{\text{max}}=200$, and $[E](0)=20$. The $R_0^{MF} = 1$ line is plotted, serving as an analytical threshold for categorizing outburst behaviour.}
    \label{fig:HomogSingle_tau_gamma_i}
\end{figure}

We also perform a parameter sweep over $\tau$ and $\gamma_i$ by running the single-level approximation model on homogeneous networks with various $(\gamma_i, \tau)$ pairs. We present the resulting heatmaps that record the final sizes of outbursts in $E$ and $P$, respectively. Figure \ref{fig:HomogSingle_tau_gamma_i} shows these results.

\section{\label{app:stability_homogeneous_SingleLevel}Stability of generalized steady-state solution to the single-level model on homogeneous networks}

For the single-level approximation model on homogeneous networks, we can analyse a more general equilibrium beyond the engagement- and protest-free solution and study its stability using a reduced system, as in \cite{Epidemics_Networks}. The discussion in this appendix provides supplementary detail to Section~\ref{subsubsec:Homog_R0_SingleLevel}, which focuses on a specific application of interest related to understanding the conditions under which outbursts occur in our system. For completeness, we present here a more general analysis of the stability of equilibrium solutions.

In the single-level approximation model on homogeneous networks \eqref{model:Homogeneous_SingleLevel}, notice that $[D] = N - [U] - [E]$, allowing us to reduce the system to a four-dimensional space, $([U], [E], [P], [R])$. The steady-state solution to the system is $([U]^*, 0, 0, [R]^*)$ where $[U]^*, [R]^* \in [0, N]$. We analyse its stability by examining the Jacobian at this equilibrium, given by
\begin{align*}
   J^{MF'}=\left[\begin{array}{cccc}
    0&-\frac{k\tau}{N}[U]^*&-\frac{\theta}{N}[U]^*&0\\
    0&\frac{k\tau}{N}[U]^*-(\eta+\gamma_i)&\frac{\theta}{N}[U]^*&0\\
    0&\eta&-\gamma_p&0\\
    0&0&\gamma_p&0    
    \end{array}\right]
\end{align*}
$J^{MF'}$ has two zero eigenvalues and two other eigenvalues, which are
\begin{align*}
    \lambda_{\pm}^{MF'}  = \frac{1}{2}\left[ \left(\frac{k\tau}{N}[U]^*-\eta-\gamma_i-\gamma_p \right)\pm \sqrt{\left(\frac{k\tau}{N}[U]^*-\eta-\gamma_i-\gamma_p \right)^2-4 \left(\eta\gamma_p+\gamma_i\gamma_p-\frac{k\tau\gamma_p}{N}[U]^*-\frac{\theta\eta}{N}[U]^* \right)} \right].
\end{align*}
The equilibrium is stable if both $\operatorname{Re} \left(\lambda_{-}^{MF'} \right) < 0$ and $\operatorname{Re} \left(\lambda_{+}^{MF'} \right) < 0$, which requires
\begin{align*}
    &\frac{k\tau}{N}[U]^*-\eta-\gamma_i-\gamma_p < 0\\
    \mbox{and} \quad &[U]^* < N\frac{\gamma_p(\eta + \gamma_i)}{\eta\theta + k\tau\gamma_p}.
\end{align*}
Notice that since $[U]^* \in [0,N]$, if the conditions in $\eqref{eq:stability_MF}$ are satisfied, then the equilibrium solution corresponding to any $[U]^*$ is stable.

\section{\label{app:solve_homogeneous_single} Explicit solutions in the single-level model on homogeneous networks}

In this appendix, we present additional information about the solutions of the online compartments, $U$ and $E$, in the single-level approximation model on homogeneous networks, offering an alternative perspective for understanding the system’s dynamics and stability.

For the homogeneous single-level model \eqref{model:Homogeneous_SingleLevel}, we can solve for $[U]$ and $[E]$ as functions of $[D]$ and $[R]$, and construct the phase portrait using methods similar to those in \cite{Epidemics_Networks}. For simplicity, we assume $[D](0) = 0$, so $[U] + [E] = N$. Note we have $\displaystyle \frac{d[U](t)}{dt} + \tau \frac{k}{N}[U][E] + \frac{\theta}{N}[U][P] = 0$. Using integrating factor, we get $\displaystyle \frac{d}{dt} \left( e^{\frac 1 N \int_0^t k\tau[E](t')+\theta[P](t') dt'} [U](t) \right) = 0$.

Notice that $\displaystyle \int_0^t [E](t') dt' = \frac{[D]}{\eta+\gamma_i}$ since $\displaystyle \frac{d[D](t)}{dt} = (\eta+\gamma_i)[E]$, and similarly, $\displaystyle \int_0^t [P](t') dt' = \frac{[R]}{\gamma_p}$ since $\displaystyle \frac{d[R](t)}{dt} = \gamma_i [P]$. Then we have $ \displaystyle \frac{d}{dt} \left( e^{\frac{k\tau}{N} \frac{[D]}{\eta+\gamma_i} + \frac \theta N \frac{[R]}{\gamma_p}} [U](t) \right) = 0$, which implies that
\[ [U] = U_0 e^{-\frac{k\tau}{N} \frac{[D]}{\eta+\gamma_i} - \frac \theta N \frac{[R]}{\gamma_p}} \quad \mbox{ where } \quad U_0 = [U](0) \]
and since $[E] = N - [U] - [D]$,
\[ [E] = N - U_0 e^{-\frac{k\tau}{N} \frac{[D]}{\eta+\gamma_i} - \frac \theta N \frac{[R]}{\gamma_p}} - [D]. \]

\begin{figure}[!htbp]
    \centering
    \begin{subfigure}[t]{0.32\textwidth}
        \centering
        \includegraphics[width=\linewidth]{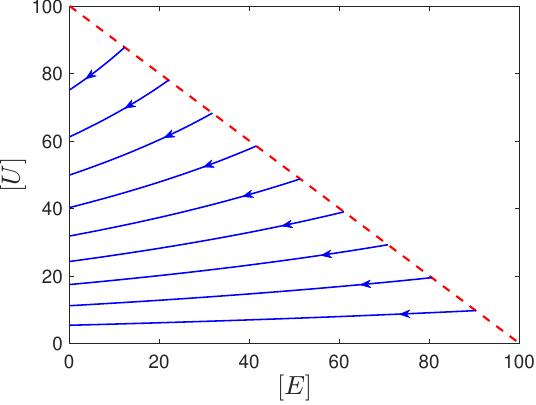} 
        \caption{$\tau = 0.05$} \label{fig:UE_phase_portraits-1}
    \end{subfigure}
    \hfill
    \begin{subfigure}[t]{0.32\textwidth}
        \centering
        \includegraphics[width=\linewidth]{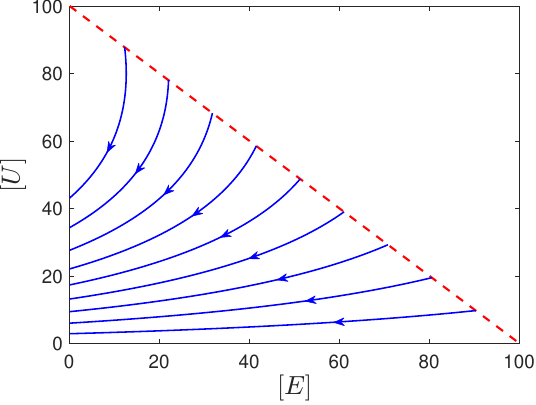}
        \caption{$\tau = 0.1$} \label{fig:UE_phase_portraits-2}
    \end{subfigure}
    \hfill
    \begin{subfigure}[t]{0.32\textwidth}
        \centering
        \includegraphics[width=\linewidth]{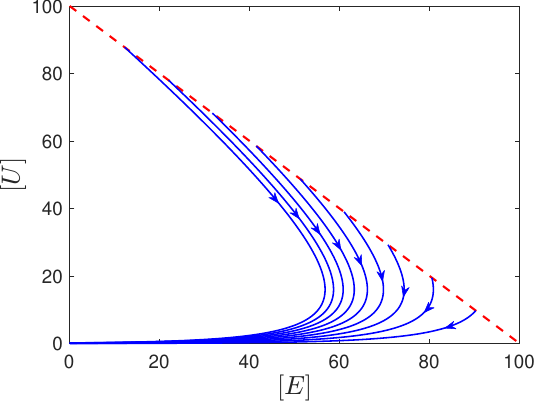}
        \caption{$\tau = 0.5$} 
        \label{fig:UE_phase_portraits-3}
    \end{subfigure}
    \caption{Phase portraits of the single-level approximation model on homogeneous networks, plotted on the $[E]$-$[U]$ plane and parametrized by $[D]$ and $[R]$. The fixed parameters are $N=100, k=5, \theta=0.01, \eta=0.2, \gamma_i=0.2, \gamma_p=0.2$, and $[R]=50$. Each of the $9$ blue lines corresponds to a different initial condition, $U_0 = 10, 20, \ldots, 90$. For each line, $[D]$ is varied over the range $[0, N]$, tracing the trajectory. Arrows indicate the direction of flow as $[D]$ increases. The area bounded by the dashed red line is the regime of physically meaningful solutions. Only the physical solutions are shown. These phase portraits illustrate the stability of solutions, showing that higher $\tau$ values lead to fewer stable solutions.}
    \label{fig:UE_phase_portraits}
\end{figure}

Using these expressions for $[U]$ and $[E]$ as functions of $[D]$ and $[R]$, we can interpret the system’s solutions in the $[E]$-$[U]$ plane. Figure~\ref{fig:UE_phase_portraits} presents example phase portraits for different values of the online transmission parameter $\tau$, with $[R]$ held fixed and $[D]$ varied.

\section{\label{app:R0_Heatmap_Homogeneous_Pairwise} Outbursts in pairwise model on homogeneous networks}

Similar to the purpose of Appendix \ref{app:R0_Heatmap_Homogeneous_SingleLevel}, this appendix presents additional discussion on the $R_0^{PW} = 1$ threshold and the online and offline outbursts in the pairwise approximation model on homogeneous networks.

\begin{figure}[H]
    \centering
    \includegraphics[width=0.45\linewidth]{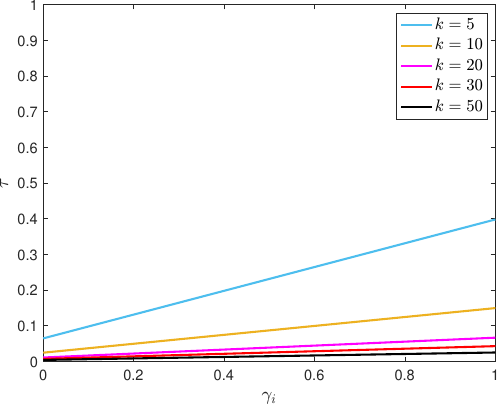}
    \caption{Analytical $R_0^{PW} = 1$ lines plotted in the $\gamma_i$-$\tau$ plane for different degree $k$. Parameters are $\theta=0.01, \eta=0.2$, and $\gamma_p=0.5$. A higher degree $k$ corresponds to a larger parameter region where outbursts can occur.}
    \label{fig:HomogPairwiseR0Lines}
\end{figure}

Figure \ref{fig:HomogPairwiseR0Lines} shows the $R_0^{PW} = 1$ lines in the $\gamma_i$-$\tau$ plane with several representative degrees.

\begin{figure}[!htbp]
    \centering
    \begin{subfigure}[t]{0.46\textwidth}
        \centering
        \includegraphics[width=\linewidth]{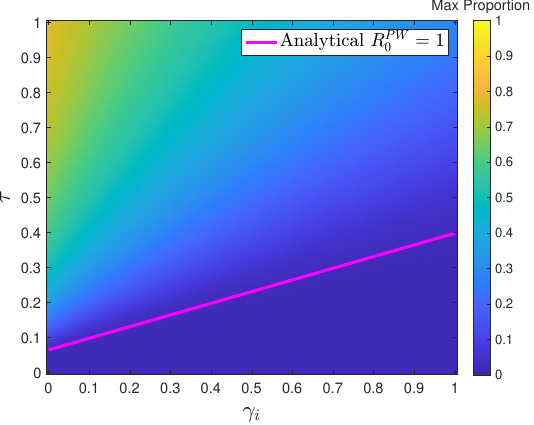} 
        \caption{Online $E$} \label{fig:HomogPairwise_tau_gamma_i_E}
    \end{subfigure}
    \hfill
    \begin{subfigure}[t]{0.46\textwidth}
    \centering
        \includegraphics[width=\linewidth]{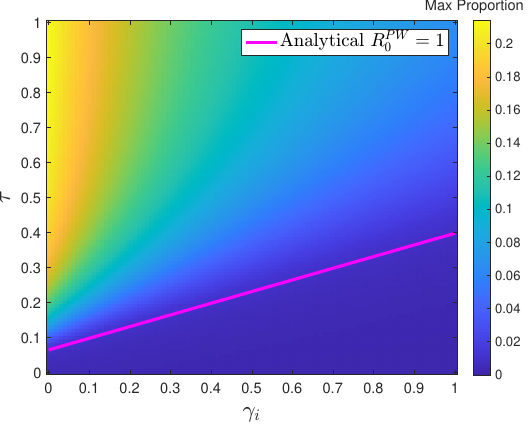} 
        \caption{Offline $P$} \label{fig:HomogPairwise_tau_gamma_i_P}
    \end{subfigure}
    \caption{Heatmaps from simulations of the pairwise approximation model on homogeneous networks showing the proportions of mean final outburst sizes in $E$ and $P$ across $(\gamma_i, \tau)$. Grid is $100 \times 100$. Parameters are $N=1\,000, k=5, \theta=0.01, \eta=0.2, \gamma_p=0.5, t_{\text{max}}=200$, and $[E](0)=20$. The $R_0^{PW} = 1$ line is plotted as an analytical threshold for outburst classification.}
    \label{fig:HomogPairwise_tau_gamma_i}
\end{figure}

We sweep over $\tau$ and $\gamma_i$ using the pairwise approximation model on homogeneous networks and present heatmaps of the final outburst sizes in $E$ and $P$. Figure \ref{fig:HomogPairwise_tau_gamma_i} shows the results.

\section{\label{app:heter_model_on_other_syn_nets} Heterogeneous models on other synthetic networks}

In this appendix, we present results and discussion comparing model dynamics on additional types of synthetic networks.

\begin{figure}[H]
    \centering
    \begin{subfigure}[b]{0.32\textwidth}  
        \centering 
        \includegraphics[width=\textwidth]{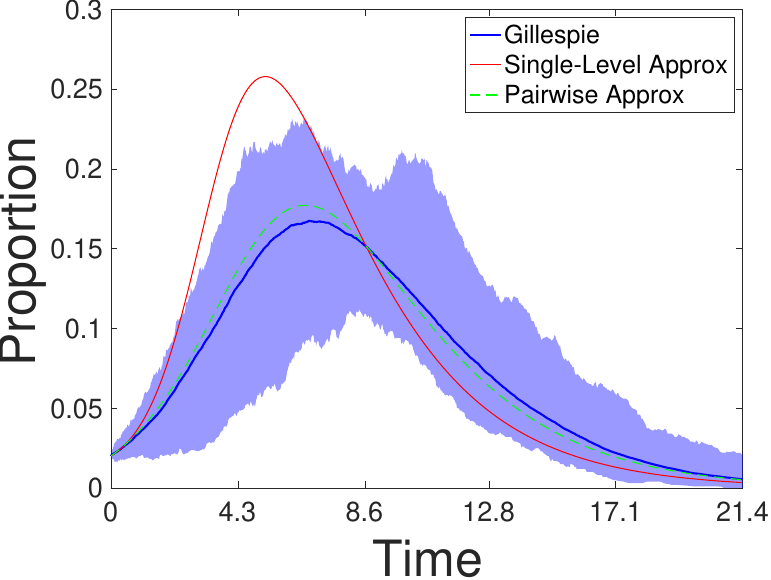}
        \caption{Online $d_1=2, d_2=8$}%
        \label{fig:BimodalOnline1}
    \end{subfigure}
    \hfill
    \begin{subfigure}[b]{0.32\textwidth}  
        \centering 
        \includegraphics[width=\textwidth]{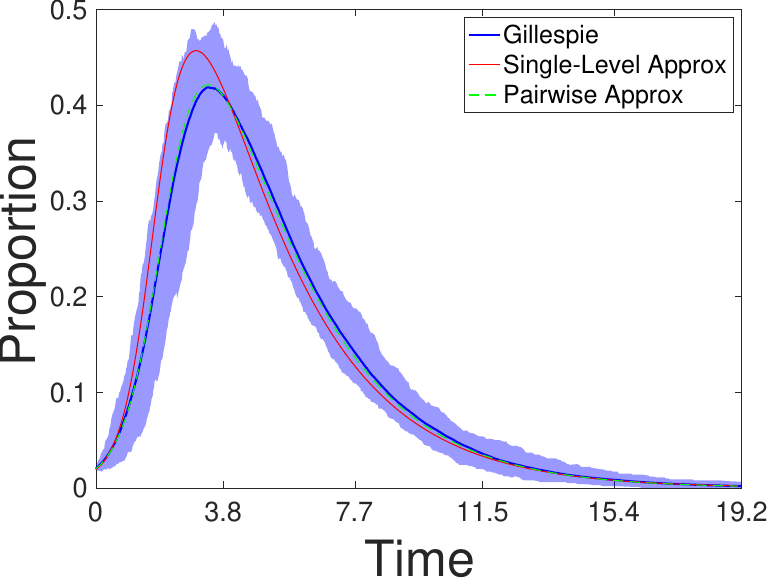}
        \caption{Online $d_1=5, d_2=15$}%
        \label{fig:BimodalOnline2}
    \end{subfigure}
    \hfill
    \begin{subfigure}[b]{0.32\textwidth}  
        \centering 
        \includegraphics[width=\textwidth]{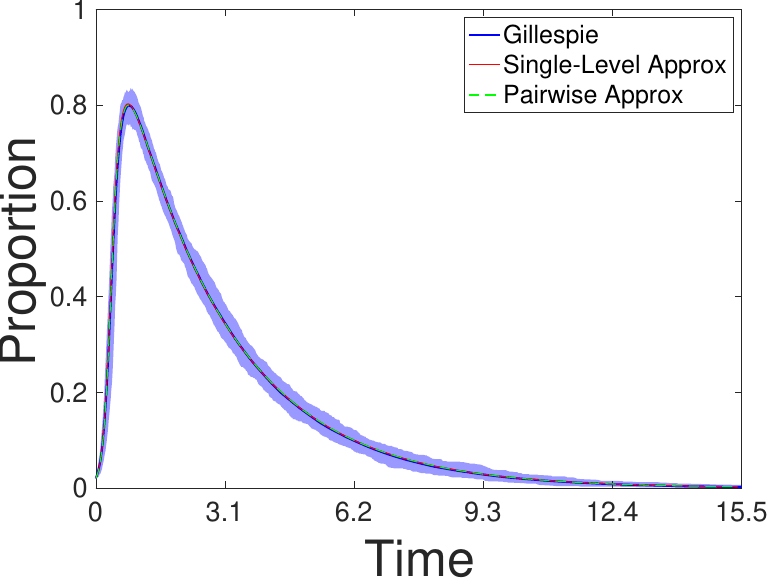}
        \caption{Online $d_1=30, d_2=70$}%
        \label{fig:BimodalOnline3}
    \end{subfigure}
    \\
    \begin{subfigure}[b]{0.32\textwidth}  
        \centering 
        \includegraphics[width=\textwidth]{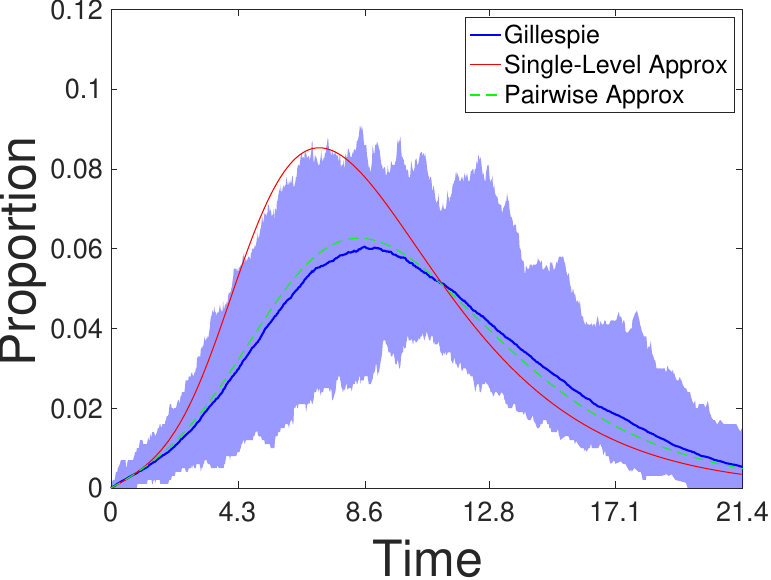}
        \caption{Offline $d_1=2, d_2=8$}%
        \label{fig:BimodalOffline1}
    \end{subfigure}
    \hfill
    \begin{subfigure}[b]{0.32\textwidth}  
        \centering 
        \includegraphics[width=\textwidth]{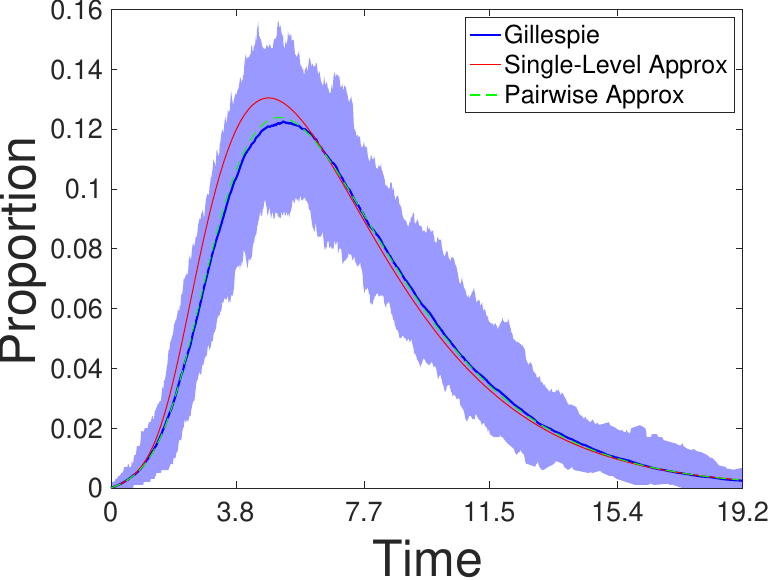}
        \caption{Offline $d_1=5, d_2=15$}%
        \label{fig:BimodalOffline2}
    \end{subfigure}
    \hfill
    \begin{subfigure}[b]{0.32\textwidth}  
        \centering 
        \includegraphics[width=\textwidth]{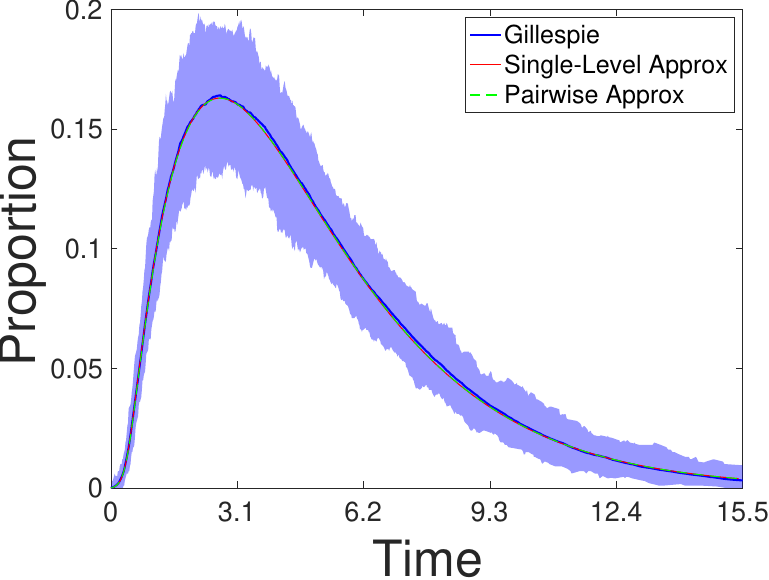}
        \caption{Offline $d_1=30, d_2=70$}%
        \label{fig:BimodalOffline3}
    \end{subfigure}
    \caption{Dynamics of online $E$ and offline $P$ on bimodal random networks with varying average degree $\langle k \rangle$. $N_1=N_2=500, \tau=0.2, \theta=0.01, \eta=0.2, \gamma_i=0.2, \gamma_p=0.5, t_{\text{max}}=200, [E](0)=20$, and $\text{iter}=100$. These results show that the pairwise approximation more accurately captures the dynamics than the single-level approximation, and that both models improve as $\langle k \rangle$ increases.}      
    \label{fig:BimodalEgs}
\end{figure}

First, we consider bimodal random networks where the degree in the network is either $d_1$ or $d_2$ and the corresponding number of nodes with such degree is $N_1$ or $N_2$, respectively. Figure \ref{fig:BimodalEgs} shows some example comparisons of dynamics on this type of networks.

\begin{figure}[H]
    \centering
    \begin{subfigure}[b]{0.32\textwidth}  
        \centering 
        \includegraphics[width=\textwidth]{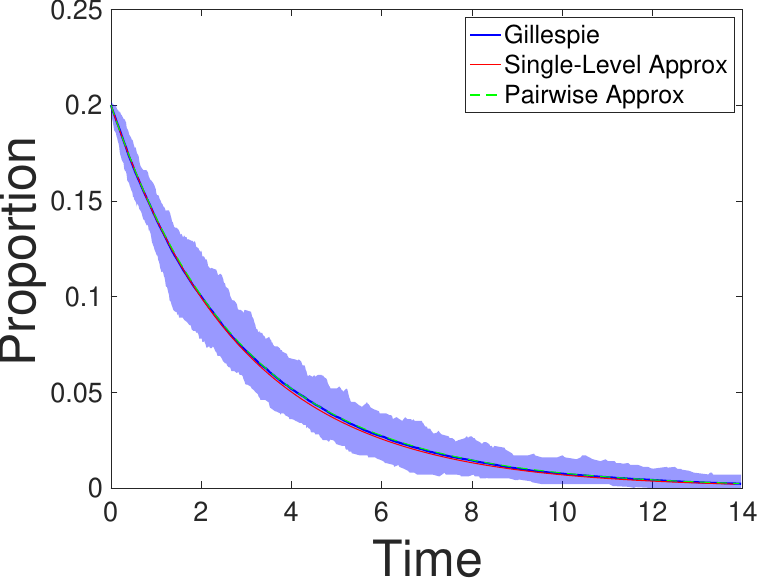}
        \caption{Online}%
        \label{fig:PLOnline1}
    \end{subfigure}
    \hfill
    \begin{subfigure}[b]{0.32\textwidth}  
        \centering 
        \includegraphics[width=\textwidth]{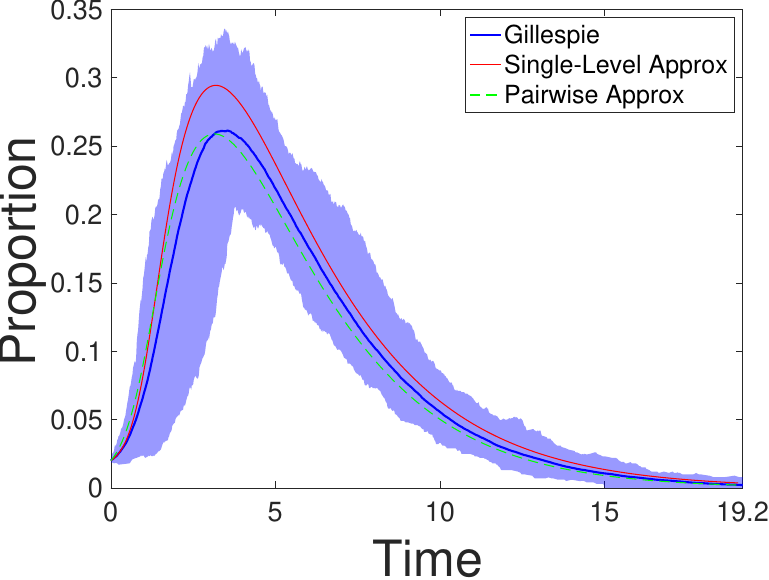}
        \caption{Online}%
        \label{fig:PLOnline2}
    \end{subfigure}
    \hfill
    \begin{subfigure}[b]{0.32\textwidth}  
        \centering 
        \includegraphics[width=\textwidth]{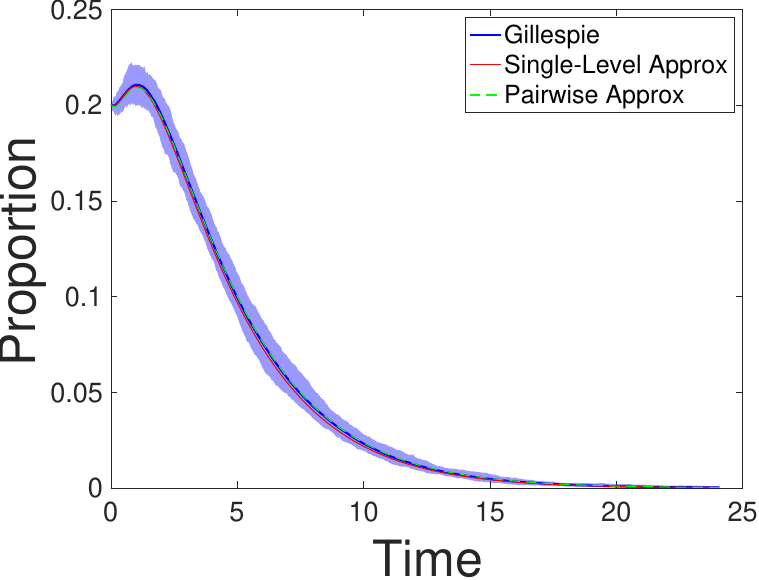}
        \caption{Online}%
        \label{fig:PLOnline3}
    \end{subfigure}
    \\
    \begin{subfigure}[b]{0.32\textwidth}  
        \centering 
        \includegraphics[width=\textwidth]{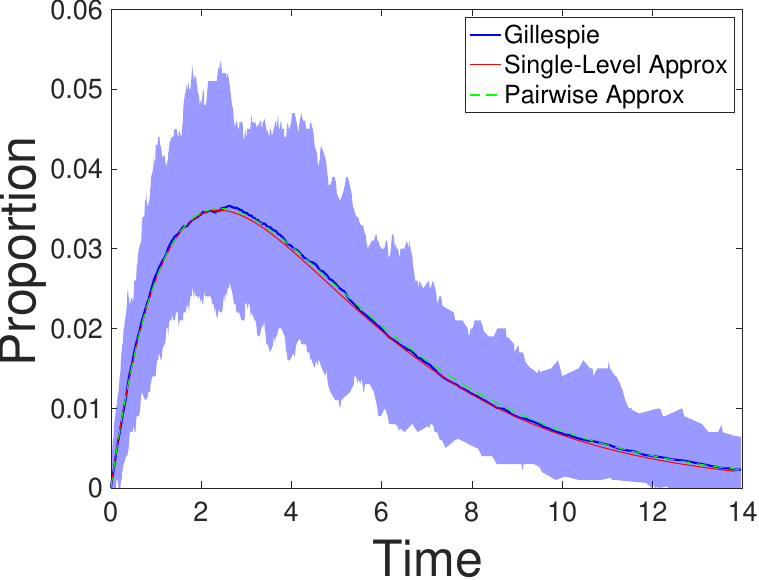}
        \caption{Offline}%
        \label{fig:PLOffline1}
    \end{subfigure}
    \hfill
    \begin{subfigure}[b]{0.32\textwidth}  
        \centering 
        \includegraphics[width=\textwidth]{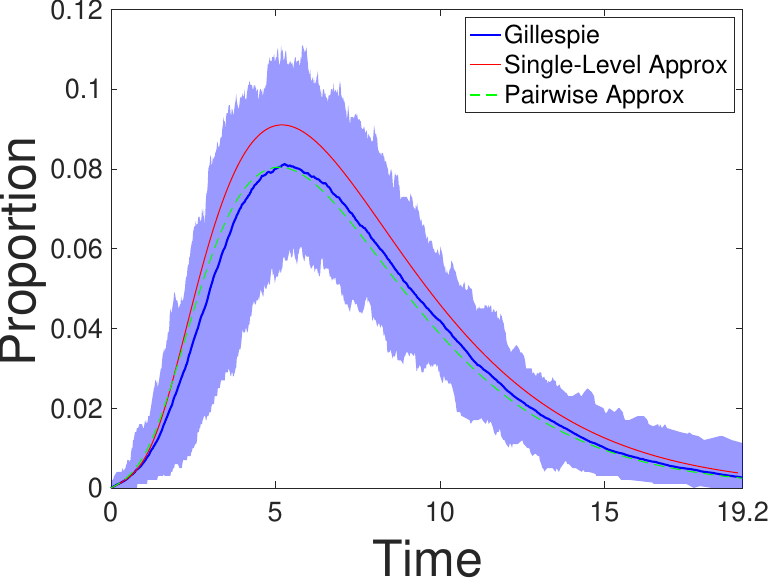}
        \caption{Offline}%
        \label{fig:PLOffline2}
    \end{subfigure}
    \hfill
    \begin{subfigure}[b]{0.32\textwidth}  
        \centering 
        \includegraphics[width=\textwidth]{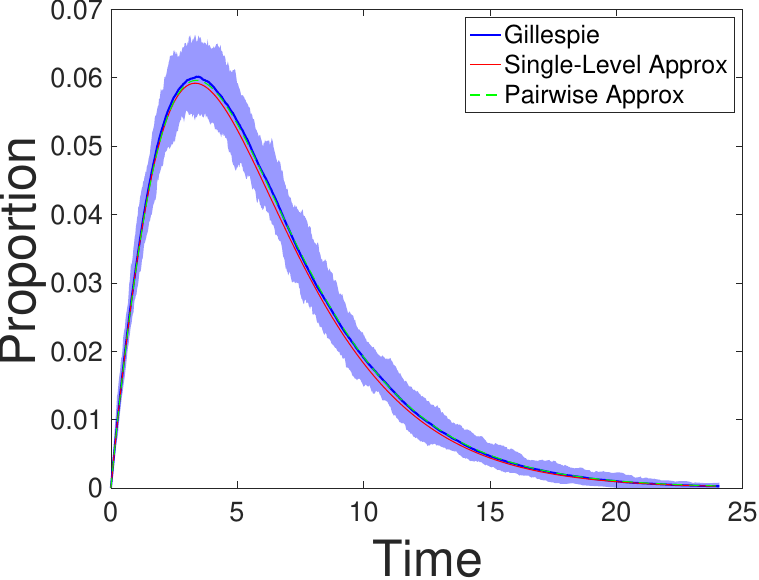}
        \caption{Offline}%
        \label{fig:PLOffline3}
    \end{subfigure}
    \caption{Dynamics of online $E$ and offline $P$ on scale-free networks. (a), (d) $N=1\,000, \langle k \rangle=5, K=50, \alpha=2, \tau=0.01, \theta=0.05, \eta=0.2, \gamma_i=0.2, \gamma_p=0.5, t_{\text{max}}=200, [E](0)=200$, and $\text{iter}=100$; (b), (e) $N=1\,000, \langle k \rangle=5, K=50, \alpha=2, \tau=0.2, \theta=0.01, \eta=0.2, \gamma_i=0.2, \gamma_p=0.5, t_{\text{max}}=200, [E](0)=20$, and $\text{iter}=100$; (c), (f) $N=10\,000, \langle k \rangle=50, K=1\,000, \alpha=1.5, \tau=0.01, \theta=0.05, \eta=0.2, \gamma_i=0.2, \gamma_p=0.5, t_{\text{max}}=200, [E](0)=2\,000$, and $\text{iter}=100$. When the online transmission rate $\tau$ is very small, $E$ strictly decays, and both approximation models closely match the stochastic mean of $E$ and $P$. In parameter regimes where $E$ exhibits an outburst, the pairwise approximation more accurately captures the dynamics than the single-level approximation.}
    \label{fig:PowerLawEgs}
\end{figure}

Another synthetic network type we consider is the scale-free random network, characterized by a power-law degree distribution $\mathbb{P}(n) \propto n^{-\alpha}$ for some exponent $\alpha > 1$. Such scale-free property is commonly observed in many real-world networks \cite{barabasi2016network}. In generating these networks, we specify the average degree $\langle k \rangle$ and the maximum degree $K$. Figure~\ref{fig:PowerLawEgs} presents example comparisons of the model dynamics in $E$ and $P$.

\begin{figure}[H]
    \centering
    \begin{subfigure}[b]{0.32\textwidth}  
        \centering 
        \includegraphics[width=\textwidth]{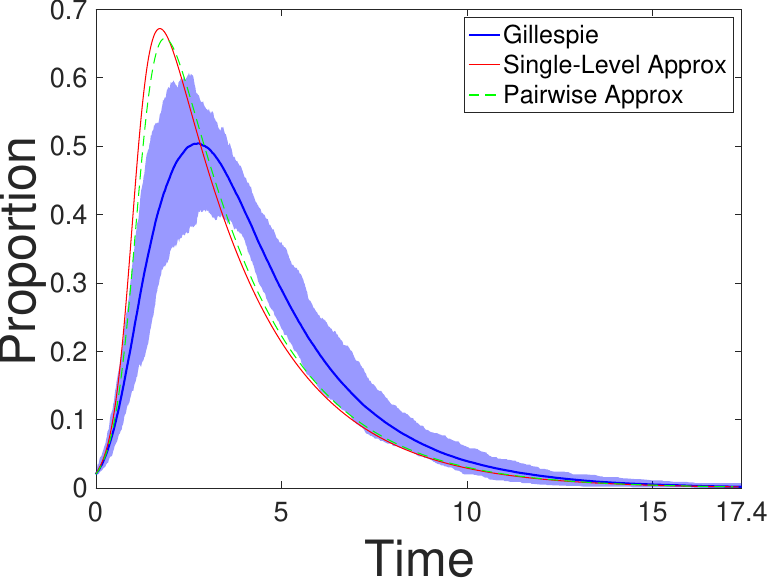}
        \caption{Online $p=0.01$}%
        \label{fig:WSOnline1}
    \end{subfigure}
    \hfill
    \begin{subfigure}[b]{0.32\textwidth}  
        \centering 
        \includegraphics[width=\textwidth]{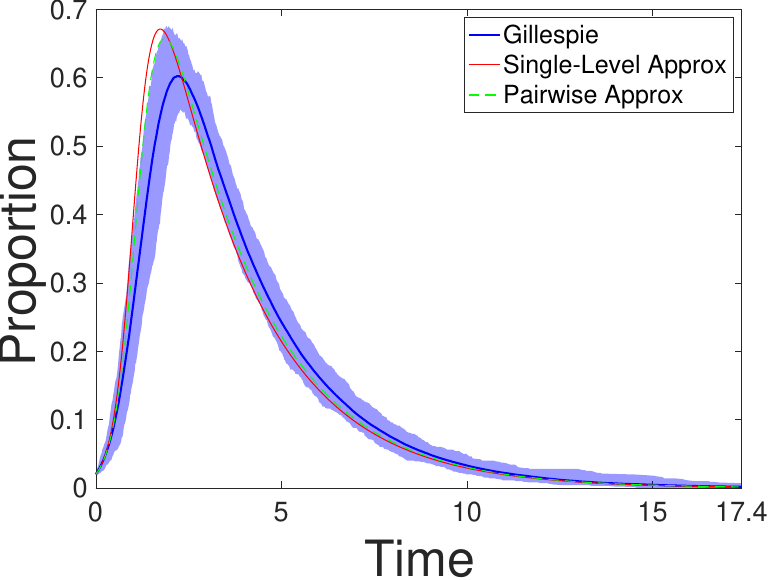}
        \caption{Online $p=0.1$}%
        \label{fig:WSOnline2}
    \end{subfigure}
    \hfill
    \begin{subfigure}[b]{0.32\textwidth}  
        \centering 
        \includegraphics[width=\textwidth]{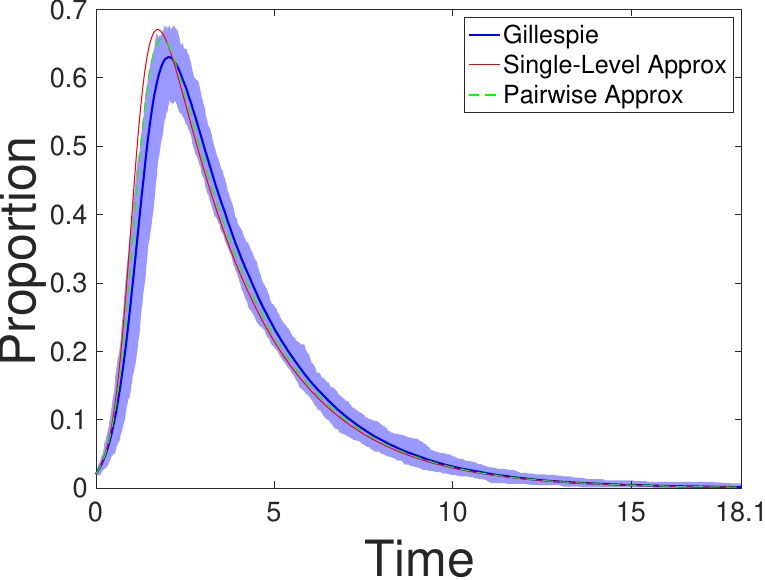}
        \caption{Online $p=0.2$}%
        \label{fig:WSOnline3}
    \end{subfigure}
    \\
    \begin{subfigure}[b]{0.32\textwidth}  
        \centering 
        \includegraphics[width=\textwidth]{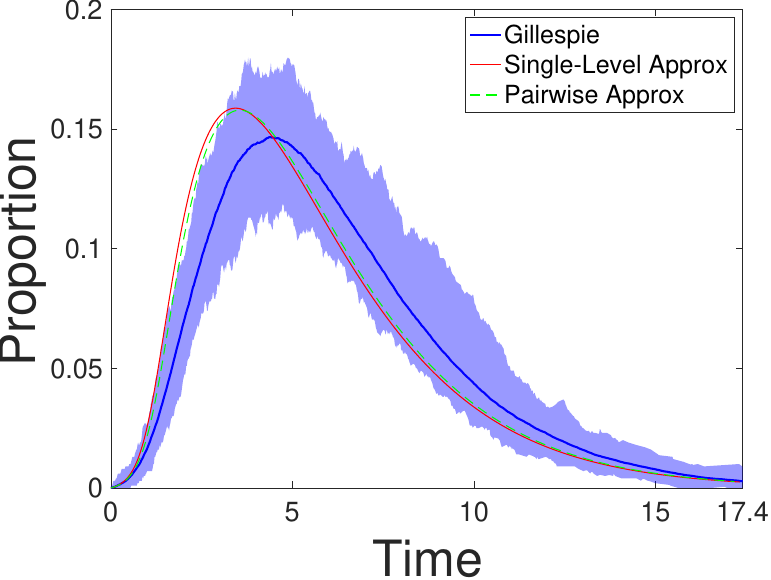}
        \caption{Offline $p=0.01$}%
        \label{fig:WSOffline1}
    \end{subfigure}
    \hfill
    \begin{subfigure}[b]{0.32\textwidth}  
        \centering 
        \includegraphics[width=\textwidth]{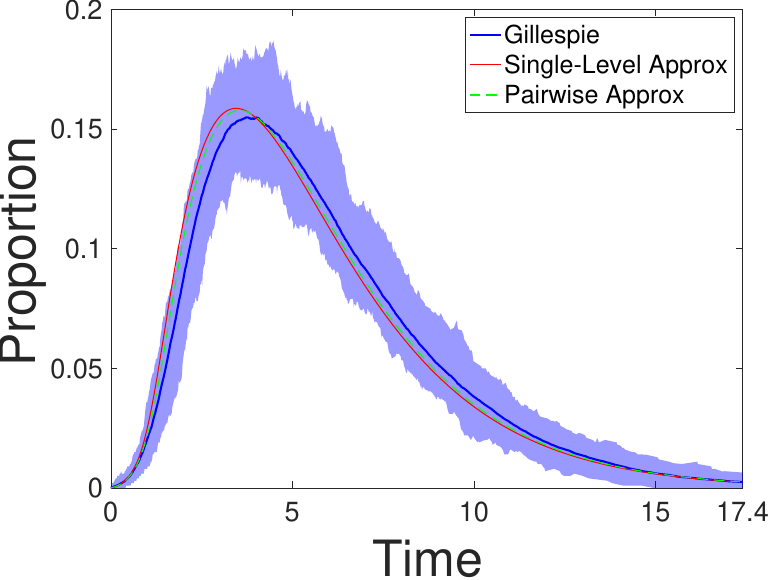}
        \caption{Offline $p=0.1$}%
        \label{fig:WSOffline2}
    \end{subfigure}
    \hfill
    \begin{subfigure}[b]{0.32\textwidth}  
        \centering 
        \includegraphics[width=\textwidth]{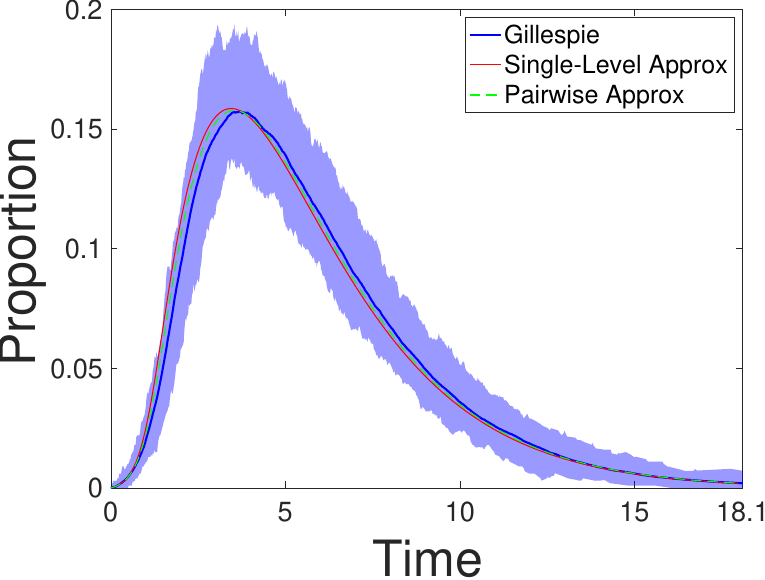}
        \caption{Offline $p=0.2$}%
        \label{fig:WSOffline3}
    \end{subfigure}
    \caption{Dynamics of online $E$ and offline $P$ on Watts-Strogatz networks with different rewiring probability $p$. The parameters are $N=1\,000, \langle k \rangle=10, \tau=0.2, \theta=0.01, \eta=0.2, \gamma_i=0.2, \gamma_p=0.5, t_{\text{max}}=200, [E](0)=20$, and $\text{iter}=100$. The pairwise approximation performs better than the single-level model, but not significantly, reflecting the limited advantage of higher-order mean-field approximations in networks with high clustering and low edge randomness.}
    \label{fig:WSEgs}
\end{figure}

While the synthetic networks discussed above capture certain structural features relevant to empirical networks, they do not incorporate the small-world effect commonly found in social systems, which motivates our inclusion of the Watts–Strogatz network for further experiments \cite{Epidemics_Networks}. The construction of such a random network works as follows: We begin with a symmetrically coupled ring graph where each node is connected to a certain number of its nearest neighbors, evenly on each side. We think of each edge as having a ``first" and ``second" node where the edge connects the first to the second in a clockwise sense. With a probability $p$, we independently rewire each edge. If an edge is rewired, we disconnect the two nodes originally connected and join the ``first" node to a randomly selected node in the network, which is not already its neighbor. Notice that for small $p$, the network is more highly clustered and the typical path length between two nodes is of order $N$. We expect the simulation of our approximated models in this scenario to be not as good since not only the clustering is high but also shortcuts are much rarer and so the model assumption that the neighbors are randomly selected is less valid. Example comparisons of the model dynamics in $E$ and $P$ on Watts–Strogatz networks are shown in Figure \ref{fig:WSEgs}.
In these simulations, the pairwise approximation shows better agreement with the stochastic mean than the single-level approximation. However, the improvement is modest compared to what we observed on other synthetic networks. This suggests that while the pairwise model captures some higher-order structural effects, its advantages are diminished in networks with strong clustering and limited randomness in connectivity.

\section{\label{app:heatmap_heterogeneous} Outbursts in heterogeneous models}

This appendix presents a discussion of online and offline outbursts in models defined on heterogeneous networks.

\begin{figure}[!htbp]
    \centering
    \begin{subfigure}[t]{0.49\textwidth}
        \centering
        \includegraphics[width=\linewidth]{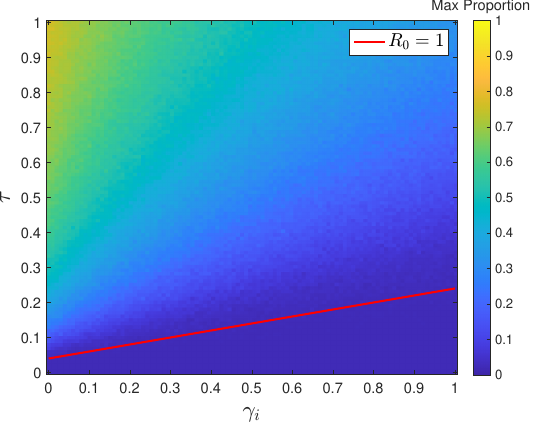} 
        \caption{Online $E$} \label{fig:HeterStoch_tau_gamma_i_E}
    \end{subfigure}
    \hfill
    \begin{subfigure}[t]{0.49\textwidth}
    \centering
        \includegraphics[width=\linewidth]{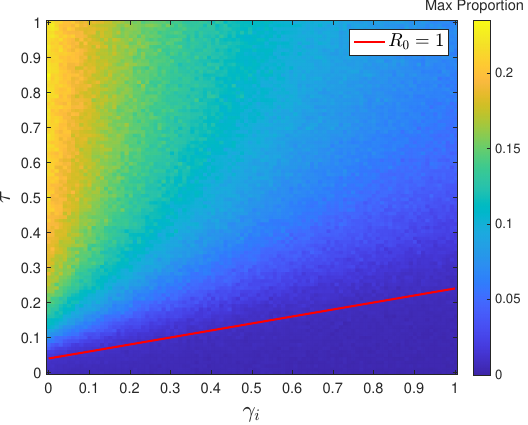} 
        \caption{Offline $P$} \label{fig:HeterStoch_tau_gamma_i_P}
    \end{subfigure}
    \caption{Heatmaps from stochastic simulations on Erd\H{o}s-R\'{e}nyi random networks for the proportions of mean final outburst sizes in $E$ and $P$, averaged over $\text{iter}=200$ runs for each parameter combination swept over $\tau$ and $\gamma_i$. Grid resolution is $100 \times 100$. Parameters are $N=1\,000, \langle k \rangle=5, \theta=0.01, \eta=0.2, \gamma_p=0.5, t_{\text{max}}=200$, and $[E](0)=20$. The analytical $R_0 = 1$ line is plotted and reflects the general trend of the threshold separating outburst regimes.}
    \label{fig:HeterStoch_tau_gamma_i}
\end{figure}

\begin{figure}[!htbp]
    \centering
    \begin{subfigure}[t]{0.49\textwidth}
        \centering
        \includegraphics[width=\linewidth]{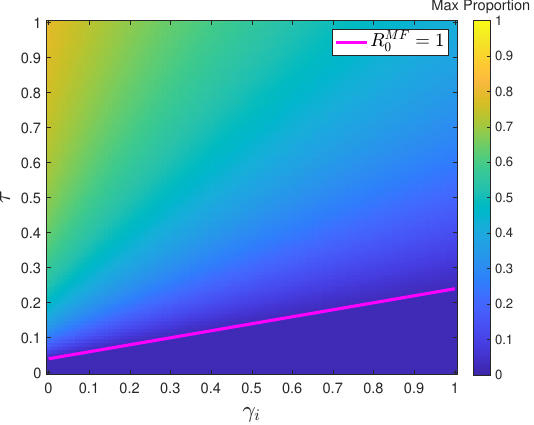} 
        \caption{Online $E$} \label{fig:HeterSingle_tau_gamma_i_E}
    \end{subfigure}
    \hfill
    \begin{subfigure}[t]{0.49\textwidth}
    \centering
        \includegraphics[width=\linewidth]{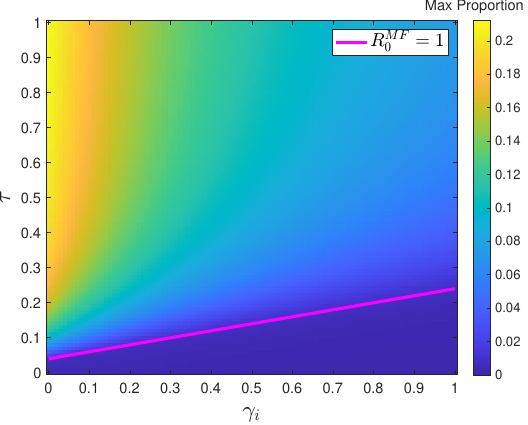} 
        \caption{Offline $P$} \label{fig:HeterSingle_tau_gamma_i_P}
    \end{subfigure}
    \caption{Heatmaps from simulations of the single-level approximation model on Erd\H{o}s-R\'{e}nyi random networks showing the proportions of mean final outburst sizes in $E$ and $P$ across a sweep of $\tau$ and $\gamma_i$. Grid resolution is $100 \times 100$. Parameters are $N=1\,000, \langle k \rangle=5, \theta=0.01, \eta=0.2, \gamma_p=0.5, t_{\text{max}}=200$, and $[E](0)=20$. The $R_0^{MF} = 1$ line from the single-level model on homogeneous networks is plotted. As in Figure~\ref{fig:HeterStoch_tau_gamma_i}, the threshold line in general captures the boundary between different outburst regimes.}
    \label{fig:HeterSingle_tau_gamma_i}
\end{figure}

\begin{figure}[!htbp]
    \centering
    \begin{subfigure}[t]{0.49\textwidth}
        \centering
        \includegraphics[width=\linewidth]{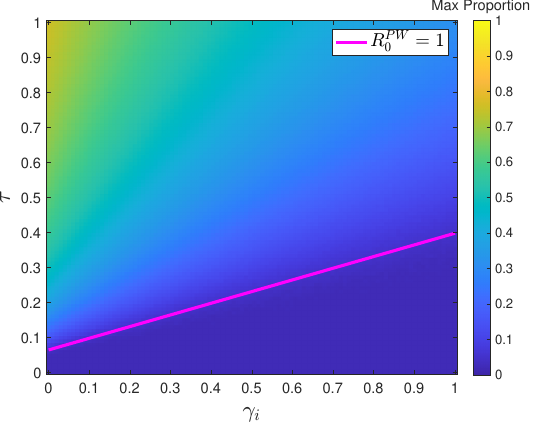} 
        \caption{Online $E$} \label{fig:HeterPairwise_tau_gamma_i_E}
    \end{subfigure}
    \hfill
    \begin{subfigure}[t]{0.49\textwidth}
    \centering
        \includegraphics[width=\linewidth]{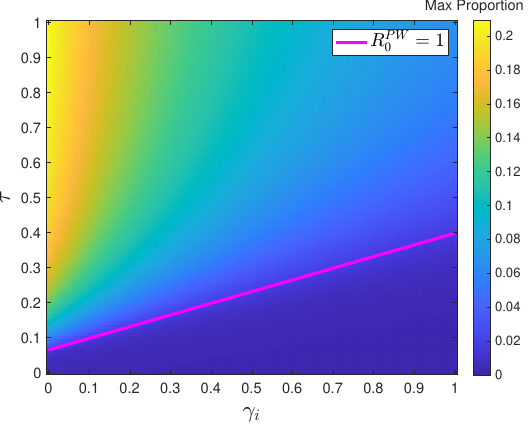} 
        \caption{Offline $P$} \label{fig:HeterPairwise_tau_gamma_i_P}
    \end{subfigure}
    \caption{Heatmaps from simulations of the pairwise approximation model on Erd\H{o}s-R\'{e}nyi random networks, showing the proportions of mean final outburst sizes in $E$ and $P$ across a sweep in $\tau$ and $\gamma_i$. Grid resolution is $100 \times 100$. Parameters are $N=1\,000, \langle k \rangle=5, \theta=0.01, \eta=0.2, \gamma_p=0.5, t_{\text{max}}=200$, and $[E](0)=20$. The $R_0^{PW} = 1$ line, derived from the pairwise model on homogeneous networks, is plotted. Compared to the $R_0 = 1$ and $R_0^{MF} = 1$ lines shown in Figures~\ref{fig:HeterStoch_tau_gamma_i} and~\ref{fig:HeterSingle_tau_gamma_i}, respectively, this threshold is steeper in the $\gamma_i$–$\tau$ plane but still provides a reasonably accurate classification of the outburst regimes.}
    \label{fig:HeterPairwise_tau_gamma_i}
\end{figure}

The reproductive number for models on heterogeneous networks is much more difficult to study analytically due to the explosion in the dimension of the system. However, numerical parameter sweeps are feasible, and the reproductive number expressions derived from the homogeneous models may remain informative to some extent. Here, we perform parameter sweeps over $\tau$ and $\gamma_i$ on Erd\H{o}s-R\'{e}nyi random networks using the stochastic model, single-level model, and the pairwise model, and plot the $R_0 = 1, R_0^{MF} = 1, R_0^{PW} = 1$ lines on them respectively. The results for the different models are shown in Figures \ref{fig:HeterStoch_tau_gamma_i}, \ref{fig:HeterSingle_tau_gamma_i}, and \ref{fig:HeterPairwise_tau_gamma_i}.

\bibliographystyle{plain}
\bibliography{bibfile}

\end{document}